\documentstyle[aps,prb,epsf,twocolumn]{revtex}
\begin{document}
\draft
\wideabs{
\title{Thermodynamics of the dissipative two-state system: a Bethe Ansatz
study}
\author
{T. A. Costi\cite{tac-email}}
\address
{Universit\"{a}t Karlsruhe, Institut f\"{u}r Theorie der Kondensierten
Materie, 76128 Karlsruhe, Germany}
\address{Theoretische Physik III, Universit\"{a}t Augsburg, D-86135
Augsburg, Germany\cite{tac-present-address}.}
\author{G. Zar\'and\cite{zar-email}}
\address
{
Research Group of the Hungarian Academy of Sciences, Institute of Physics,
TU Budapest, P.O. Box 91, H-1521 Hungary
}
\address
{Physics Department, University of California, Davis, 
1 Shields Avenue, 95616, U.S.A\cite{zar-present-address}}
\maketitle
\begin{abstract}
The thermodynamics of the dissipative two-state system is calculated exactly
for all temperatures and level asymmetries for the case of Ohmic dissipation.
We exploit the equivalence of the two-state system to the anisotropic Kondo
model and extract the thermodynamics of the former by solving the thermodynamic
Bethe Ansatz equations of the latter. The universal scaling functions for the
specific heat $C_{\alpha}(T)$ and static dielectric susceptibility 
$\chi_{\alpha}(T)$ are extracted for all dissipation strengths $0<\alpha<1$
for both symmetric and asymmetric two-state systems. 
The logarithmic corrections to these quantities at high temperatures are found
in the Kondo limit $\alpha\rightarrow 1^{-}$, whereas for $\alpha< 1$ we find
the expected power law temperature dependences with the powers being functions
of the dissipative coupling $\alpha$. The low temperature behaviour is always
that of a Fermi liquid.
\end{abstract}
}
\section{INTRODUCTION}

The low temperature properties of many physical systems, such as 
quantum tunneling between  flux states in a SQUID, \cite{Leggett84}, 
two-level atoms coupled to the electromagnetic field in quantum optics, 
\cite{LeClair}, or tunneling dislocations and point defects in 
solids\cite{disl} can be described by the dissipative two-state system model 
\cite{leggett.87,weiss.99}. All these systems
have the common feature that at low energies the subsystem 
investigated (the flux states of the SQUID, the atom, or the crystal 
defect in the  examples above) can  occupy only two distinct quantum states, 
which  are  coupled to a continuum of excitations 
(the electromagnetic field, the phonons or the conduction electrons), 
leading to the appearance of dissipation in the system 
\cite{caldeira+leggett}. 

In the dissipative two state system (DTSS)  model
the two distinct quantum states above are described in terms  of a
pseudospin $\sigma_i$, with $\sigma_z = \pm 1$ corresponding to the two
states of the subsystem. Generally, these two states have slightly 
different energy with an energy difference $\varepsilon$ (also called asymmetry
energy)  and the decoupled two-state system (TSS) can tunnel
between them with a tunneling amplitude $\Delta$. 
The heat bath is modeled by a continuum of independent quantum
oscillators with density $\varrho(\omega)$ coupled linearly
to $\sigma_z$ with a frequency dependent coupling $g(\omega)$. 
A detailed analysis shows that the dynamical properties of 
the two-state system are uniquely determined by the environment's 
spectral function,  
$J(\omega) \sim \varrho(\omega) g^2(\omega)\sim \omega^s$ \cite{weiss.99}.

Here, we only concentrate on Ohmic dissipation
corresponding to $s=1$, i.e. $J(\omega) \sim
\omega$. This includes, for instance, the important case of a tunneling defect
in a metal, where the low-energy bosonic excitations are 
electron-hole pairs close to the Fermi surface. These excitations 
have a linear dispersion, and in a first approximation their
coupling to the defect is energy-independent, leading to $J(\omega)
= 2\pi \alpha \omega$ for $\omega < \omega_c$ with $\omega_c$ a
high-energy cutoff in the model of the order of the Fermi energy
$E_F$. The dynamical behaviour of the TSS 
as a function of the coupling strength $\alpha$ has been the subject of
extensive studies during the past two decades and it  is 
well understood by now \cite{general-dynamics}. To distinguish between the
different cases it is useful to introduce the zero temperature 
spin correlation function $S(t) \equiv
\Im \langle\sigma_z(0)\sigma_z(t)\rangle$. 
{\it (a)} For $\alpha < 1/2 $  the TSS oscillates between the two
states $\sigma_z= \pm 1$ and $S(t)$ has an oscillatory behaviour.
However, the environment introduces 
some decoherence in the system, reflected in the 
exponential decay of the envelope of $S(t)$. It also renormalizes the
tunneling amplitude: $\Delta \to \Delta_r < \Delta$.
{\it (b)} In the parameter range
$1/2<\alpha < 1$ the coherent oscillations\cite{note-coherence} 
become completely
suppressed, and $S(t)$ shows an exponential behaviour without a change
of sign. {\it (c)} Finally, for $\alpha>1$ and a finite level asymmetry 
the TSS becomes localized in the lowest quantum state (at $T=0$)
\cite{localization}. In this case $S(t)$ tends to a {\it finite} 
value as $t\to \infty$. 
It is important to note that the localized   state
obtained is immediately  destroyed once  assisted tunneling or  assisted pair
tunneling is included in the DTSS model \cite{Zawa,MF}. 

\begin{figure}[t]
\centerline{\epsfysize 6.1cm 
{\epsffile{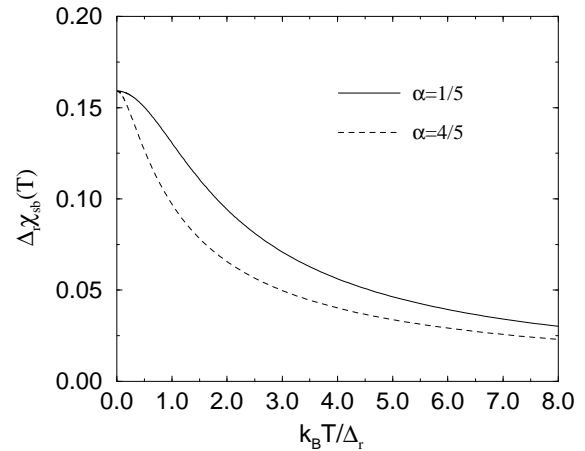}}}
\vspace{0.1cm}
\caption{
The dielectric susceptibility, $\chi_{sb}$, on a linear temperature scale
for the symmetric case at weak and strong dissipations.
}
\label{linear-scale-chi}
\end{figure}

In the present paper we study the {\it thermodynamics} of a
dissipative TSS model in the parameter range $0<\alpha < 1$.  
To this purpose we exploit a mapping between the spin anisotropic 
Kondo  model (AKM) describing an impurity spin coupled to the spin density of
the conduction electrons via an anisotropic exchange interaction
(see Sec.~IIb) and the dissipative TSS model, \cite{leggett.87,weiss.99}, 
and study  the Bethe Ansatz equations \cite{tsvelik.83} for the former 
both numerically and analytically. As  discussed in Sec.~\ref{ss:equiv} and
Appendix~\ref{app:equiv}, within this mapping the tunneling amplitude 
maps to the spin flip scattering amplitude, $\Delta \leftrightarrow 
J_\perp$, the asymmetry energy $\varepsilon$ corresponds to a
local magnetic field $h$ applied to the impurity spin, and the  
dissipation strength $\alpha$ is related to the coupling $J_z$ in the
Kondo model. It is very remarkable that the $\alpha$ values 
separating the three different regions of the DTSS model are mapped to
some special points in the parameter space of the Kondo model. 
The point $\alpha=1$ turns out to correspond to the case $J_z=0$, 
separating the ferromagnetic ($J_z < 0 \Leftrightarrow\alpha > 1$) and the 
antiferromagnetic ($J_z > 0 \Leftrightarrow \alpha < 1$) regimes in the 
Kondo model. While in the first case the Kondo model scales 
to a finite fixed point, in the second  the Kondo fixed point 
turns out to be at infinite coupling. The crossover to the 'strong
coupling' regime happens at the  so-called Kondo energy, $T_K$,\cite{Hewson,GrunerZawa}
which can be  identified with the  
renormalized tunneling amplitude in the DTSS model, $\Delta_r \sim T_K$.
The other special point, $\alpha = 1/2$, can be shown to be equivalent 
to the Toulouse line\cite{toulouse.69} of the AKM. Along this line the 
Bethe Ansatz (BA) equations simplify enormously, and the model 
can be described  by a simple resonant level model without interaction. 
\begin{figure}[t]
\centerline{\epsfysize 6.1cm 
{\epsffile{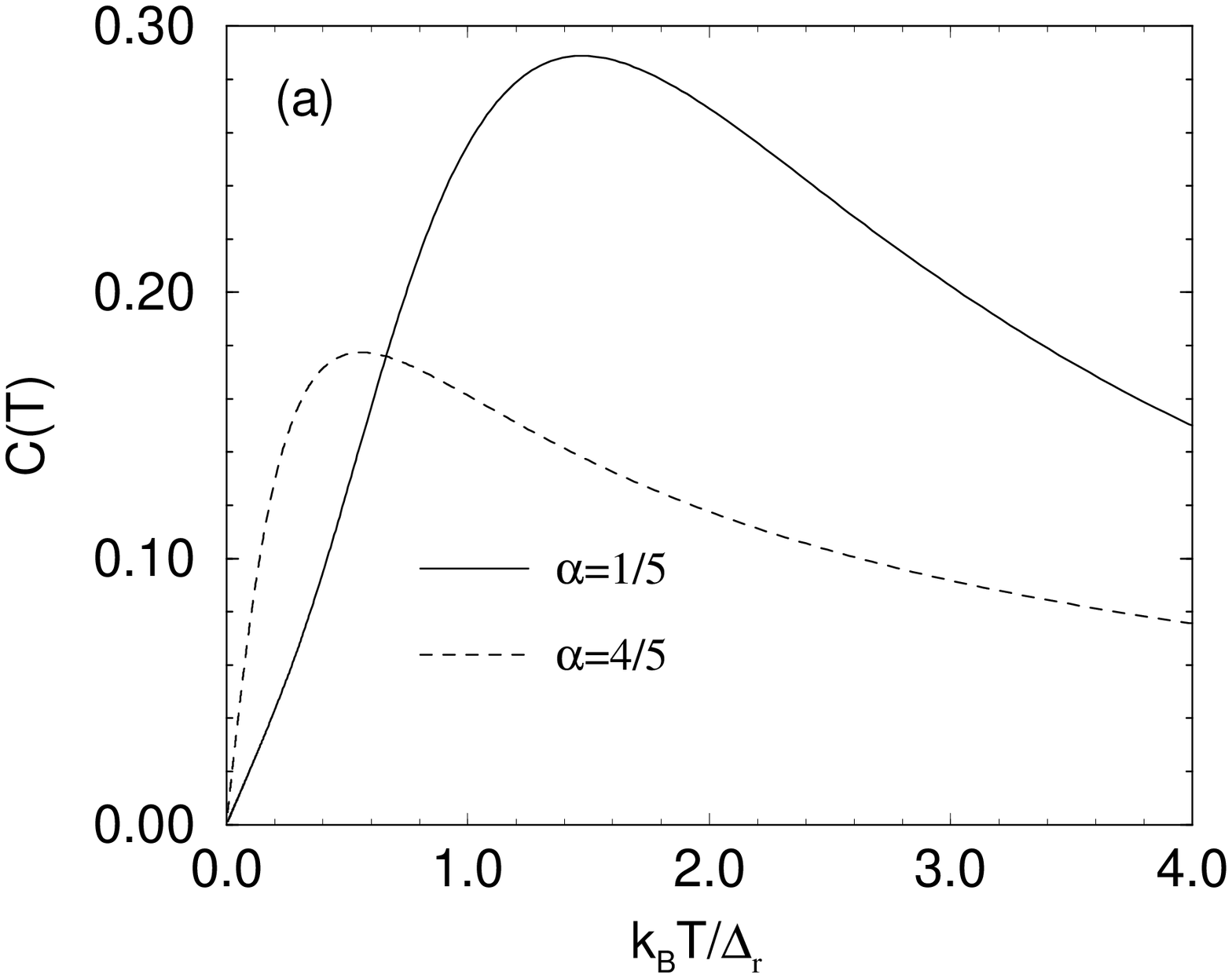}}}
\vspace{0.1cm}
\centerline{\epsfysize 6.1cm 
{\epsffile{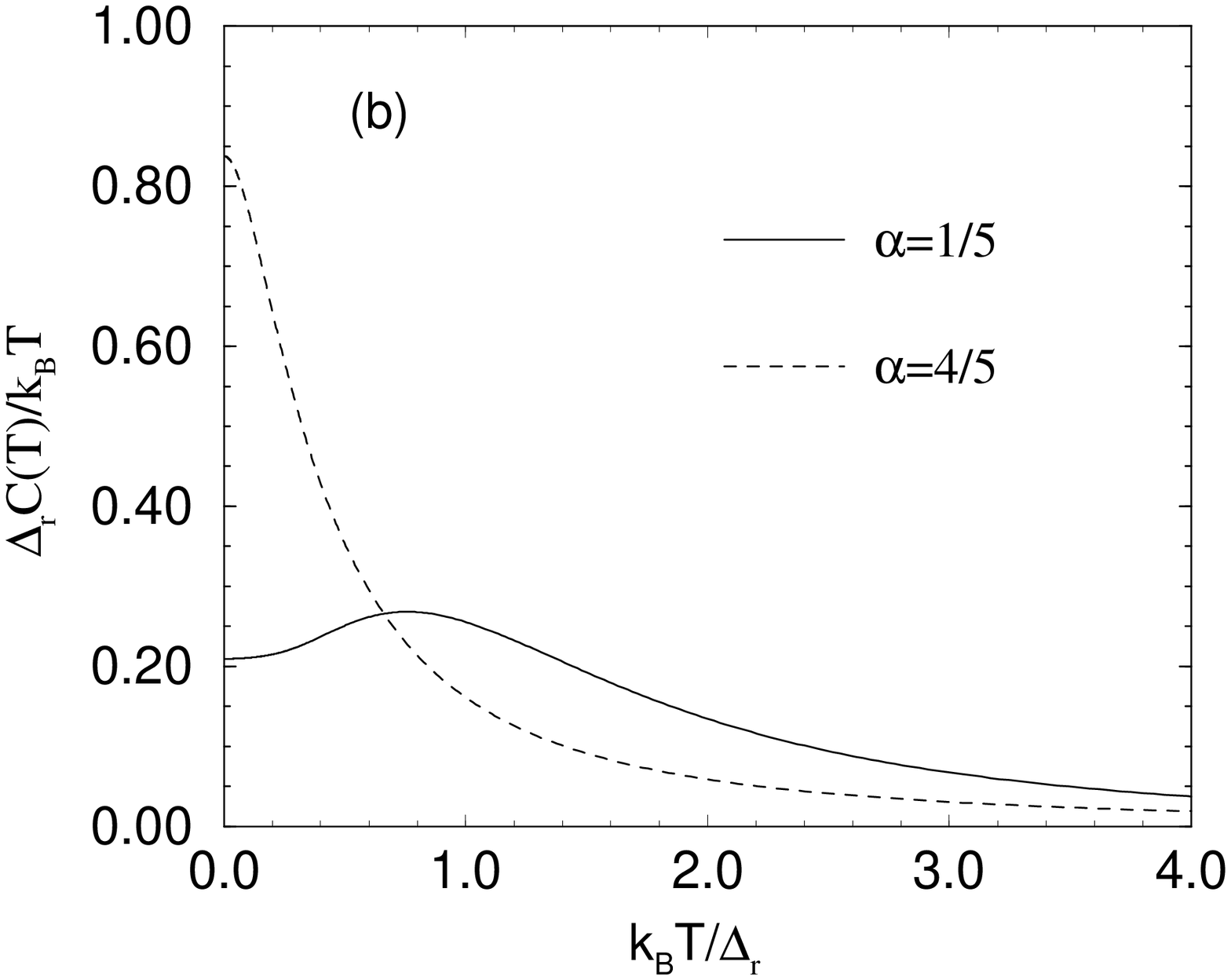}}}
\vspace{0.1cm}
\caption{(a) The specific heat $C(T)$
on a linear temperature scale for the symmetric case at weak and 
strong dissipations. The position of the maximum of the Schottky 
peak in $C(T)$ is of order $\Delta_{r}$ but
its exact location changes with $\alpha$, as does the peak height.
(b) The quantity $\Delta_{r}C(T)/k_{B}T$ shows non-monotonic behaviour
for weak dissipations, and monotonic behaviour for $\alpha \ge 1/3$.
}
\label{linear-scale-heats}
\end{figure}
We give a detailed description of the thermodynamics of the DTSS model, 
solving the BA equations for arbitrary asymmetry and temperature 
for both $\alpha > 1/2$ and $\alpha < 1/2$. 
For the sake of completeness and clarity, we also included a 
detailed analysis of the  two models to demonstrate, how the 
different concepts such as scaling, strong coupling limit, energy
scales,  etc. appear in the AKM and the  DTSS model.
As we shall see, there are clear indicators in the specific heat, $C(T)$, 
for distinguishing weak from strong dissipation limits. 
This is not evident in $C(T)$ directly, which shows a Schottky anomaly at 
$k_{B}T\sim \Delta_{r}$ for all dissipation strengths $\alpha<1$ as depicted
in Fig.\,\ref{linear-scale-heats}a . However, as seen in 
Fig.\,\ref{linear-scale-heats}b, and as we shall discuss in detail
later, the quantity $C(T)/T$, shows quite different behaviour at weak 
and strong dissipations. For dissipations $\alpha < 1/3$, 
with no asymmetry, $C(T)/T$ is found to have a peak at 
$k_{B}T\sim \Delta_{r}$ indicating the expected tendency towards
activated behaviour of the two-level system as $\alpha\rightarrow 0$.
A quite different behaviour is found for $\alpha\ge 1/3$ where we find
that $C(T)/T$ is monotonically decreasing with increasing temperature.
The tendency towards activated behaviour, signaled by a peak at approximately
$\sqrt{\Delta_{r}^{2}+\varepsilon^{2}}$ in $C(T)/T$, is also found at 
all dissipation strengths for sufficiently large (typically of order 
$\Delta_{r}$) asymmetries $\varepsilon$.
In the Kondo language this corresponds to the Zeeman splitting of the 
Kondo resonance due to a local magnetic field. In contrast, the 
dielectric susceptibility,
$\chi_{sb} = -\partial^{2} F/ \partial \varepsilon^{2}$, with $F(T)$ the 
two-level system free energy, shows
only a monotonically decreasing behaviour with increasing temperature 
for all dissipation strengths $0<\alpha<1$ in the symmetric case (see
Fig.\,\ref{linear-scale-chi} and for further details Sec.~\ref{sec-susc}). 
A peak in $\chi_{sb}(T)$ at finite temperature only appears for a 
sufficiently large level asymmetry. The low temperature behaviour corresponds
to that of a renormalized Fermi liquid at all $\alpha < 1$ with the 
renormalizations increasing with $\alpha$. Beyond these 
overall features we also discuss the detailed form of the universal 
scaling functions of the dissipative two-state system for all $\alpha<1$ 
and $\varepsilon$.

The only previous detailed studies of the thermodynamics of the 
Ohmic two-state system which we are aware of are, (a), the numerical 
renormalization group study \cite{costi.98} and, (b), the work 
of G\"{o}rlich and Weiss \cite{goerlich.88}. 
The latter authors used a path integral method \cite{leggett.87} to calculate
the partition function of the dissipative two-state system for both Ohmic
and non-Ohmic dissipation. Their results are restricted to weak level 
asymmetries $\varepsilon\ll \Delta_{r}$ and no results are presented for
the finite temperature dielectric susceptibility. For the Ohmic case
and $\alpha\ll 1$ they recover the linear $T$ behaviour
of the specific heat at low temperature $k_{B}T\ll \Delta_{r}$ and the
correct high temperature behaviour at all $\alpha$, however reliable results
for strong dissipation $1/2<\alpha<1$ could not be obtained within their
perturbative approach. The numerical renormalization group calculations in
\cite{costi.98} are non-perturbative and gave the specific heat accurately
for all temperatures at both weak and strong dissipations. A drawback of
this method, however, is the logarithmic discretization 
\cite{wilson.75,kww.80} of
the fermionic environment. This limits the ability of the method to resolve
finite temperature features, such as the peak in $C(T)/T$ at 
$k_{B}T\sim \Delta_{r}$ for $\alpha\ll 1$. The calculation of
the dielectric susceptibility by this method \cite{costi.98} is also 
problematical at sufficiently low temperatures $k_{B}T \leq \Delta_{r}$ 
\cite{problematical-nrg-chi}.
As explained in \cite{costi.98}, accurate results for $\chi_{sb}$ at $T=0$
required an analysis of the strong-coupling fixed point Hamiltonian 
together with the leading irrelevant deviations. Thus, this method
gave accurate results for $\chi_{sb}$ at $T=0$ (from the fixed point analysis)
and for $k_{B}T\ge \Delta_{r}$, but it was not possible within this method 
to determine equally accurately the bahaviour of $\chi_{sb}$ for 
$0<k_{B}T\leq \Delta_{r}$. As we shall see the Bethe Ansatz method we use
in this paper overcomes all the above difficulties.

The paper is organized as follows: in Sec.~\ref{models-sec} we introduce the
model of the dissipative two-state system and outline its equivalence to 
the anisotropic Kondo model for the case of Ohmic dissipation. 
Some implications of the Anderson-Yuval scaling picture
of the anisotropic Kondo model for the Ohmic two-state system are briefly
discussed. This gives a qualitative understanding of the physics of the 
latter in both the tunneling and localized regimes in terms of the fixed 
points, their stability and their associated low energy scales. Finally 
we show the connection between the scaling picture and the renormalization 
group flow obtained from the exact solution of the AKM via the Bethe Ansatz. 
The correspondence between the models via bosonization, described in
Appendix~\ref{app:equiv}, 
is then used to translate the thermodynamic Bethe Ansatz 
equations for the anisotropic Kondo model, derived by Tsvelik and Wiegman, 
into the language of the Ohmic two-state system in Sec.~\ref{tba-sec} 
for both strong dissipation, $\alpha>1/2$, (or weak anisotropy in 
the Kondo model) and weak dissipation, $\alpha < 1/2$ 
(or large anisotropy in the Kondo model) and at any level 
asymmetry $\varepsilon$ (or local magnetic field in the Kondo model).
Analytic results are then presented for the specific heat and dielectric
susceptibility of the two-state system at high and low temperatures and
arbitrary dissipation in Sec.~\ref{sec-asymptotic} and at all 
temperatures at the Toulouse point ($\alpha=1/2$) in Sec.~\ref{sec:Toulouse}.
The Wilson ratio for the Ohmic two-state system is discussed
in Sec.~\ref{Wilsonr}. Sec.~\ref{num-sec} gives the numerical solution of the 
thermodynamic Bethe Ansatz equations at all temperatures for both weak 
and strong dissipation and for both symmetric and asymmetric two-level 
systems. Our conclusions are summarized in Sec.~\ref{sec-conclusions}. 
Appendix~\ref{tba-derivation} contains
some details on the Bethe Ansatz solution of the AKM and the corresponding
thermodynamic Bethe Ansatz (TBA) equations which we solved in this paper.
Appendix~\ref{num-procedure} 
gives details of the numerical procedure used to solve the TBA
equations and Appendix~\ref{wd-univ-eq} contains the 
universal TBA equations for weak
dissipation (large anisotropies in the AKM), with some corrections made to
those found originally in Ref.~\onlinecite{tsvelik.83}

\section{MODELS}
\label{models-sec}
\subsection{The dissipative two-state system}
\label{sec-dtss}
The model of the dissipative two-state system is given by
\begin{eqnarray}
H_{SB} & = &-\frac{1}{2}\hbar\Delta \sigma_{x}+\frac{1}{2}\varepsilon\sigma_{z}
        +\sum_{i} \omega_{i}(a_{i}^{\dagger}a_{i}+\frac{1}{2})\nonumber\\
        &+&\frac{1}{2}q_{0}\sigma_{z}\sum_{i}
\frac{C_{i}}{\sqrt{2m_{i}\omega_{i}}}(a_{i}+a_{i}^{\dagger})\label{eq:SB}.
\end{eqnarray}
Here $\sigma_{i},i=x,y,z$ are Pauli spin matrices, the two states of the
system correspond to $\sigma_{z}=\uparrow$ and $\sigma_{z}=\downarrow$.
$\Delta$ is the bare tunneling matrix element and $\varepsilon$ is a bias.
The environment is represented by an infinite set of 
harmonic oscillators (labeled by the index $i$) 
with masses $m_{i}$ and 
frequency spectrum $\omega_{i}$ coupling
linearly to the coordinate $Q=\frac{1}{2}q_{0}\sigma_{z}$ 
of the two-level system via a
term characterized by the couplings $C_{i}$.
The environment spectral
function is given in terms of these couplings, oscillator masses 
and frequencies by
 $J(\omega)=\frac{\pi}{2}
\sum_{i}(\frac{C_{i}^{2}}{m_{i}\omega_{i}})
\delta(\omega-\omega_{i})$.
In the case of an Ohmic heat bath, of interest to us here, we have 
$J(\omega)=2\pi\alpha\omega$, for $\omega << \omega_{c}$, where $\omega_{c}$
is a high energy cut-off and $\alpha$ is a dimensionless parameter 
characterizing the strength of the dissipation. 
The Ohmic two-state model (also called the Ohmic spin-boson model) 
has been intensively studied (for reviews we 
refer the reader to \cite{leggett.87,weiss.99}). The model has
a low energy scale, $\Delta_r<\Delta$ for $\Delta << \omega_c$, 
which depends on the dissipation
strength $\alpha$, and which may be interpreted as a renormalized tunneling
{\em amplitude}. For $\alpha=0$ the two-level system is decoupled from the 
environment and $\Delta_r=\Delta$, whereas with increasing coupling to the 
environment, this energy scale is strongly renormalized:
$\Delta_{r}/\omega_c \sim (\Delta/\omega_c)^{1/(1-\alpha)}$.
Another scale, the frequency of tunneling oscillations, 
$\Omega(\alpha,\Delta_{r})=Q(\alpha)\Gamma(\alpha,\Delta_{r})$, is relevant
for time dependent quantities. 
Here $\Gamma(\alpha,\Delta_{r})\sim \Delta_{r}$ is the
decay rate and $Q(\alpha)=\cot(\frac{\pi}{2}\frac{\alpha}{1-\alpha})$ is the 
quality factor of the oscillations \cite{leggett.87}. The latter vanishes 
at the Toulouse point $\alpha=\frac{1}{2}$, where the tunneling oscillations 
vanish (the ``coherence-decoherence'' crossover). 
For $0<\alpha<1/2$ the dynamics corresponds to damped 
oscillations of frequency $\Omega(\alpha,\Delta_{r})$
\cite{leggett.87,weiss.99,lesage.98}. 
This is sometimes called the ``coherent''
\cite{note-coherence} regime. The system exhibits phase coherence throught
this regime, albeit with damped oscillatory contributions to 
real time dynamical quantities. A smooth crossover to 
``incoherent''behaviour occurs at $\alpha=1/2$. The
tunneling amplitude remains finite in the ``incoherent'' 
regime $1/2\le \alpha < 1$, but there is no phase coherence in time dependent
dynamical quantities: $\Omega(\alpha,\Delta_{r})=0$ for $\alpha\ge 1/2$. 
Another physically relevant value of the dissipation
strength is $\alpha=1/3$, where an inelastic peak, present
in the neutron scattering cross-section for $\alpha<1/3$, vanishes and gives 
rise to a quasielastic peak for $\alpha>1/3$ 
\protect{\cite{costi.96,lesage.96,voelker.98}} (see also the discussion in
Sec.~\ref{sec-symmetric-heats}). Finally, for sufficiently strong
dissipation $\alpha\rightarrow 1^{-}$, the 
renormalized tunneling amplitude vanishes
giving rise to the phenomenon of ``localization'' or ``self-trapping'' for 
$\alpha > \alpha_{c}\approx 1$ ($\alpha_{c}$ depends also on the precise 
value of $\Delta$). In this paper we will be interested only in the 
thermodynamics of the dissipative two-state system. For such quantities
the exact solution shows that, in the tunneling regime 
($0<\alpha<1$) \cite{note-tunneling}, the only relevant scale 
is $\Delta_{r}$. 

\subsection{Equivalence to the anisotropic Kondo model}
\label{ss:equiv}
The equivalence of the Ohmic two-state system to the anisotropic Kondo 
model (AKM) has been shown at the Hamiltonian level via bosonization 
\cite{guinea.85b} as outlined in Appendix~\ref{app:equiv}.
This equivalence was believed to be valid in the region $\alpha > 1/2$, which 
corresponds (see below for the precise statement of the equivalence) 
to the region in the parameter space of the AKM between weak-coupling 
($\rho J_{\parallel} \ll 1$) and the Toulouse point 
($\rho J_{\parallel}\approx 1$). 
Recent work \cite{costi.96} shows 
that the equivalence extends beyond the Toulouse point into the region 
describing weak dissipation $0<\alpha<1/2$ 
(or large antiferromagnetic $J_{\parallel}$ in the AKM, see also 
\cite{kotliar.96}). 
The AKM is given by \cite{anderson.70}
\begin{eqnarray}
H &=& \sum_{k,\sigma} \varepsilon_{k}c_{k\sigma}^{\dagger}c_{k\sigma} + 
\frac{J_{\perp}}{2}\sum_{kk'}
        (c_{k\uparrow}^{\dagger}c_{k'\downarrow}S^{-} +
         c_{k\downarrow}^{\dagger}c_{k'\uparrow}S^{+})\nonumber\\
  &+& \frac{J_{\parallel}}{2}\sum_{kk'}
         (c_{k\uparrow}^{\dagger}c_{k'\uparrow} -
          c_{k\downarrow}^{\dagger}c_{k'\downarrow})S^{z} 
+ g\mu_{B}hS_{z}.\label{eq:AKM}
\end{eqnarray}
The first term represents non-interacting conduction electrons and the
second and third terms represent an exchange interaction between a localized
spin $1/2$ and the conduction electrons with strength 
$J_{\perp},J_{\parallel}$. The last term in Eq.~(\ref{eq:AKM})
is a local magnetic field, $h$, coupling only 
to the impurity spin. The correspondence 
between $H$ and $H_{SB}$, outlined in Appendix~\ref{app:equiv}, requires that
\begin{eqnarray}
\varepsilon&=&g\mu_{B}h\\
\frac{\Delta}{\omega_{c}}&=& \rho J_{\perp}\\
\alpha&=&(1+ \frac{2 \delta}{ \pi})^{2},
\end{eqnarray}
where $\tan{\delta}= -\frac{ \pi \rho J_{\parallel}}{4}$, $\delta$ 
is the phase shift for scattering of electrons from a 
potential $J_{\parallel}/4$ and $\rho=1/2D$ is the conduction 
electron density of states per spin at the Fermi level for a 
flat band of width $2D=\omega_{c}$ \cite{leggett.87,costi.96}. 
We note that weak dissipation ($\alpha\rightarrow 0$) in the Ohmic two-state
model corresponds to large antiferromagnetic coupling 
($J_{\parallel}\rightarrow\infty$) in the Kondo model whereas dissipation 
strength $\alpha>1$ in the Ohmic two-state model corresponds to ferromagnetic
coupling $J_{\parallel}<0$ in the Kondo model.

\subsection{Renormalization group flow}
\label{sec-rgflow}
The renormalization group flow of the Ohmic two-state system can be obtained
by making use of the above equivalence and the Anderson-Yuval scaling 
equations\cite{anderson.70} for the AKM. These equations hold to lowest 
order in $\rho J_{\perp}$ but for all $\rho J_{\parallel}$: 
$-\infty<\rho J_{\parallel}<+\infty$. They are therefore valid for all
$0\le \alpha \le 4$ provided $\Delta/\omega_{c}\ll 1$. The Anderson-Yuval 
scaling equations extend the validity of the well known Poor Man's scaling 
equations to the whole $J_{\parallel}$ axis and reduce to those when
$\rho J_{\parallel}\ll 1$. In terms of the dimensionless quantities 
$\rho J_\perp$ and 
\begin{equation}
\tilde{\epsilon}=-8\frac{\delta}{\pi} 
(1+\frac{\delta}{\pi})
\end{equation} where $\delta$ was defined above, 
the Anderson-Yuval scaling equations read
\cite{anderson.70,leggett.87}
\begin{eqnarray}
\frac{d\tilde{\epsilon}}{d\ln D} & = & (\tilde{\epsilon}-2)(\rho J_{\perp})^{2} 
+ {\cal O}(\rho J_{\perp})^{4}
\nonumber\\
\frac{d\rho J_{\perp}}{d\ln D} & = & -\frac{\tilde{\epsilon}}{2}\rho J_{\perp} 
+ {\cal O}(\rho J_{\perp})^{3}
\label{eq.YuvalAnd}
\end{eqnarray}
By using the correspondence between the models given above, and noting that 
$\tilde{\epsilon}=2(1-\alpha)$, we obtain the following scaling equations for
the Ohmic two-state system
\begin{eqnarray}
\frac{d\alpha}{d\ln \omega_{c}} & = & \alpha(\frac{\Delta}{\omega_{c}})^{2} 
+ {\cal O}(\frac{\Delta}{\omega_{c}})^{4}\label{eq:scaling1}\\
\frac{d(\Delta/\omega_{c})}{d\ln \omega_{c}} & = & -(1-\alpha)
(\frac{\Delta}{\omega_{c}}) + {\cal O}(\frac{\Delta}{\omega_{c}})^{3}\label{eq:scaling2}
\end{eqnarray}
Note that in these equations $\alpha$ and  $\Delta$ are running variables 
which are functions of the running cut-off, $\omega_{c}$. The equations have 
to be supplemented by specifying initial conditions
\begin{eqnarray*}
\alpha(\omega_{c}&=&\omega_{0})=\alpha_{0}\nonumber\\
\Delta(\omega_{c}&=&\omega_{0})=\Delta_{0}.
\end{eqnarray*}
where $\alpha_{0},\Delta_{0},\omega_{0}$ are now the parameters appearing 
in the bare Hamiltonian (where they appeared 
as $\alpha,\Delta,\omega_{c}$).
We shall use this notation for the remainder of this section.

\begin{figure}[t]
\centerline{\epsfysize 6.1cm {\epsffile{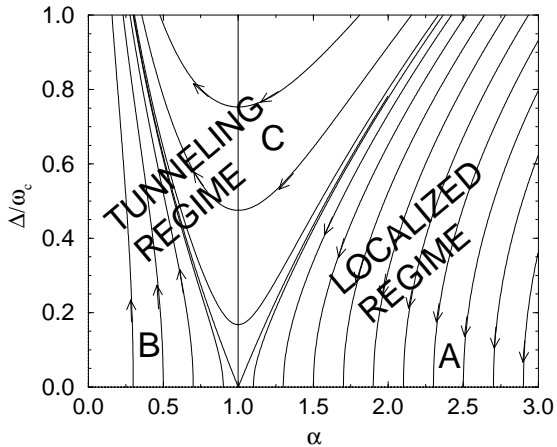}}}
\vspace{0.1cm}
\caption{
The scaling trajectories of the Ohmic two-state system obtained 
from the Anderson-Yuval scaling equations for the AKM. Only the region 
$0<\alpha<3$ is shown. The left and right separatrices at 
$\alpha=1,\Delta/\omega_{c}=0$ define the regions labeled $A$, $B$ and $C$ and
the arrows indicate the direction of decreasing $\omega_{c}$.
}
\label{scaling-trajectories}
\end{figure}
From (\ref{eq:scaling1}--\ref{eq:scaling2}) there is a line
of fixed points at $\Delta_{0}/\omega_{0}=0$ for $\alpha_{0} \ge 0$. Their 
stability to a finite $\Delta_{0}/\omega_{0}$ 
follows from (\ref{eq:scaling2}), 
which states that $\Delta/\omega_{c}$ is relevant, marginal or irrelevant 
depending on whether the dissipation strength $\alpha_{0}$ is less than, 
equal to or larger than 1 \cite{note1}. Hence, the line of fixed points 
at $\Delta_{0}/\omega_{0}=0$ for $\alpha_{0}>1$ are stable low energy fixed
points, whereas the line of fixed points at $\Delta_{0}/\omega_{0}=0$ for 
$\alpha_{0}\le 1$ are unstable high energy fixed points. The scaling 
trajectories can be calculated by dividing the
two equations (\ref{eq:scaling1}--\ref{eq:scaling2}) and integrating 
the resulting equation from $\omega_{0}$ down to $\omega_{c}$:
\begin{equation}
\frac{1}{2}\left[(\frac{\Delta}{\omega_{c}})^{2}-
(\frac{\Delta_{0}}{\omega_{0}})^{2}\right] = -((\ln\alpha - \alpha) - 
(\ln \alpha_{0}-\alpha_{0}))\label{eq:trajectories}
\end{equation}
They are shown in Fig.(\ref{scaling-trajectories}). 
The arrows indicate the direction of decreasing 
$\omega_c$. When the flow is to strong coupling, the scaling trajectories
will be quantitatively correct only for $\Delta/\omega_{c}\ll 1$.
The scaling diagram is divided into three regions
by two separatrices meeting at $\alpha=1,\Delta/\omega_{c}=0$.

The regime A ($\alpha_{0}>1$) corresponds to the localized regime
of the Ohmic two-state system (or the ferromagnetic sector of the
AKM). The dimensionless tunneling amplitude $\Delta/\omega_{c}$ is 
irrelevant and the flow is to a line of fixed 
points $(\alpha=\alpha^{\ast}$ and $\Delta/\omega_{c}=0)$. 
This case is easily analyzed since the scaling 
equations remain valid as $\omega_{c}\rightarrow 0$. Since $\Delta/\omega_{c}$ 
decreases as $\omega_{c}\rightarrow 0$, it follows from (\ref{eq:scaling1}) 
that $\alpha$ remains almost unrenormalized: 
$\alpha\rightarrow \alpha^{*}\approx \alpha_{0}$ as $\omega_{c}\rightarrow 0$.
Integrating (\ref{eq:scaling2}) gives a renormalized tunneling amplitude 
$\Delta_{r}\equiv\Delta(\omega_{c})=
\Delta_{0}(\omega_{c}/\omega_{0})^{\alpha_{0}}$
which vanishes at $T=0$ at low energies. Quantum mechanical tunneling is
absent for $\alpha_{0}>1$ at $T=0$ and for sufficiently small 
$\Delta_{0}/\omega_{0}$. 
\begin{figure}[t]
\centerline{\epsfysize 6.1cm {\epsffile{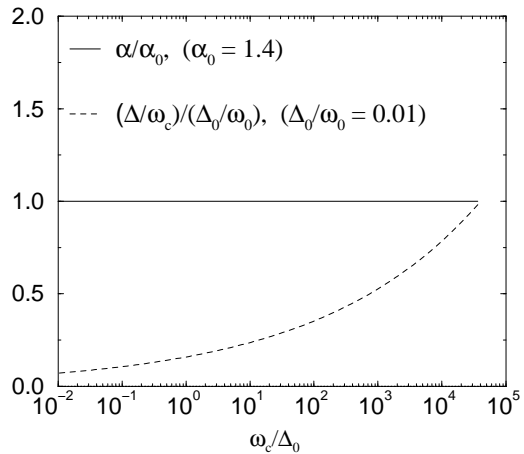}}}
\vspace{0.1cm}
\caption{Flow of running coupling constants in the localized
region $A$ of Fig.\,\ref{scaling-trajectories}, for $\alpha_{0}=1.4$ 
and $\Delta_{0}/\omega_{0}=0.01$.
}
\label{running-couplings-localized}
\end{figure}

At finite temperature the low energy
cut-off $\omega_{c}$ is replaced by $k_{B}T$ resulting in the well known 
temperature temperature dependent tunneling amplitude
$\Delta_{r}(T)=\Delta_{0}(k_{B}T/\omega_{0})^{\alpha_{0}}$ in the strong 
dissipation limit.
Fig.\,\ref{running-couplings-localized}
shows the flow of the dimensionless coupling constants for a typical 
case in the localized regime. These were obtained by 
integrating (\ref{eq:scaling1}--\ref{eq:scaling2}) using the Runge-Kutta 
algorithm for 1st order differential equations. 

\begin{figure}[t]
\centerline{\epsfysize 6.1cm {\epsffile{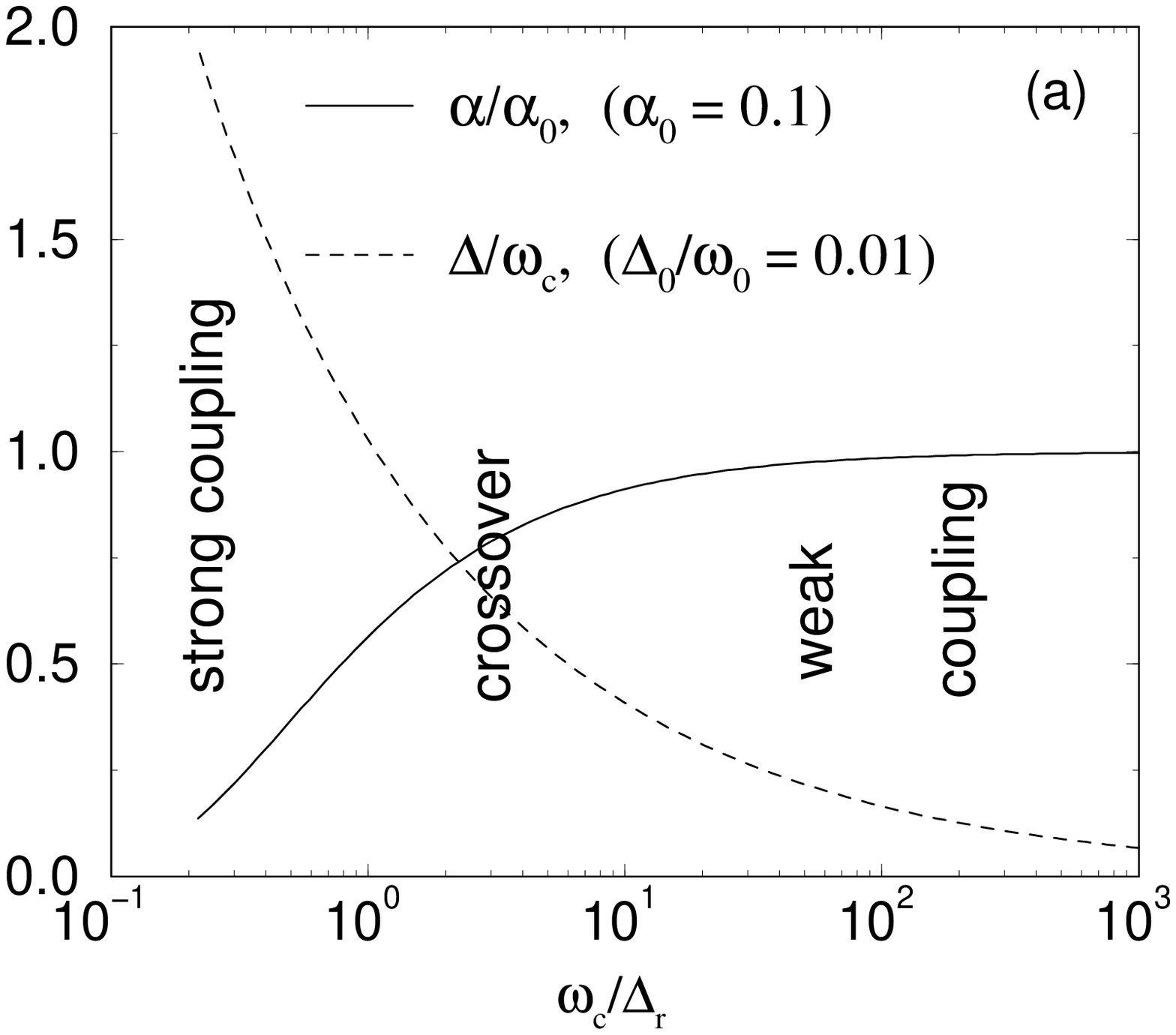}}}
\vspace{0.1cm}
\centerline{\epsfysize 6.1cm {\epsffile{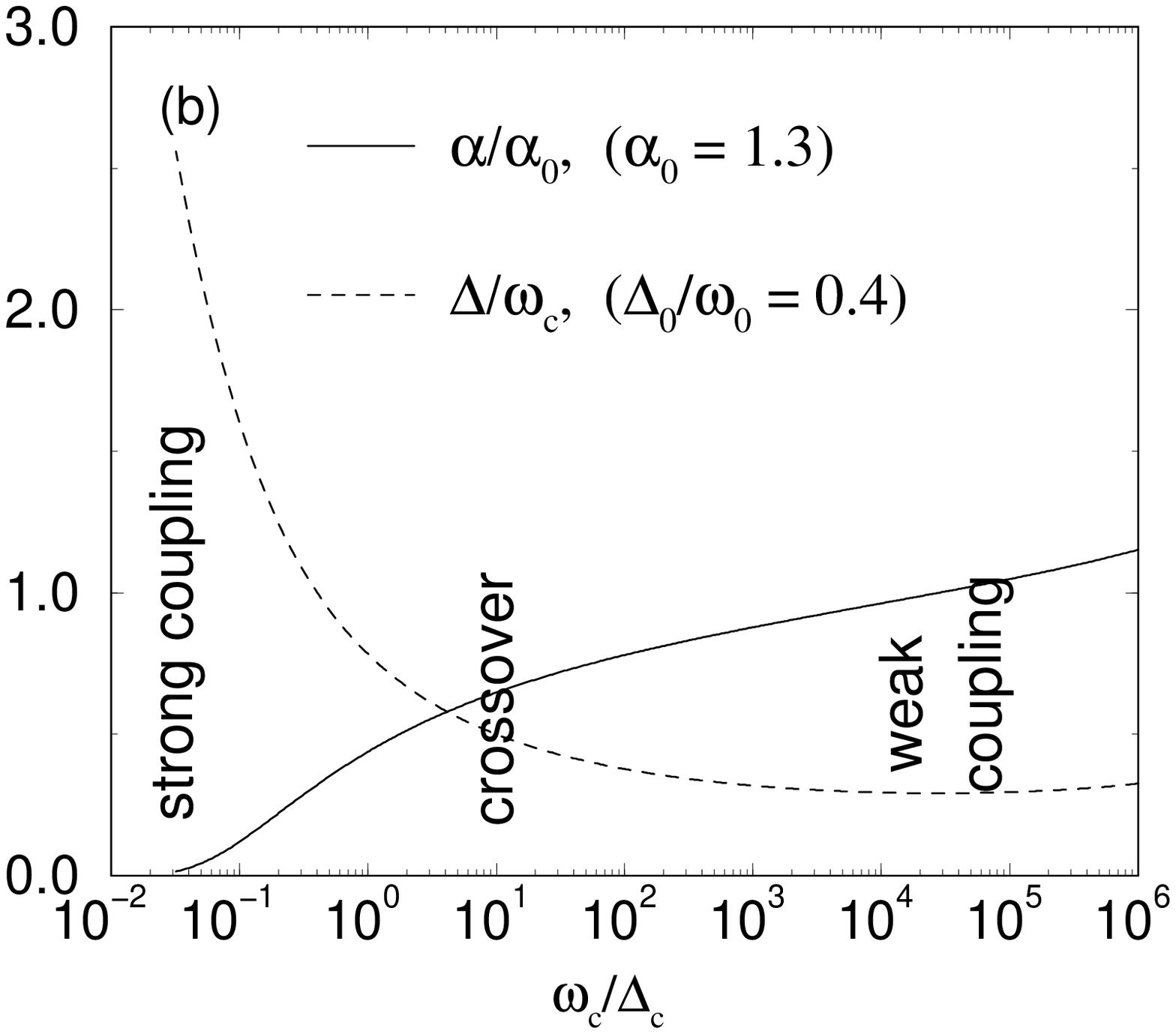}}}
\vspace{0.1cm}
\caption{Flow of running coupling constants for (a) $\alpha_{0}=0.1$ 
and $\Delta_{0}/\omega_{0}=0.01$ corresponding to the tunneling 
region $B$ in Fig.\,\ref{scaling-trajectories},
and (b) $\alpha_{0}=1.3$ and $\Delta_{0}/\omega_{0}=0.4$ 
corresponding to the tunneling region $C$ in Fig.\,\ref{scaling-trajectories}.
In (b), $\Delta_{c}$ is a crossover scale at which $\Delta$ becomes of O(1). 
}
\label{running-couplings-tunneling}
\end{figure}

The regime B ($\alpha<1$) corresponds to the 
tunneling regime of the Ohmic two-state system (or the antiferromagnetic 
sector of the AKM): in this regime $\Delta/\omega_{c}$ is relevant
and the flow for $0< \alpha_{0}\le 1$ is away from the line of high 
energy fixed points at $\Delta/\omega_{c}=0$ towards the strong coupling 
fixed point at $\alpha=0$ and $\Delta/\omega_{c}=\infty$. 
This is shown in the numerical solution of 
(\ref{eq:scaling1}--\ref{eq:scaling2}), in 
Fig.\,\ref{running-couplings-tunneling}a.
The scaling analysis, of course, breaks down when  $\Delta/\omega_{c}=O(1)$,
however, other methods, such as the numerical renormalization group and 
the Bethe Ansatz, show that the low energy fixed point is at 
$\Delta/\omega_{c}=\infty$ and $\alpha=0$. In this regime,
$\Delta(\omega_c)$ tends to a finite renormalized tunneling amplitude, 
$\Delta_{r}$ as $\omega_{c}\rightarrow 0$. In the AKM this low energy scale
is the Kondo scale, generalized to the anisotropic case.
We can estimate the $T=0$ renormalized tunneling amplitude as the crossover 
scale separating weak ($\Delta/\omega_{c}\ll1$) and strong coupling 
($\Delta/\omega_{c}\gg1$) regimes of the model. 
Define $\Delta_{r}=\Delta(\tilde{\omega}_{c})$ where $\tilde{\omega}_{c}$ is
the crossover scale such that 
$\Delta(\tilde{\omega}_{c})/\tilde{\omega}_{c}=1$.
Integrating (\ref{eq:scaling2}) down to this crossover scale 
\begin{equation}
\int_{1}^{\frac{\Delta_{0}}{\omega_{0}}} 
\frac{d(\Delta/\omega_{c})}{\Delta/\omega_{c}} = 
-\int_{\tilde{\omega}_{c}}^{\omega_{0}}(1-\alpha) d\ln \omega_{c}\nonumber
\end{equation}
and approximating $\alpha$ by $\alpha_{0}$ over this energy range gives
\begin{equation}
\Delta_{r}/\omega_{0}=
(\Delta_{0}/\omega_{0})^{\frac{1}{1-\alpha_{0}}},
\end{equation}
the correct low energy scale for the Ohmic two-state system, up to 
prefactors depending on $\alpha_{0}$. 

Finally, there is a region C, of strong dissipation, which also corresponds
to the tunneling regime of the Ohmic two-state system. However, the flow
to the strong coupling fixed point is such that, for $\alpha_{0}>1$, 
$\Delta/\omega_{c}$ initially decreases with decreasing $\omega_{c}$, signaling
a tendency to localization for strong dissipation, but eventually, as a 
result of a strong renormalization of $\alpha$ to below 1, $\Delta/\omega_{c}$ 
becomes relevant and then increases to strong coupling.
The flow is to the same strong coupling fixed point as for region B and
therefore region C belongs to the tunneling regime of the model.
Due to the strong renormalization of $\alpha$, 
Fig.\,\ref{running-couplings-tunneling}b, it is difficult 
to estimate the form of the low energy scale $\Delta_{r}$ in this case.

The renormalization group flow described above consists of a one parameter 
family of scaling trajectories labeled by a parameter $C=C(\alpha,\Delta)$
which takes a constant value along each trajectory. This constant of the 
motion is called a scaling invariant and can be found from 
(\ref{eq:trajectories}):
\begin{equation}
C(\alpha,\Delta)=-1-\frac{1}{2}(\frac{\Delta}{\omega_{c}})^{2}
-(\ln\alpha - \alpha)\label{eq:constant-of-motion}
\end{equation}
Corresponding to this one parameter family of scaling trajectories we expect
the scaling functions for physical quantities to consist of a 
one parameter family labeled by $C$, with different scaling functions for each
scaling trajectory. The scaling invariant $C$ is not unique but depends on the
cut-off scheme of the theory. A scaling trajectory may 
be specified differently (i.e. by a different function $C$) depending on 
the cut-off scheme . One then has to identify the scaling trajectories 
within the different schemes if one wishes to compare results. 
The Bethe Ansatz solution of the AKM, which 
we want to use in the next section, is an example where such a different 
scheme is used. We state here how we identify the scaling trajectories of 
the Bethe Ansatz solution with those we discussed above for the Ohmic 
two-state system (or equivalently the AKM with a finite bandwidth 
$2D=\omega_{c}$) and leave the details to Appendix~\ref{tba-derivation}. 
As discussed in the appendix, the Bethe Ansatz solution 
yields a renormalization
group flow depending on two functions, $\mu$ and $f$, of the dimensionless
couplings of the AKM. The function $\mu$ is the scaling invariant 
and specifies the scaling trajectories and the function $f$ sets the 
low energy scale 
\begin{equation}
T_{K}={2D}\exp(-f/(\mu/\pi)). \label{eq:tk-definition}
\end{equation}
Comparing  the high-temperature behaviour of the two models  we find  
the correspondence 
\begin{equation}
\mu/\pi = 1-\alpha_{0}\;.
\end{equation}  
We shall show in Sec.~\ref{Wilsonr} that the static susceptibility, 
$\chi_{BA}$, calculated from the Bethe Ansatz solution is equal
to the dielectric susceptibility, $\chi_{sb}$, of the Ohmic two-state
system and that, at $T=0$, these are related to the Kondo temperature, as 
defined above, by
\begin{equation}
\chi_{BA}=\chi_{sb}=
\frac{1}{2\pi(1-\mu/\pi)T_{K}}=\frac{1}{2\pi\alpha_{0}T_{K}}.
\label{eq:chisb-tk}
\end{equation}
This suggests that we define the renormalized tunneling 
amplitude $\Delta_{r}$ in terms of $T_{K}$ by \cite{note-factor-alpha}
\begin{equation}
\Delta_{r} = \alpha_{0}T_{K}\label{eq:deltar-tk}
\end{equation}
so that the local $T=0$ dielectric susceptibility, $\chi_{sb}$, of the Ohmic 
two-state system is given in terms of $\Delta_{r}$ by:
\begin{equation}
\chi_{sb}=\frac{1}{2\pi\Delta_{r}}.\label{eq:chi-deltar}
\end{equation}
The above relations fix $f$ in terms of $\mu$ and $T_{K}$ 
(or equivalently in terms of $\alpha_{0}$ and $\Delta_{r}$) and will prove 
useful in translating the Bethe Ansatz results of Sec III into results 
for the Ohmic two-state system. 

\section{THERMODYNAMIC BETHE ANSATZ EQUATIONS}
\label{tba-sec}
\subsection{Thermodynamic Bethe Ansatz equations for $\alpha=1/\nu$ and 
$\alpha=1-1/\nu$}
\label{tba-section}
The thermodynamic Bethe Ansatz equations for the anisotropic Kondo model have
been derived by Tsvelik and Wiegman \cite{tsvelik.83} 
(for a short overview of the derivation see Appendix~\ref{tba-derivation}). 
For a general anisotropy they consist of an infinite set, 
$n=1,2,\dots$, of coupled  integral equations for the ``excitation
energies'' 
$\epsilon_{n}(\lambda)$ (defined in Appendix~\ref{tba-derivation}). 
The thermodynamics is calculated from the impurity
free energy which depends explicitly only on $\epsilon_{1}$. As discussed 
in \cite{tsvelik.83}, the infinite set of integral equations for the 
$\epsilon_{n}$ decouple to a finite set for values of the anisotropy 
corresponding to rational values of the scaling invariant $\mu/\pi$ of the
anisotropic Kondo model, and hence to rational values of the dissipation 
strength $\alpha=1-\mu/\pi$ in the Ohmic two-state system. In particular 
for anisotropies given by $\mu/\pi=1/\nu$ and 
$\mu/\pi=1-1/\nu$ with $\nu=3,4,\dots$, there are only $\nu$ coupled integral 
equations for the $\nu$ quantities $\epsilon_{1},\epsilon_{2},\dots,
\epsilon_{\nu}$. These two cases allow us to study both
the weak dissipation ($\alpha=1/\nu < 1/2$) and the strong dissipation 
($\alpha=1-1/\nu > 1/2$) limits of the Ohmic two-state system. Explicitly, 
the $\nu$ coupled integral equations in Eq.6.2.11 of \cite{tsvelik.83} are
\begin{eqnarray}
\frac{\epsilon_{j}(\lambda)}{T} & = & 
s\ast\bigl[\ln(1+\exp{\frac{\epsilon_{j-1}}{T}})
(1+\exp{\frac{\epsilon_{j+1}}{T}})\;,\nonumber\\
& + & \delta_{j,\nu-2}\ln(1+\exp{-\frac{\epsilon_{\nu}}{T}})\bigr] 
+ \delta_{j,1}D(\lambda)
\nonumber\\
\frac{\epsilon_{\nu-1}(\lambda)}{T} & = & x_{0} +
s\ast\ln(1+\exp{\frac{\epsilon_{\nu-2}}{T}})\;,
\label{eq:BAepsilonint}\\
\frac{\epsilon_{\nu}(\lambda)}{T} & = & x_{0} - 
s\ast\ln(1+\exp{\frac{\epsilon_{\nu-2}}{T}})\;.\nonumber
\end{eqnarray}
Here $x_{0}=\frac{1}{2}\frac{g\nu\varepsilon}{T}$, $\varepsilon$ is the bias, 
$g=1$ is a $g$-factor in the Kondo problem and $D(\lambda)$ is the 
driving term which is given explicitly by
\begin{equation}
D(\lambda) = -\mbox{sign}[\alpha-\frac{1}{2}]
\frac{\omega_c}{T}\arctan(\exp(\pi\lambda))\;.
\label{eq:driving}
\end{equation}
with $\omega_c$ a high energy cut-off (corresponding to the band width 
cutoff, $D$, in the Kondo model).
The integral operator $s\ast$ is defined as in \cite{tsvelik.83}:
\begin{equation}
s\ast f(\lambda) = \int_{-\infty}^{+\infty}\frac1
{2\cosh(\pi(\lambda-\lambda'))}{f(\lambda')}\;d\lambda'.
\label{eq:integral-operator}
\end{equation}
The $\epsilon_{j}$'s ($j=1,\dots,\nu$) satisfy boundary 
conditions at $\lambda=\pm\infty$ that can be obtained easily 
from Eq.~(\ref{eq:BAepsilonint}). 
The impurity contribution to the free energy can be simply expressed as
\begin{equation}
F = - T \int_{-\infty}^{+\infty} s(\lambda + \frac f\mu) \;\ln\bigl(1 + 
e^{\varepsilon_1(\lambda)/T}\bigr)\;d\lambda\;.
\label{eq.impfreeen}
\end{equation}
Eqs.~(\ref{eq:BAepsilonint}) and (\ref{eq.impfreeen}) describe the 
complete thermodynamics of the model.
Note that the field term $x_0$ in Eq.~(\ref{eq:BAepsilonint}) 
represents a {\it global} magnetic field in the AKM coupling to both the
impurity and conduction electrons. 
However, as we discuss in Sec.~\ref{Wilsonr}, due to electron-electron 
interactions introduced to assure integrability , the impurity susceptibility
of the BA solution, $-\partial^{2}F/\partial \varepsilon^{2}$, coincides
with the susceptibility of the Ohmic two-state system (i.e. with the
impurity susceptibility of the AKM with a field coupling {\em only} to the
impurity). Therefore the dielectric susceptibility, 
$\chi_{sb}(T,\varepsilon)$, and specific heat, $C(T,\varepsilon)$, 
of the Ohmic two-state system can be simply calculated as:
\begin{eqnarray}
\chi_{sb}(T,\varepsilon) &=&  -\partial^{2}F/\partial \varepsilon^{2}\;,\\
C(T,\varepsilon)&=&-T\partial^{2}F/\partial T^{2}\;.
\end{eqnarray}

\subsection{Analytic results}
\subsubsection{Scaling and universality}
A careful analysis of Eqs.~(\ref{eq:BAepsilonint}) and (\ref{eq.impfreeen}) 
makes immediately transparent the meaning of universality. 
For $\mu < \pi/2$ ($\alpha > 1/2$) the impurity free energy 
can be shown to be  dominated by contributions from the region 
$\lambda \ll 0$. In this limit the driving term can be approximated as
$D(\lambda)\approx -\frac{\omega_{c}}T e^{\pi\lambda}$ and the explicit
cutoff dependence can be transformed out of the equations by a simple
shift, $\lambda\to \lambda + \frac1\pi \ln \frac T{\omega_c}$.
In this way one arrives at the following universal equations for the 
quantities $\varphi_{j}(\lambda)\equiv \epsilon_{j}(\lambda + 
\frac1\pi \ln \frac T{\omega_c}) /T$:
\begin{eqnarray}
&&{\varphi_{j}(\lambda)}  =  
s\ast\bigl[\ln(1+e^{\varphi_{j-1}})
(1+e^{\varphi_{j+1}})\nonumber\\
&&\phantom{nnnn} +  \delta_{j,\nu-2}\ln(1+e^{-\varphi_{\nu}})\bigr]
- \delta_{j,1}e^{\pi\lambda}
\nonumber \\
&&\varphi_{\nu-1}(\lambda)  =  x_{0} +
s\ast\ln(1 + e^{\varphi_{\nu-2}})
\label{eq:univ} \\
&&\varphi_{\nu}(\lambda)  =  x_{0} - 
s\ast\ln(1+e^{\varphi_{\nu-2}}),
\nonumber\;, \\
&& F = - T \int_{-\infty}^{+\infty}
s(\lambda + \frac 1 \pi \ln \frac T {T_K}) 
\;\ln\bigl(1 + 
e^{\varphi_1(\lambda)}\bigr)\;d\lambda\;,
\label{eq:impfree}
\end{eqnarray}
where the Kondo temperature has been introduced as
\begin{eqnarray}
T_K & = & \omega_c  \exp(-\pi f/ \mu) \nonumber \\
& \equiv & \Delta_r/\alpha \;.
\label{eq:T_K}
\end{eqnarray}
For $\mu > \pi/2$ ($\alpha<1/2$) the seemingly innocent sign change 
of the driving term alters the structure of the solutions completely. 
The derivation of the universal equations becomes  much more 
complicated  in this case, but is still possible \cite{tsvelik.83}. 
These were not used
in obtaining the numerical results described below, but they served as a
useful check on the simpler set of equations (\ref{eq:BAepsilonint}). 
We reproduce them, correcting some minor typos in 
\cite{tsvelik.83}, in Appendix~\ref{wd-univ-eq}, 
together with a comparison of 
numerical  results obtained from them and the simpler set of equations 
(\ref{eq:BAepsilonint}). 
The equations above clearly show that the thermodynamic quantities depend 
only on the ratios $\varepsilon/T$ and $T/T_K \sim T/\Delta_r$. Note,
however, that while the parameter $f$ only influences  $T_K$,
for each $\mu$ ($\alpha$) one obtains a different set of equations 
and therefore different thermodynamic behaviour. Thus two
models have essentially the same universal behaviour if and only if
their parameter $\mu$ ($\alpha$) is the same. From these considerations
immediately follows that the usual RG scaling trajectories
correspond to the lines $T_K={\rm const}$ and $\mu = {\rm const}$. 
(The latter
requirement also follows from the fact that $2\mu$ turns out to 
be the anomalous dimension characterizing the high temperature
behaviour, which should be scale invariant.) In the small coupling
limit one immediately obtains the  usual
leading logarithmic scaling equations by expanding $f$ and $\mu$ 
in $\rho J_\perp\approx I_\perp$  and $\rho J_z\approx I_z$:
\begin{eqnarray}
{d {\rho J_z} \over d \ln \omega_c } & =& - (\rho J_\perp)^2\;,\\
{d {\rho J_\perp} \over d \ln \omega_c } &=& - \rho  J_\perp \rho J_z \;,
\end{eqnarray}
in agreement with Eq.~(\ref{eq.YuvalAnd}).

\subsubsection{Asymptotic properties}
\label{sec-asymptotic}
The asymptotic behaviour of various physical quantities can be
determined by analyzing Eq.~(\ref{eq:BAepsilonint}). 
Rewriting Eq.~(\ref{eq:BAepsilonint}) in terms of the  quantities 
$\xi_{j}$, $j=1,\dots,\nu$ defined by
\begin{eqnarray}
\xi_{j} & = &\ln[1+\exp(\frac{\epsilon_{j}}{T})]\;\;\;\;\;(j=1,\dots,\nu-1),
\nonumber\\
\xi_{\nu} & = &\ln[1+\exp(-\frac{\epsilon_{\nu}}{T})]\;,
\label{eq:xi}
\end{eqnarray}
one can easily show that in the $\lambda\to-\infty$ limit the
asymptotic solution of the BA equations behaves as
\begin{eqnarray}
\xi_j(\lambda \to -\infty)& =& \xi^{-}_j(x_0) + b^{-}_j(x_0)
e^{ \tau \lambda} \;.
\end{eqnarray}
with $\tau  =  2\mu$. On the other hand, for $\lambda\gg \ln(T/\omega_c)$
$ \xi_1$ vanishes extremely  fast  and 
can be approximated by 0. 
Substituting these  into Eq.~(\ref{eq:impfree}) and using
$\mu/\pi=1-\alpha$ and $\Delta_{r}\sim T_{K}$ one can 
immediately extract the leading behaviour of the impurity free energy.
In the high temperature limit one obtains:
\begin{eqnarray}
&&F(T\gg \Delta_{r}, \varepsilon)  \approx -T \Bigl\{\ln\frac{\sinh(\varepsilon g/T)}
{\sinh(\varepsilon g/2T)} + \nonumber \\
&& \phantom{nnnnn} - \bigl(\frac{\Delta_{r}}T\bigr)^{2-2\alpha}
\Bigl( A + B \bigl({\varepsilon \over T }\bigr)^2\Bigr) \Bigr\}\;,
\\
&&\chi_{sb}(T\gg \Delta_{r};\;\varepsilon=0) 
= -\frac{\partial^{2}F}{\partial
\epsilon^{2}}\approx \nonumber \\
&& \phantom{nnnnn} \approx\frac1T \bigl( {g^2 \over 4 } - 2 B
\bigl(\frac{\Delta_{r}}T\bigr)^{2-2\alpha} \bigr)\;,\label{eq:chi-hightemp}
\\
&&C (T\gg \Delta_{r};\;\varepsilon=0) \sim \bigl(
\frac{\Delta_{r}}T\bigr)^{2-2\alpha}\;,\label{eq:c-hightemp}
\end{eqnarray}
where the constants $A$ and $B$ depend only on $\mu$. From these equations 
it is clear that in order to recover the free
impurity spin at high temperatures --- unlike the choice $g =
(\nu-1)/\nu$ of Ref.~\onlinecite{tsvelik.83} --- we have to take the bare
value $g=1$ for the electronic g-factor. This special choice will
influence the Wilson ratio discussed below. 

Similarly, for the low temperature regime we obtain:
\begin{eqnarray}
&&F(\varepsilon \ll T  \ll \Delta_{r})  \approx {T^2\over \Delta_{r}}  
\Bigl( \frac{\alpha\pi} 6 + {g^2\over 4\pi}{\varepsilon^2 \over T^2 }
\Bigr)\;,
\label{eq:fimp}
\\
&&\chi_{sb}(T\ll \Delta_{r};\;\varepsilon=0) \approx  \frac1{\Delta_{r}}
{g^2\over 2\pi}\;, 	
\label{eq:chi-fliq}
\\
&&C(T\ll \Delta_{r};\;\varepsilon=0) \approx 
{\pi\over 3}\frac {T\alpha}{\Delta_{r}}\;,
\label{eq:c-fliq}
\end{eqnarray}
where the numerical constants have been calculated  following the 
same lines as in Ref.~\onlinecite{tsvelik.83}.
 Thus at low temperatures the well-known Fermi liquid behaviour is 
recovered \cite{Nozieres,Yamada}.

\subsubsection{Susceptibility and Wilson ratio for the Ohmic two-state system}
\label{Wilsonr}
As discussed by Wiegmann and Tsvelik,\cite{tsvelik.83} to ensure
integrability, an artificial electron-electron interaction has to 
be introduced.  While this interaction has no effect in the course 
of the solution of the isotropic model, in the anisotropical model it 
renormalizes the  electronic and impurity $g$-factors:
\begin{equation}
g =1 \to \tilde g = {1\over \sqrt{\alpha}} = {1\over \sqrt{1-\mu/\pi}}\;.
\end{equation}
This can be most easily checked by calculating the host susceptibility of the
AKM , $\chi_{host}= -\partial^2 F_{host}/\partial h^2$, 
following  the lines of Ref.~\onlinecite{tsvelik.83}. 
(Here $F_{host}$ is the free energy of the electrons, and is 
given by an expression 
similar to Eq.~(\ref{eq.impfreeen}) with $f=0$.) After a tedious calculation
one obtains the result that: 
\begin{eqnarray} 
\chi_{host} = {\chi_{free}\over1-\mu/\pi} = {\chi_{free}\over
\alpha}\;,
\label{chi_host}
\end{eqnarray}
where $\chi_{free} = L/(4\pi N)$ denotes
the free electron susceptibility  ($L$ is the length
of the system, $N$ the number of electrons, and $v_F = k_B = 2\mu_B =
1$). Note that the specific heat, $C_{host} = C_{free} = TL\pi /N 3$,
is  completely unaffected by the  electron-electron interaction. 

We now prove the statement made in Sec.~\ref{tba-section}, that the 
impurity contribution to the global susceptibility of the AKM,
{\em obtained from the Bethe Ansatz calculations}, is identical to the 
susceptibility of the Ohmic two-state system. We denote the bare 
g-factors of the impurity and conduction electrons in the AKM 
by $g_{i}$ and $g_{e}$ and the corresponding susceptibility of the AKM by 
$\chi(g_{i},g_{e})$ \cite{note-SCREAM}. 
Now in the Bethe Ansatz solution, starting 
with bare values $g_{i}=g_{e}=1$, the renormalizations discussed above
imply that the BA susceptibility, 
$\chi_{BA}\equiv -\partial^{2}F/\partial\varepsilon^{2}$, 
is given by
\begin{eqnarray}
\chi_{BA}=\chi(\tilde{g}_{i}=\tilde{g}_{e}=1/\sqrt{\alpha})\equiv 
\chi(g_{i}=g_{e}=1)/\alpha.\label{eq:chi-ba}
\end{eqnarray}
The dielectric susceptibility of the Ohmic two-state system, $\chi_{sb}$, 
measures the response to a local electric field and is equal to the impurity
susceptibility of the AKM, $\chi(g_{i}=1,g_{e}=0)$, with the magnetic
field coupling only to the impurity spin: 
\begin{equation}
\chi_{sb}=\chi(g_{i}=1,g_{e}=0). 
\label{eq:chi-sb}
\end{equation}
This follows from the equivalence of the two models discussed in 
Appendix~\ref{app:equiv}. 
We now make use of Eq.21 of Ref.~\onlinecite{vigmann.78} 
(valid for arbitrary $T$), connecting the impurity susceptibilities of the 
AKM with arbitrary g-factors. This states that
\begin{eqnarray}
\chi(g_i = g_e = 1) & = & (1 + 2{\delta\over \pi})^2\chi(g_i = 1;g_e = 0)\;,
\label{eq:w+f}
\end{eqnarray}
where the phase shifts have been defined in 
Appendix~\ref{app:equiv} (note the sign change 
with respect to Ref.~\onlinecite{vigmann.78}). 
Hence, using (\ref{eq:chi-ba}-\ref{eq:chi-sb}) and Eq.~(\ref{eq:alpha})  
$\alpha= (1 + 2{\delta\over \pi})^2$, we find that the BA susceptibility 
is just the susceptibility of the Ohmic two-state system,
\begin{equation}
\chi_{BA}(T,\varepsilon) \equiv   -\partial^{2}F/\partial
\varepsilon^{2}
 = \chi(\tilde g_i = \tilde g_e = {1/\sqrt{\alpha}}) =  \chi_{sb}\;,
\end{equation}
as stated earlier. This can be further checked by calculating 
the high temperature Curie susceptibility and the Wilson ratio 
(from Eqs.~(\ref{eq:chi-fliq}) and (\ref{eq:c-fliq}))
for which we find 
\begin{eqnarray}
\chi_{sb} = \chi_{BA} (T\gg T_K\sim \Delta_{r}) \approx  {1\over 4T}\;, \\
R_{sb} = R_{BA} \equiv \lim_{T\to0}{C_{free}(T)\over \chi_{free}}
{\chi_{BA}\over C(T)} = 2/\alpha\;, 
\end{eqnarray}
in agreement with exact results obtained for the spin-boson 
model.\cite{sassetti.90}
Furthermore, one can easily check by a bosonization procedure
along the Toulouse line\cite{JanZar}  that  Eq.~(\ref{Toulouse}) is identical
with the impurity contribution  to the free energy in the presence 
of a {\em local} magnetic field  applied at the impurity. 
We note that the above result for the Wilson ratio of the Ohmic two-state
system holds for all level asymmetries (local magnetic field in the AKM).
Finally, in order to prevent confusion, we state the connection of the above
Wilson ratio for the Ohmic two-state system to that usually encountered
in the Kondo model. The former is defined using the susceptibility of
the Ohmic two-state system $\chi_{sb}$ which we showed was equal to
the susceptibility $\chi_{BA}$ resulting from the BA calculation on the
AKM (with an electron-electron interaction to ensure integrability). The
susceptibility used in defining the Wilson ratio for magnetic impurities
is, however, not $\chi_{BA}=\chi_{sb}=\chi(g_{i}=1,g_{e}=0)$ but 
$\chi(g_{i}=1,g_{e}=1)$. Therefore, in terms of this susceptibility, 
and using Eq.~(\ref{eq:w+f}), the corresponding Wilson ratio, $R_{akm}$,
is given by
\begin{equation}
R_{akm} \equiv \lim_{T\to0}{C_{free}(T)\over \chi_{free}}
{\chi(g_{i}=g_{e}=1)\over C(T)} = 2\;.
\end{equation}
The enhancement over the non-interacting value $R=1$, indicates
that the quasiparticles are strongly interacting at low temperatures 
\cite{Hewson}.

\subsubsection{The Toulouse point: $\alpha=1/2$}
\label{sec:Toulouse}
The AKM possesses a so-called Toulouse 
line\cite{toulouse.69,vigmann.78,JanZar}.
Along this line the model can be mapped by a simple unitary
transformation to a resonant level model without interaction
and can be solved by refermionization. For a dissipative two-state 
system this line has been shown to correspond to the special 
value $\alpha = 1/2$ ($\mu = \pi/2$, $\nu=2$ in the BA solution) 
separating the coherent and incoherent tunneling
regimes.

Along this line the BA solution simplifies enormously too: For
$\nu=2$ only 'one-strings' with parity $v=\pm$ are allowed, and 
as one can immediately check by using Eq.~(\ref{eq:spinBA}), the rapidities
of the spin-excitations are completely decoupled. Therefore,
in the first of Eqs.~(\ref{eq:BAepsilonint}) only the driving term remains
and one obtains in the scaling limit:
\begin{eqnarray}
\frac1T\epsilon_1(\lambda)& = & -\frac{\omega_c}T \;e^{\pi\lambda} 
	+ {g\varepsilon\over T}\;,\nonumber \\
\frac1T\epsilon_2(\lambda)& = & \frac{\omega_c}T \;e^{\pi\lambda} 
	+ {g\varepsilon\over T}\;.
\end{eqnarray}
Substituting  these expressions into Eq.~(\ref{eq.impfreeen}) one
immediately arrives at
\begin{eqnarray}
F &=& -{T\over \pi} \int_{0}^{\infty} dk {T_K\over
k^2 + T_K^2} \ln \bigl\{1 + 2\cosh(g\varepsilon/T) e^{-k/T}
\nonumber \\
&+&e^{-2k/T}\bigr\}\;,
\label{Toulouse}
\end{eqnarray}
which coincides with the resonant level result
(note the slight difference with respect to the formula 6.2.15 of 
Ref.~\cite{tsvelik.83}). 
It is straightforward to verify that in the limit $T\rightarrow 0$
$$
\chi_{sb} = g^2 / (\pi T_K)\equiv g^{2}/2\pi\Delta_{r},
\;\;\;\alpha=1/2
$$
and
$$
C = \pi T /(3 T_K)\equiv \pi T/6\Delta_{r},\;\;\;\alpha=1/2
$$
giving the expected Wilson ratio, $ R_{sb}=2/\alpha=4$, for the 
Ohmic two-state system at $\alpha=1/2$, with $R_{sb}$ as 
defined above. The high temperature limits 
$S=\ln 2$ and $\chi_{BA}(T\gg T_{K}\sim \Delta_{r})=g^{2}/4T$ 
are also easily verified.

\section{NUMERICAL RESULTS AT ALL TEMPERATURES}
\label{num-sec}
\subsection{Numerical procedure}
\label{subsec-num-proc}
The closure of the infinite set of thermodynamic Bethe Ansatz equations to 
a finite set at rational values of the dissipation strength $\alpha$ can
be used to obtain highly accurate results for the thermodynamics. This avoids 
the truncation errors associated with solving these equations at other 
values of $\alpha$. In particular
at $\alpha=1/\nu$ and $\alpha=1-1/\nu$ we have $\nu$ equations. In the
numerical procedure we found it more convenient to set up integral equations
for new quantities $\xi_{j}, j=1,\dots,\nu$ defined by eq.~(\ref{eq:xi})
\begin{eqnarray}
\xi_{j} & = &\ln[1+\exp(\frac{\epsilon_{j}}{T})]=\ln[1+\kappa_{j}],\;\; j=1,\dots,\nu-1\nonumber\\
\xi_{\nu} & = &\ln[1+\exp(-\frac{\epsilon_{\nu}}{T})]=\ln[1+\kappa_{\nu}],
\label{eq:kappa}
\end{eqnarray}
where the functions $\kappa_{j},j=1,\dots,\nu$ are introduced for later 
convenience. 

The TBA equations then take the form,
\begin{eqnarray}
\xi_{j}(\lambda) & = & \ln[1+\exp(\delta_{j,1}D(\lambda)\nonumber\\
&+& s\ast(\xi_{j-1}+\xi_{j+1} + \delta_{j,\nu-2}\xi_{\nu}))],\nonumber\\
\xi_{\nu-1}(\lambda) & = & \ln[1+\exp(x_{0} + s\ast\xi_{\nu-2})],\; 
\label{eq:tba-ksi}\\
\xi_{\nu}(\lambda) & = & \ln[1+\exp(-x_{0} + s\ast\xi_{\nu-2})].\nonumber
\end{eqnarray}
The impurity free energy is given by
\begin{equation}
F(T,\varepsilon) = -k_{B}T\int_{-\infty}^{+\infty}s(\lambda + f/\mu)
\xi_{1}(\lambda,T,\varepsilon)d\lambda
\end{equation}
where $s(\lambda)=(2\cosh(\pi\lambda))^{-1}$ and $f/\mu$ is 
related to the low energy scale, $T_{K}$, of the AKM by Eq.~(\ref{eq:T_K})

The Kondo temperature, $T_{K}$, is related to the renormalized tunneling 
amplitude, $\Delta_{r}$, by $\Delta_{r}=\alpha T_{K}$ as discussed in Sec.IIc.
The entropy, specific heat and dielectric susceptibility can be obtained 
by numerically differentiating the Free energy:
\begin{eqnarray*}
S(T,\varepsilon) &= & - \frac{\partial F}{\partial T}
=\int s(\lambda + f/\mu)\;\frac{\partial\; T\xi_{1}
(\lambda,T,\varepsilon)}{\partial T}\;d\lambda\nonumber\\
C(T,\varepsilon) &= & - T\frac{\partial^{2} F}{\partial T^{2}}
=\int s(\lambda + f/\mu)\;\frac{\partial^{2}\; T\xi_{1}
(\lambda,T,\varepsilon)}{\partial T^{2}}\;d\lambda\nonumber\\
\chi(T,\varepsilon) &= &-\frac{\partial^{2} 
F}{\partial \varepsilon^{2}} = 
\frac{g^{2}}{4T}\int s(\lambda + f/\mu) 
\frac{\partial^{2} \xi_{1}(\lambda,T,\varepsilon)}{\partial x_{0}^{2}}
d\lambda\nonumber
\end{eqnarray*}
A more accurate procedure is to set up integral
equations for the derivatives $\partial\;(T\xi_{j})/\partial T$, 
$\partial\xi_{j}/\partial x_{0}$ and $\partial^{2}\xi_{j}/\partial x_{0}^{2}$.
More precisely, we set up integral equations for a new set of functions, 
$E_{j},F_{j}$ and $G_{j}=\partial\;(T\xi_{j})/\partial T = 
\xi_{j}+T\partial\xi_{j}/\partial T$ 
where $E_{j}$ and  $F_{j}$ are the first and second field derivatives of 
$\xi_{j}$,
\begin{eqnarray}
E_{j} &\equiv &\frac{\partial\xi_{j}}{\partial x_{0}} = \frac{\partial\kappa_{j}/
\partial x_{0}}{1+\kappa_{j}},\label{ej-definition}\\
F_{j} &\equiv &
\frac{\partial^{2} \xi_{j}}{\partial x_{0}^{2}}=
\frac{\partial^{2}\kappa_{j}/\partial x_{0}^{2}}{1+\kappa_{j}} -
\left[\frac{\partial\kappa_{j}/
\partial x_{0}}{1+\kappa_{j}}\right]^{2},\label{fj-definition}
\end{eqnarray}
and the functions $\kappa_{j}$ where defined in (\ref{eq:kappa}).
Each set of functions $E_{j},F_{j},G_{j},j=1,\dots,\nu$ then obey 
coupled linear inhomogeneous integral equations. 
The equations are, 
\begin{eqnarray}
E_{j} & = & (1-e^{-\xi_{j}})\,
s*(E_{j-1}+E_{j+1}+\delta_{j,\nu-2}E_{\nu})\nonumber\\
E_{\nu-1} & = & (1-e^{-\xi_{\nu-1}})\,(s*E_{\nu-2}+1),\;\label{eq:ejba}\\
E_{\nu} & = & (1-e^{-\xi_{\nu}})\,(s*E_{\nu-2}-1)\nonumber
\end{eqnarray}
for the $E_{j}$, with the inhomogeneity appearing in the last two equations.
For the $F_{j}$ we have
\begin{eqnarray}
F_{j} & = & Q_{j}+
(1-e^{-\xi_{j}})\,s*(F_{j-1}+F_{j+1}+\delta_{j,\nu-2}F_{\nu})\nonumber\\
F_{\nu-1} & = & Q_{\nu-1} + (1-e^{-\xi_{\nu-1}})\,s*F_{\nu-2},\;\label{eq:fjba}\\
F_{\nu} & = & Q_{\nu} + (1-e^{-\xi_{\nu}})\,s*F_{\nu-2},\nonumber
\end{eqnarray}
with an inhomogeneous term,
$$
Q_{j}  =  E_{j}^{2}e^{-\xi_{j}}(1-e^{-\xi_{j}})^{-1}.\nonumber
$$
and for the $G_{j}$ we find
\begin{eqnarray}
G_{j} & = & S_{j}+(1-e^{-\xi_{j}})\,
s*(G_{j-1}+G_{j+1}+\delta_{j,\nu-2}G_{\nu})\nonumber\\
G_{\nu-1} & = & S_{\nu-1}+ (1-e^{-\xi_{\nu-1}})\,s*G_{\nu-2},\;\label{eq:gjba}\\
G_{\nu}   & = & S_{\nu} + (1-e^{-\xi_{\nu  }})\,s*G_{\nu-2}\nonumber
\end{eqnarray}
with an inhomogeneous term,
$$
S_{j}  =  -e^{-\xi_{j}}\ln e^{-\xi_{j}}-(1-e^{-\xi_{j}})\ln (1-e^{-\xi_{j}}).
\nonumber
$$

\begin{figure}[t]
\centerline{\epsfysize 4.0cm {\epsffile{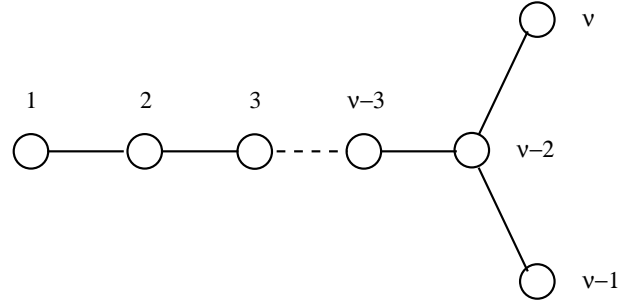}}}
\vspace{0.1cm}
\caption{
Graphical representation of the integral equations for the 
$\xi_{j},E_{j},F_{j},G_{j},j=1,\dots,\nu$.
}
\label{tba-graph}
\end{figure}

The numerical procedure is to first solve iteratively 
for the $\xi_{j},j=1,\dots,\nu$. Using this solution one then 
iteratively solves 
for the functions $E_{j},F_{j},G_{j},j=1,\dots,\nu$ in turn. 
Fig.\,\ref{tba-graph} shows
a graphical representation of these integral equations.
The entropy
and susceptibility are thereby obtained without the need to take any numerical
derivatives. Only one derivative is required to obtain the specific heat from
the entropy, so we did not set up separate integral equations for the second
temperature derivative of $T\xi_{1}$. Such a procedure has been used for zero 
field static susceptibilities in\cite{desgranges.85}. This approach also 
overcomes difficulties at large fields (level asymmetries) and low 
temperatures found in early treatments of similar TBA equations for 
Kondo systems \cite{rajan.82} provided one deals with the exponential 
decrease in the $j=1$ ($j=\nu$) functions for strong (weak) dissipation 
respectively. This and further details of the numerical 
procedure and its accuracy are given in Appendix~\ref{num-procedure}.

\subsubsection{Choice of parameters}
The TBA equations were solved for weak dissipation at $\alpha=1/6,1/5,1/4,1/3$,
and for strong dissipation at $\alpha=2/3,3/4,4/5$. The exact closed solution 
was used to obtain results at the Toulouse point $\alpha=1/2$. The 
thermodynamics was calculated at temperatures 
$t_{m}=\alpha k_{B}T_{m}/\Delta_{r}=2^{m/2}$ with 
$m=-20,-19,\dots,+19,+20$, and for level asymmetries 
$\tilde{\varepsilon}_{n}=\alpha \varepsilon_{n}/\Delta_{r}=2^{n}$
with $n=-4,-3,\dots,+3,+4$ and for the symmetric case $\varepsilon=0$.

\subsection{Entropy and Specific heat}
\subsubsection{Symmetric case: $\varepsilon=0$}
\label{sec-symmetric-heats}
\begin{figure}[t]
\centerline{\epsfysize 6.1cm {\epsffile{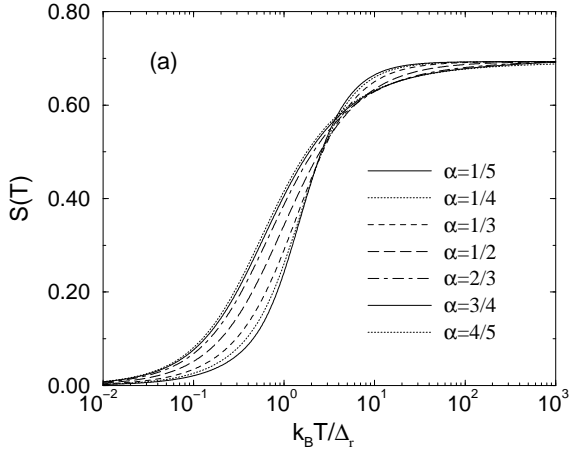}}}
\vspace{0.1cm}
\centerline{\epsfysize 6.1cm {\epsffile{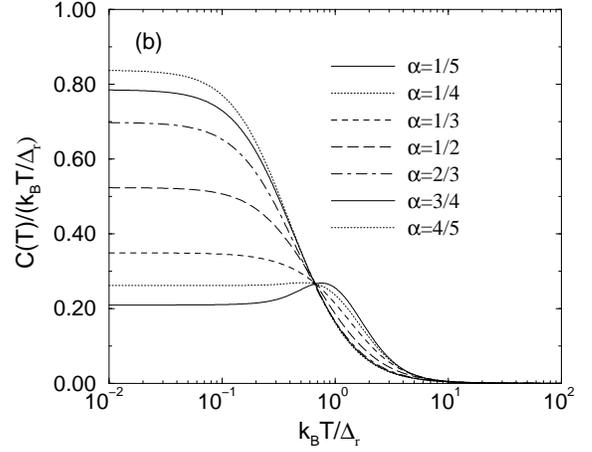}}}
\vspace{0.1cm}
\caption{
(a) Entropy, $S(T)$, and (b) specific heats, $C(T)/T$, 
for the symmetric two-state 
system ($\varepsilon=0$) for weak ($\alpha<1/2$) and strong ($\alpha>1/2$) 
dissipation cases. The $T^{3}$ coefficient in $C(T)/T$ is negative for 
$\alpha>1/3$ and positive for $\alpha<1/3$.
}
\label{s+c-symmetric-all-alpha}
\end{figure}

The entropy of the symmetric two-state system is shown in 
Fig.\,\ref{s+c-symmetric-all-alpha}a as a 
function of temperature for several values of the dimensionless dissipation
strength, $\alpha$, ranging from weak to strong dissipation. 
The correct value of the entropy, $S=\ln 2$, is recovered at
high temperature for all $\alpha$. 
Fig.\,\ref{s+c-symmetric-all-alpha}b
shows the universal specific heat curves for the dissipative two-state
system. As in other strongly correlated systems \cite{vollhardt.97} 
we observe a characteristic crossing point for the specific heat curves 
at a temperature $k_{B}T^{+}/\Delta_{r}=0.66\pm 0.02$. 
At low temperature 
the specific heat is given by 
$C(T)=\alpha\tilde{\gamma} (T/\Delta_{r}) +b(\alpha)(T/\Delta_{r})^{3}+\dots$ 
with a linear coefficient of specific heat 
$\gamma=\alpha\tilde{\gamma}/\Delta_{r}$ 
which vanishes as $\alpha\rightarrow 0$. 
From the definition of the low
energy scale $\Delta_{r}$ in terms of the zero temperature susceptibility and
the Wilson ratio (to be discussed below) it follows that 
$\tilde{\gamma}=\pi/3$ for all $\alpha$, a useful check on the 
numerical results.
The coefficient $b(\alpha)$ of 
the $T^{3}$ term is negative for $\alpha\geq 1/3$, a special point in the 
parameter space of the
dissipative two-state system. 
\label{subsec-entropy+heat-symm}
The significance of $\alpha=1/3$ is best seen
in the context of the dynamics of the two-state system, where it corresponds
to the value of the dissipation strength at which the frequency, 
$\Omega(\alpha)$, of tunneling oscillations (manifested in real time 
correlation functions) becomes equal to the decay rate, $\Gamma(\alpha)$, 
of these oscillations, i.e. $\Omega (\alpha=1/3)=\Gamma (\alpha=1/3)$ (or
the quality factor $Q(\alpha)=\Omega(\alpha)/\Gamma(\alpha)$ becomes unity)
\cite{leggett.87,voelker.98}.
For dissipation $\alpha < 1/3$  we 
have $\Omega (\alpha)>\Gamma (\alpha)$ and
the well defined oscillatory mode appears to be reflected in the 
characteristic peak in the specific heat $C(T)/T$ \cite{note-ct-peak}. 
\begin{figure}[t]
\centerline{\epsfysize 6.1cm {\epsffile{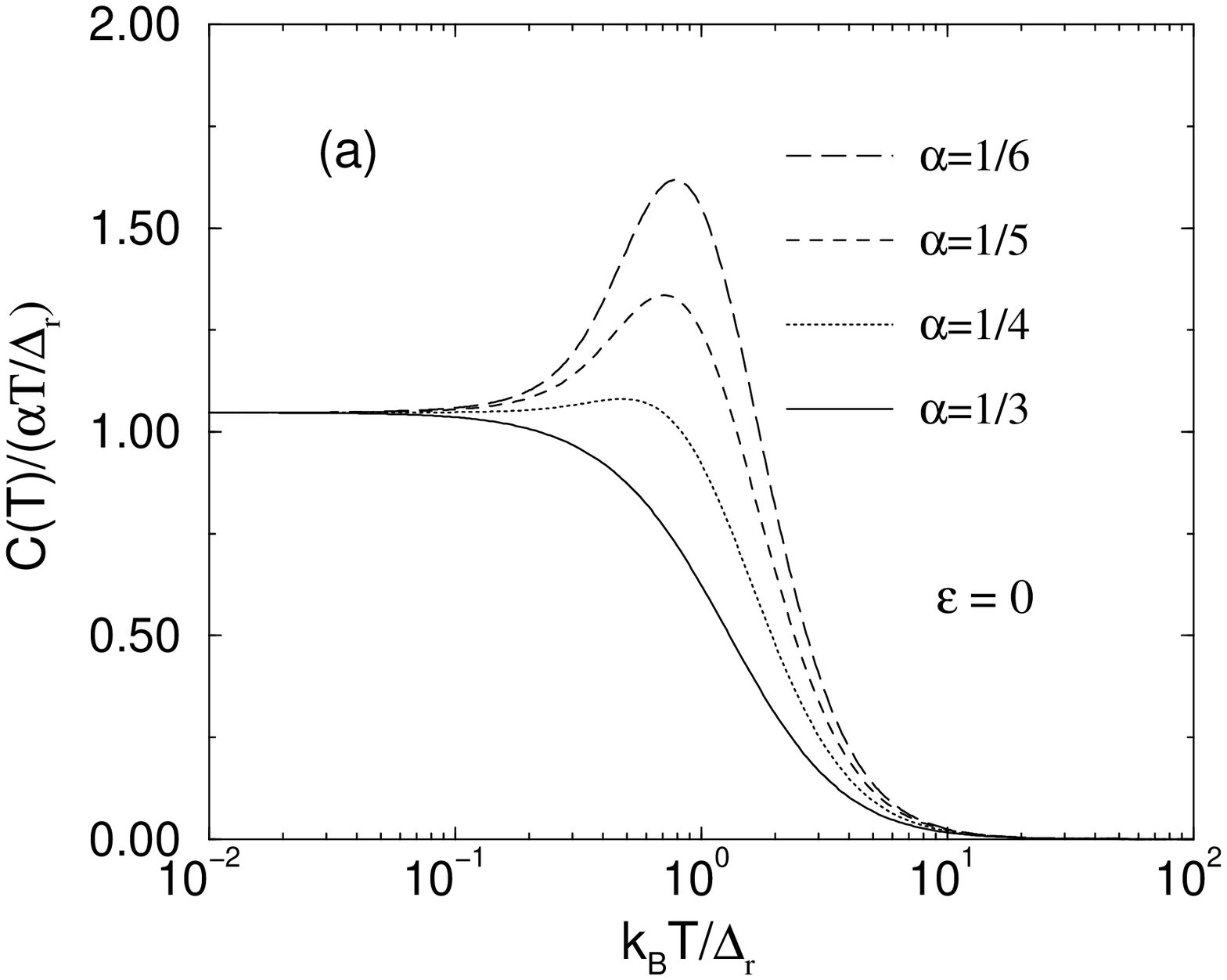}}}
\vspace{0.2cm}
\centerline{\epsfysize 6.1cm {\epsffile{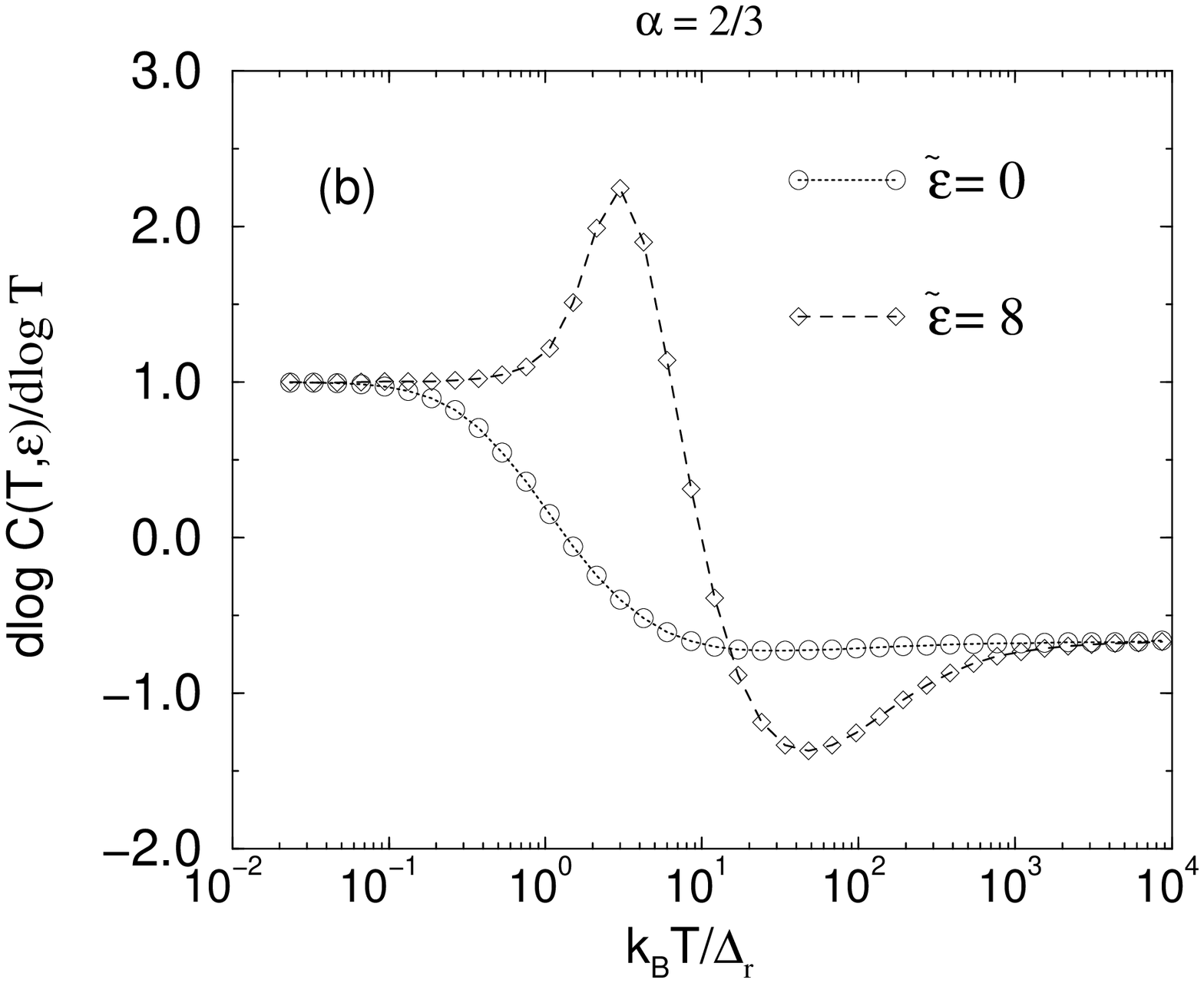}}}
\vspace{0.1cm}
\caption{
(a) Specific heat, $C(T)/(\alpha T/\Delta_{r})$, showing 
the development of the peak at $k_{B}T\approx \Delta_{r}$ for 
$\alpha < 1/3$ and $\varepsilon=0$. $\lim_{T\rightarrow 0}
C(T)/(\alpha T/\Delta_{r})=\tilde{\gamma}=\pi/3 = 1.04719755$
is recovered to 5 decimal places. (b) the 
logarithmic derivative, $d\log\; C(T,\varepsilon)/d\log\;T$,  at $\alpha=2/3$
for the symmetric ($\varepsilon=0$) and asymmetric 
($\varepsilon/\Delta_{r}=8$) cases. The approach to the expected power
law $C(T)\sim (\Delta_{r}/k_{B}T)^{\delta}$ with $\delta={2\alpha-2}=2/3$ for
$\alpha=2/3$ is found at high temperatures for both cases. For the asymmetric
system the power law arises at higher temperature corresponding to the
higher low energy scale behaving as 
$\sqrt{\varepsilon^{2}+\Delta_{r}^{2}}>\Delta_{r}$ for $\epsilon\gg 
\Delta_{r}$.
}
\label{schottky-peak+high-temp-limit}
\end{figure}
The peak in $C(T)/T$ for $\alpha< 1/3$ is shown in more detail in 
Fig.\,\ref{schottky-peak+high-temp-limit}a and is reminiscent of the
activated behaviour seen in non-interacting two-level systems. 
Since the excitation 
spectrum of the Ohmic two-state system is gapless for all
$\alpha>0$, there is no exponential suppression of $C(T)/T$ at low 
$k_{B}T< \Delta_{r}$ as with non-interacting two-level systems. The
linear specific heat persists at low temperature and the system remains
strongly interacting down to $T=0$ as is also clear from the value of
the Wilson ratio (see later). At high temperatures $k_{B}T \gg \Delta_{r}$
the specific heat vanishes according to a power law, 
$C(T)\sim (\Delta_{r}/k_{B}T)^{2-2\alpha}$ with an $\alpha$ dependent power in
accordance with the asymptotic result given by Eq.(\ref{eq:c-hightemp}) and
in \cite{goerlich.88}. 
This is shown in Fig.\,\ref{schottky-peak+high-temp-limit}b
and we see that, as for low temperatures, the behaviour
at high temperatures is again drastically different to the behaviour of 
a non-interacting two-level system which shows 
$C(T) \sim (\Delta_{0}/k_{B}T)^{2}$ for $k_{B}T\gg \Delta_{0}$ with $\Delta_{0}$
the bare tunneling matrix element. The limit $\alpha\rightarrow 1^{-}$ 
corresponds to the weak coupling Kondo model. For weak coupling, 
$\rho J_{\perp}, \rho J_{\parallel}\ll 1$, the exchange operators in the AKM 
become marginal (see above discussion of the 
Anderson-Yuval scaling equations) 
and consequently the high temperature properties acquire logarithmic 
corrections leading to a slow approach of the thermodynamic quantities to 
their high temperature limits \cite{costi.98} 
( this can be seen in the entropy in Fig.\,\ref{s+c-symmetric-all-alpha}a
and more clearly in the results for the susceptibility 
to be described in the next section).

\subsubsection{Asymmetric case: $\varepsilon>0$}
\label{subsec-entr+heat-asymm}
\begin{figure}[b]
\centerline{\epsfysize 6.1cm 
{\epsffile{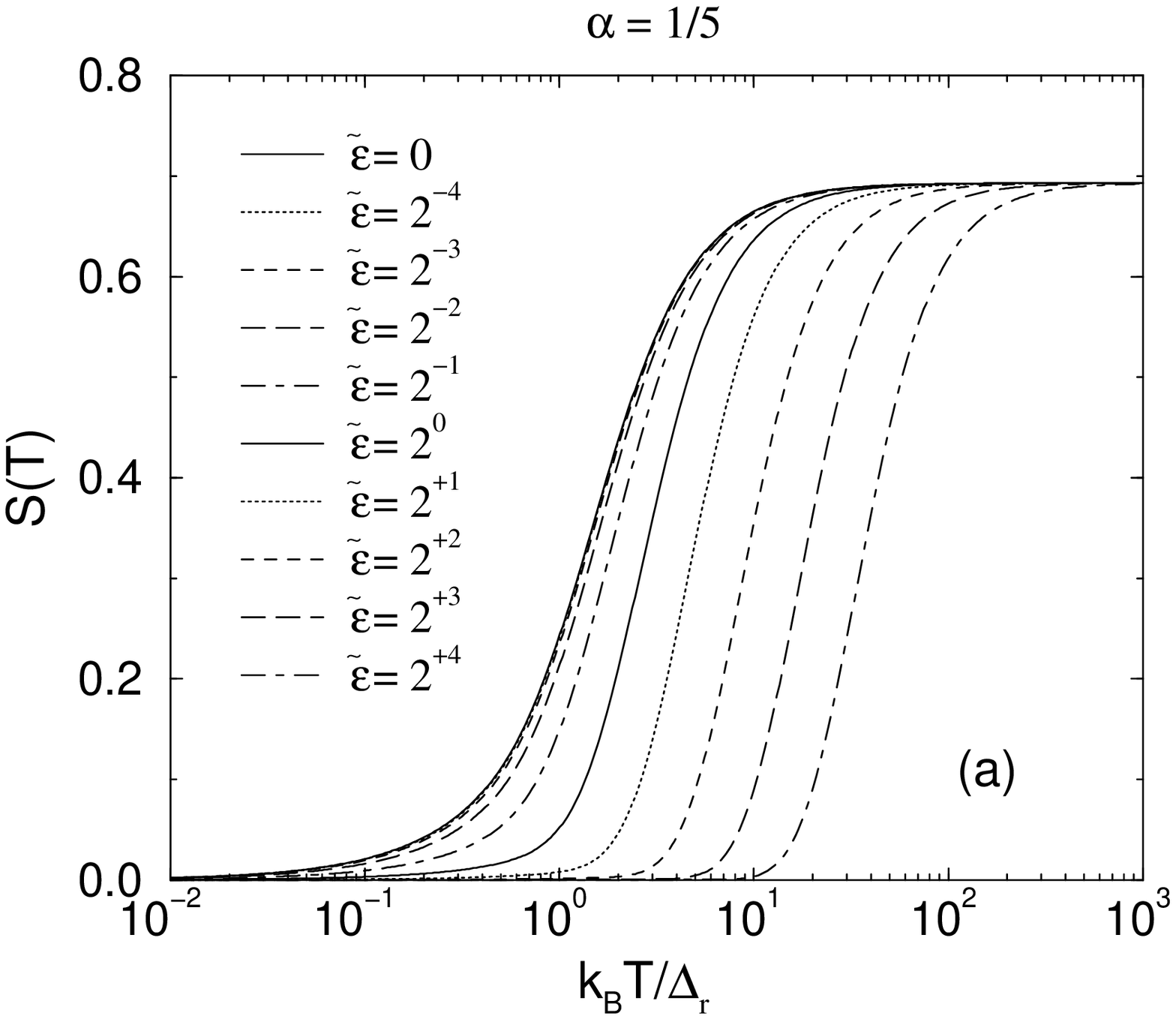}}}
\vspace{0.1cm}
\centerline{\epsfysize 6.1cm 
{\epsffile{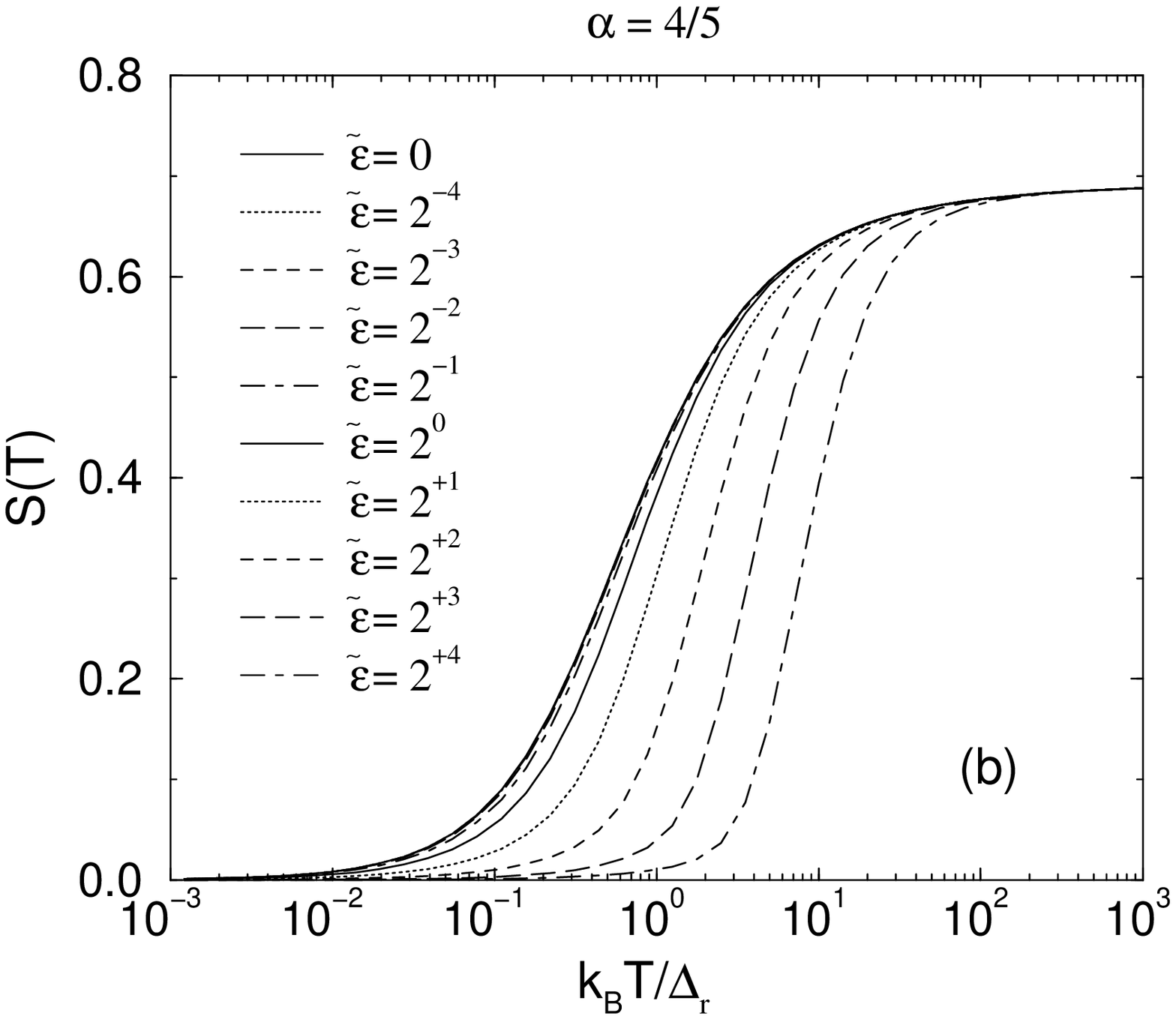}}}
\vspace{0.1cm}
\caption{
Entropy, $S(T)$, for the asymmetric two-state 
system for a range of asymmetries at (a) $\alpha=1/5$ and (b) $\alpha=4/5$. 
}
\label{entropy-asymmetric}
\end{figure}

We now turn to the asymmetric two-state system. 
Fig.\,\ref{entropy-asymmetric} 
shows the temperature dependence of the entropy
for different level asymmetries at $\alpha=1/5$ and $\alpha=4/5$. 
The correct high temperature limit $S=\ln 2$ is recovered for all
level asymmetries and dissipation strengths. As for
the symmetric case, we see again that the entropy 
approaches its high temperature
limit more slowly for strong dissipation than for weak dissipation, again
a result of increasing marginality of the interactions with 
increasing $\alpha$ at the high energy fixed point. 

\begin{figure}[t]
\centerline{\epsfysize 6.1cm 
{\epsffile{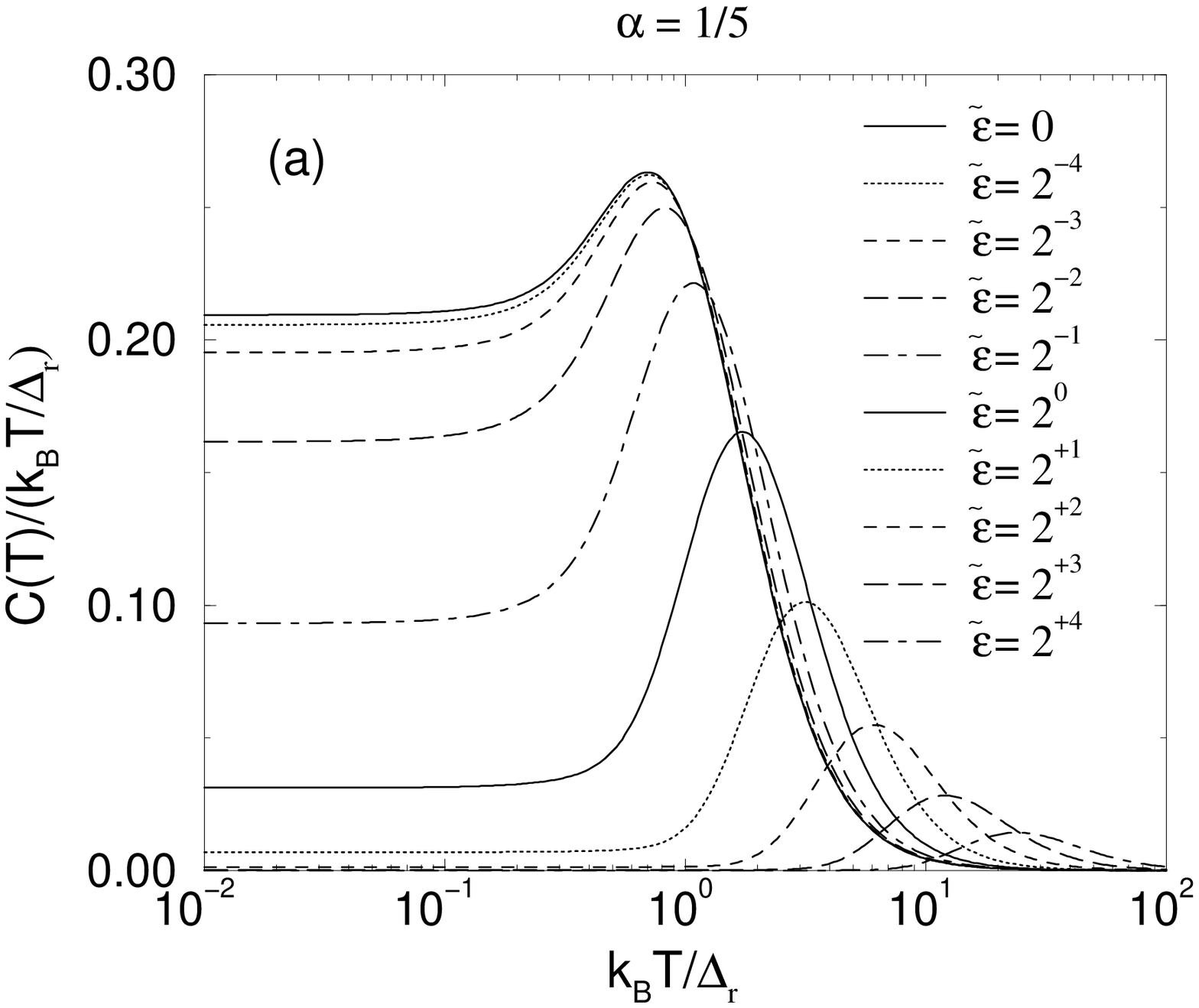}}}
\vspace{0.1cm}
\centerline{\epsfysize 6.1cm 
{\epsffile{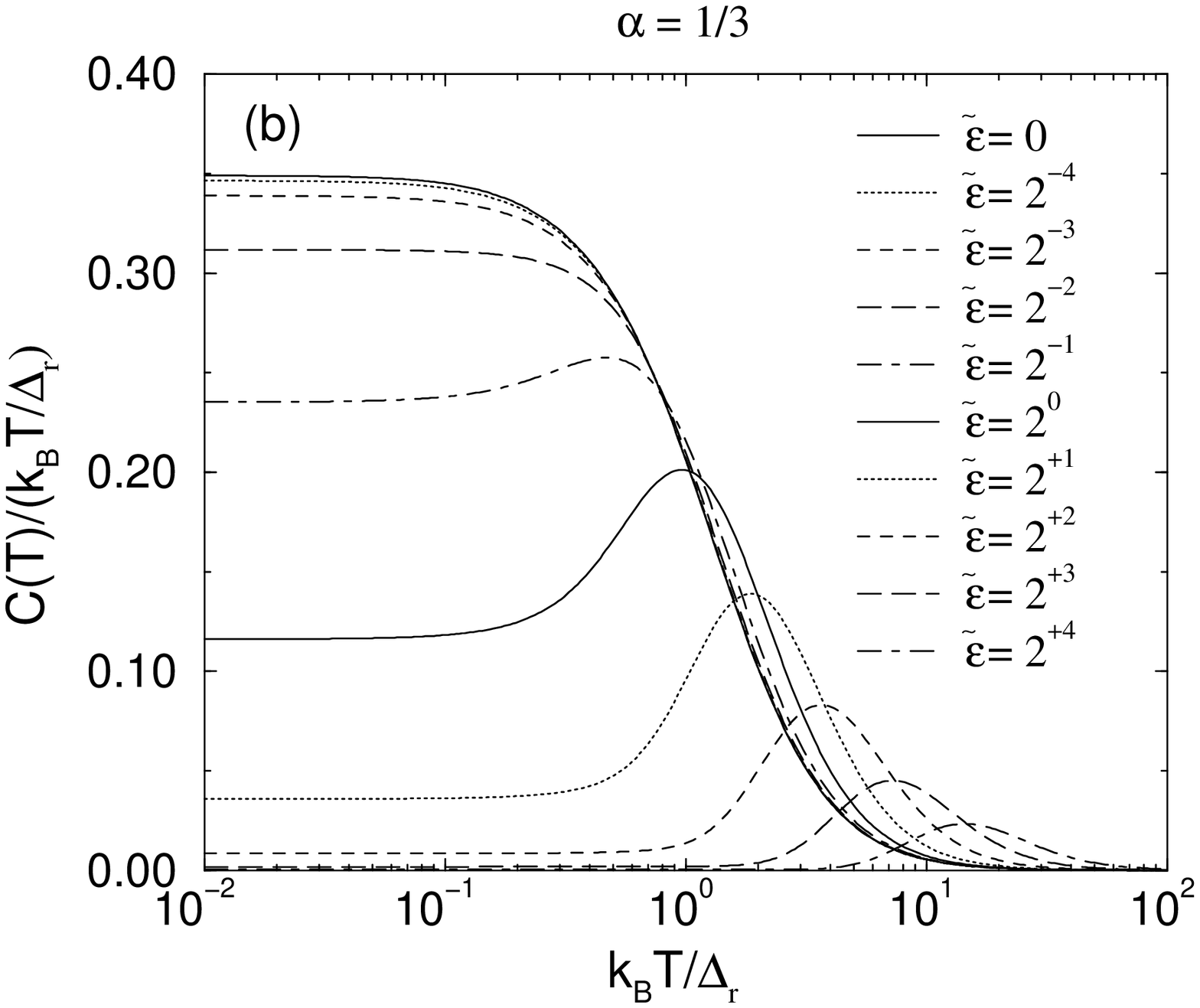}}}
\vspace{0.1cm}
\caption{
Specific heats, $C(T)/T$, for the asymmetric two-state 
system for a range of asymmetries, and some typical cases for weak 
dissipation, (a) $\alpha=1/5$, (b) $\alpha=1/3$.
}
\label{specific-heats-asymmetric-weak}
\end{figure}

\begin{figure}[t]
\centerline{\epsfysize 6.1cm 
{\epsffile{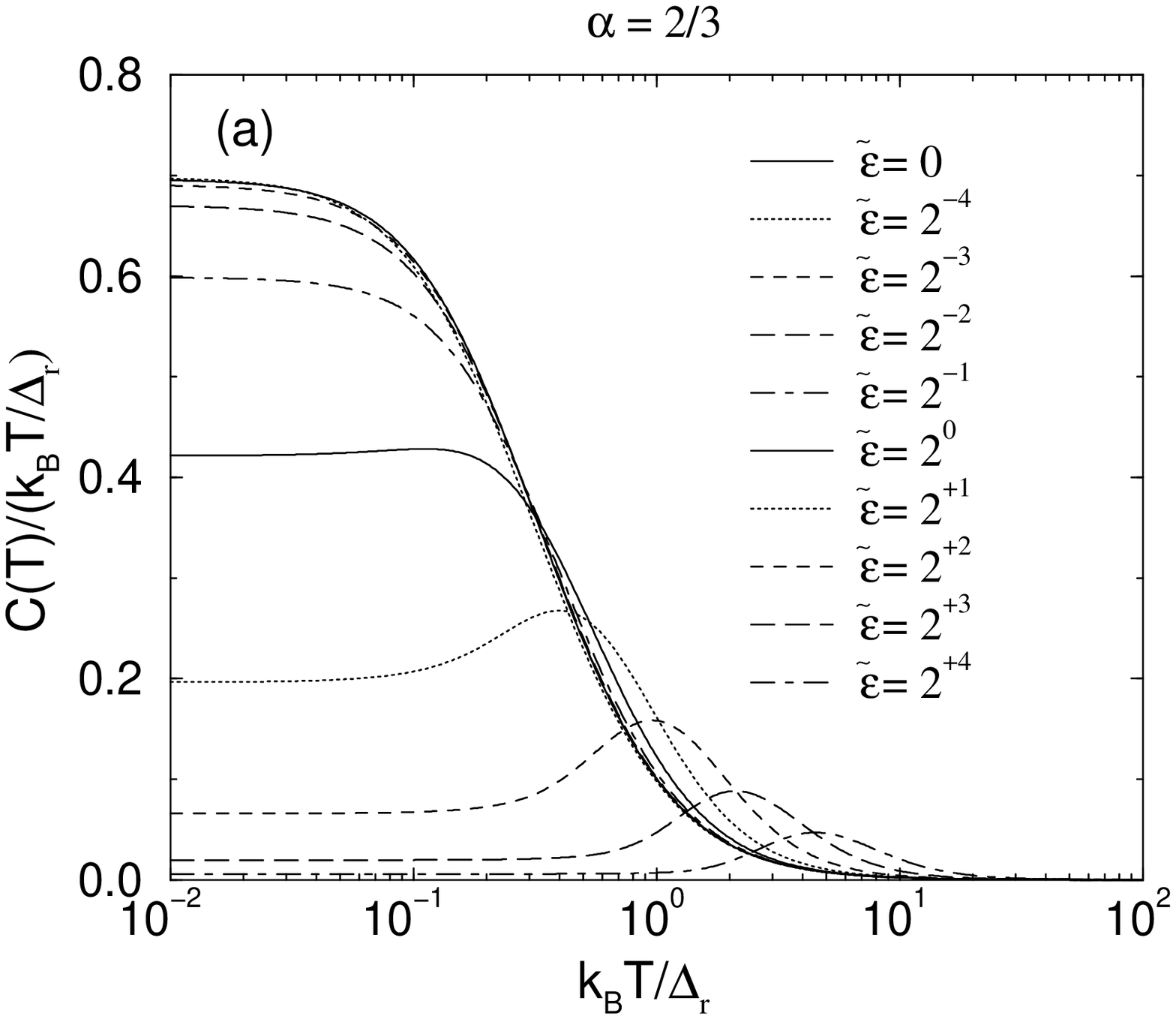}}}
\vspace{0.1cm}
\centerline{\epsfysize 6.1cm 
{\epsffile{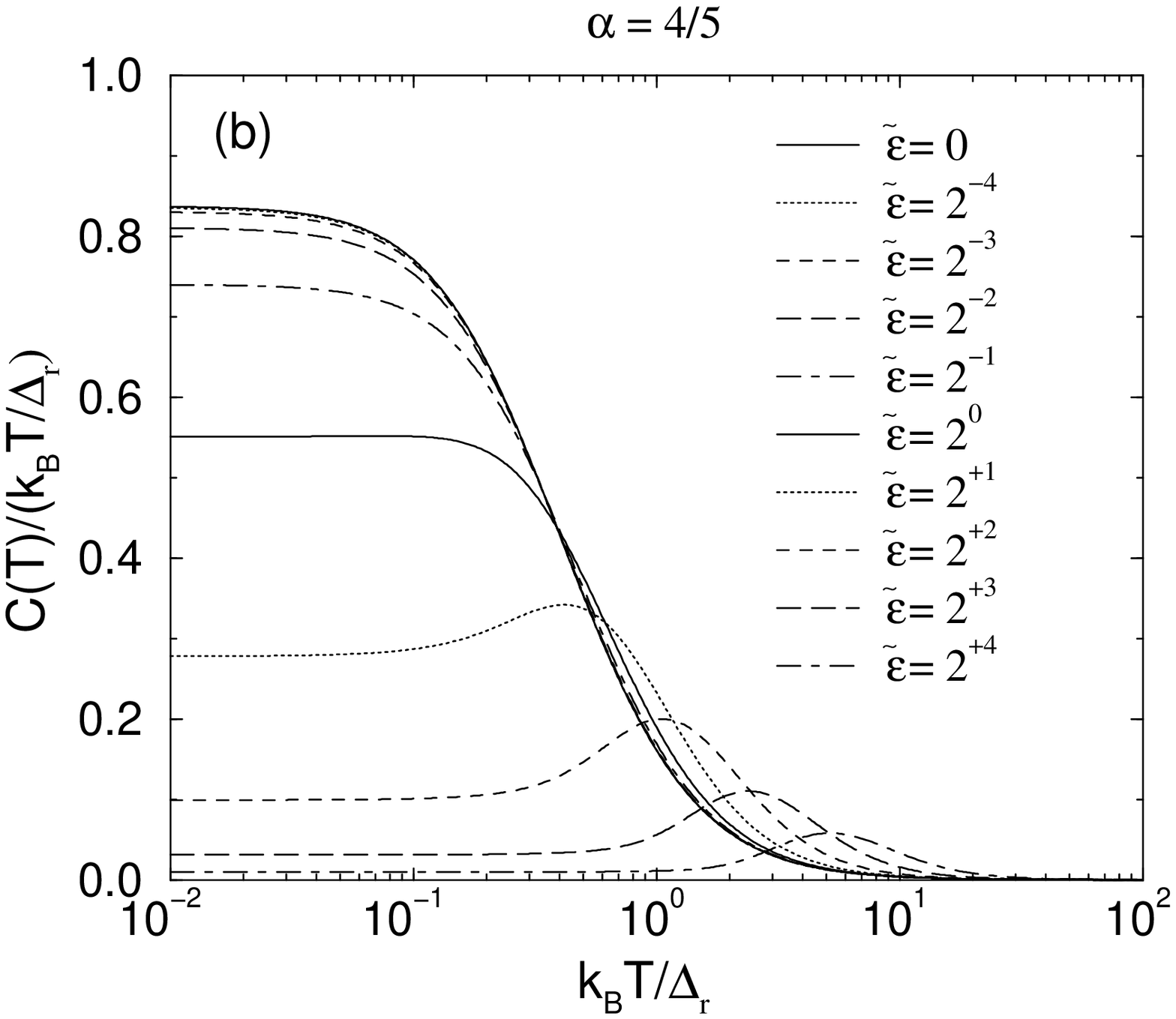}}}
\vspace{0.1cm}
\caption{
Specific heats, $C(T)/T$, for the asymmetric two-state 
system for a range of asymmetries, and some typical cases for strong
dissipation, (a) $\alpha=2/3$, (b) $\alpha=4/5$.
}
\label{specific-heats-asymmetric-strong}
\end{figure}

In Fig.\,\ref{specific-heats-asymmetric-weak}--
\ref{specific-heats-asymmetric-strong} 
we show the specific heats for different level asymmetries and
dissipation strengths. The specific heat remains linear at low temperature,
$C(T,\varepsilon)\sim \gamma T$, for all level asymmetries. The
linear coefficient $\gamma\sim 1/\Delta_{r}$ is reduced with 
increasing asymmetry $\varepsilon$, a consequence of the increasing low
energy scale with increasing $\varepsilon$, $\Delta_{r}\rightarrow \sqrt{
\Delta_{r}^{2}+\varepsilon^{2}}$. We see that a sufficiently large asymmetry
eventually leads to a peak in $C(T)/T$ for all dissipation strengths, but
that for such a peak to form requires a sizeable asymmetry for $\alpha>1/3$.
It is important to note that even for large asymmetries, the shape of the
specific heat curves is still different to those of a non-interacting 
two-level system: at low temperature the specific heat remains linear rather
than exponential and at high temperature the asymptotic behaviour of $C(T)$
is not the non-interacting  $1/T^{2}$ but instead behaves as 
$1/T^{2-2\alpha}$ as shown analytically for $\varepsilon=0$ and numerically
for $\varepsilon>0$ in Fig.\,\ref{schottky-peak+high-temp-limit}. 
Only for $\alpha\ll 1$ do we expect the
specific heat to be reasonably described by the non-interacting result, and 
then only outside the Fermi liquid regime $k_{B}T> \Delta_{r}$.

\subsection{Dielectric Susceptibility}
\label{sec-susc}

\subsubsection{Symmetric case: $\varepsilon=0$}
\label{susec-susc-symm}
\begin{figure}[t]
\centerline{\epsfysize 6.1cm 
{\epsffile{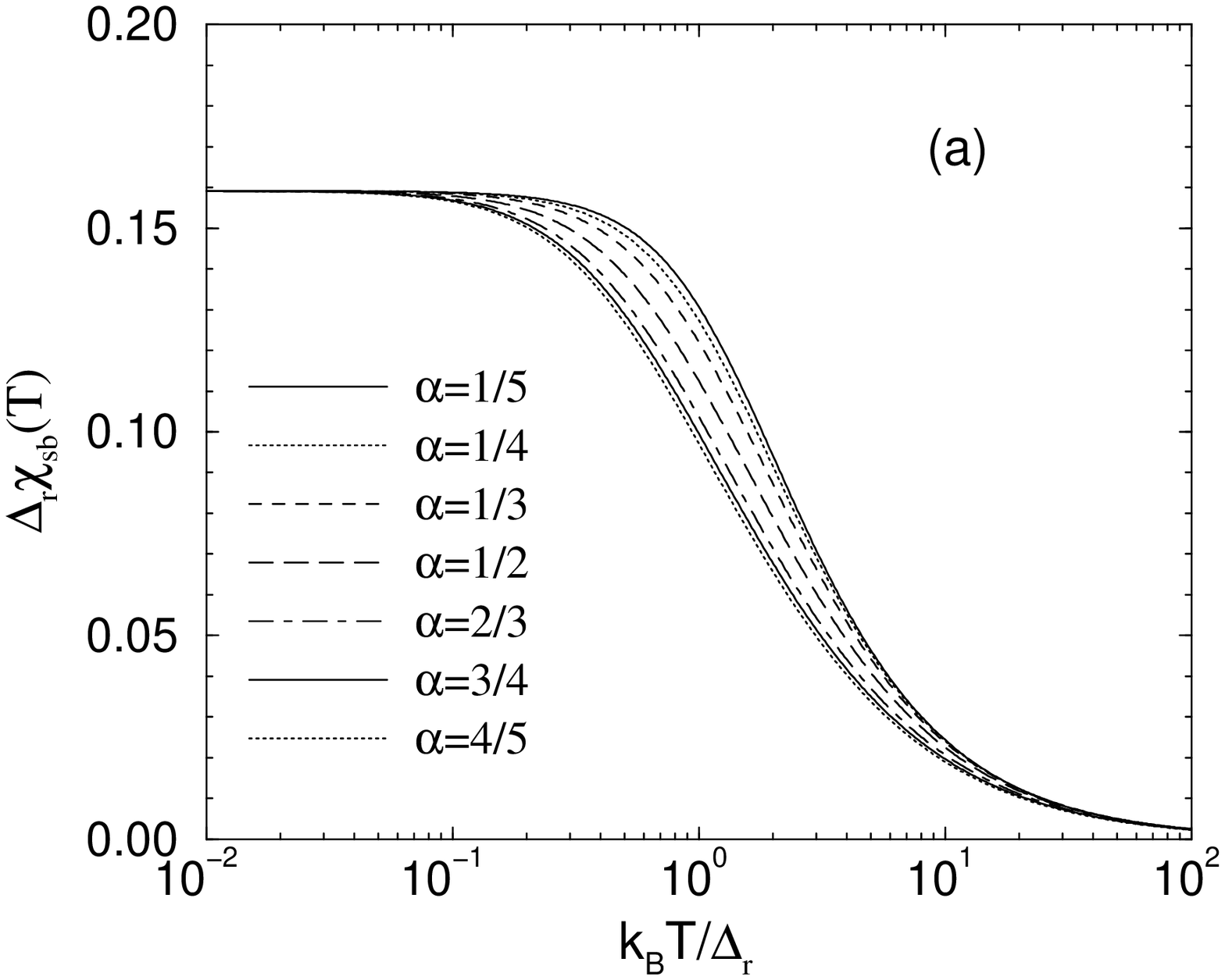}}}
\vspace{0.1cm}
\centerline{\epsfysize 6.1cm 
{\epsffile{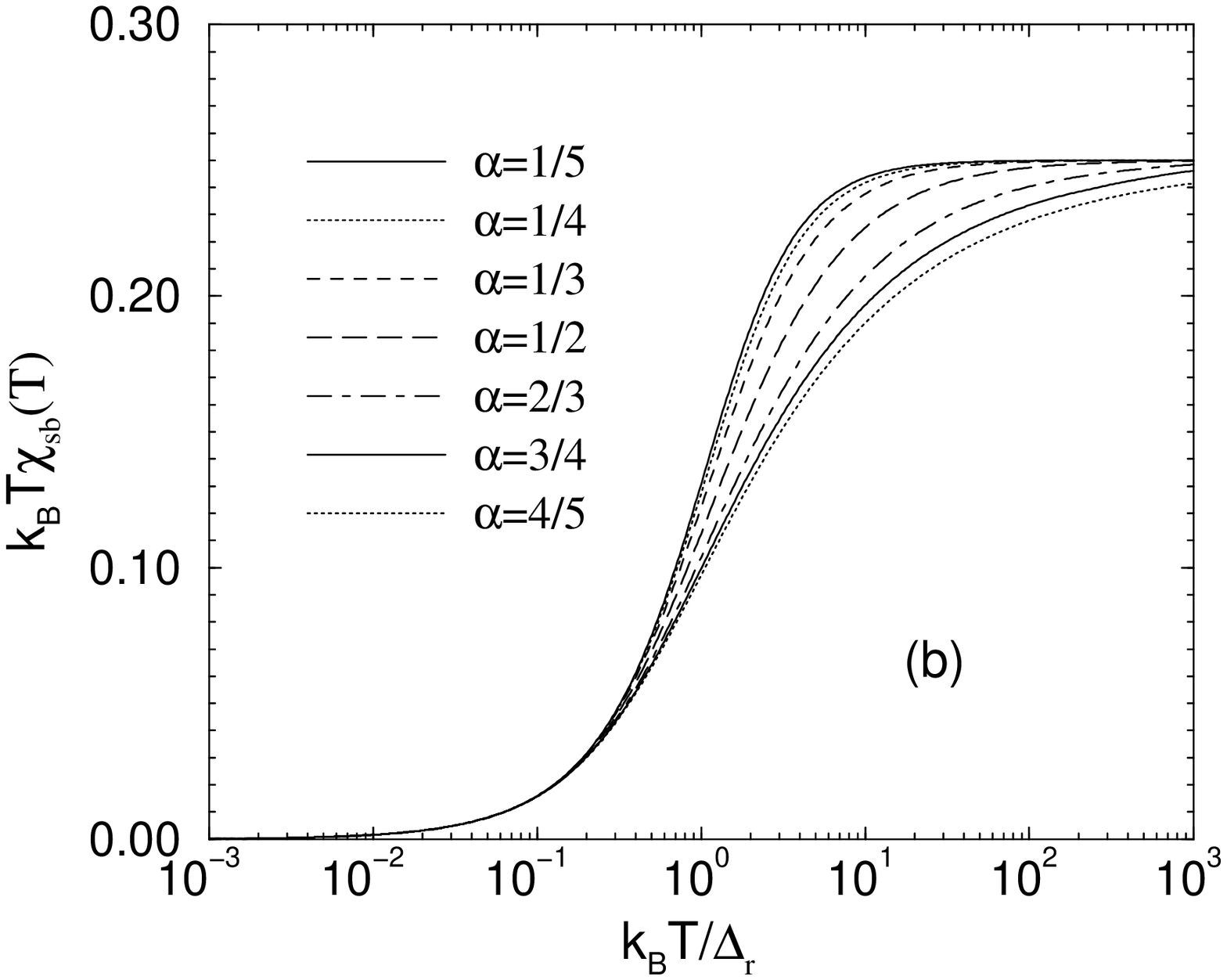}}}
\vspace{0.1cm}
\caption{
Dielectric susceptibility, $\chi_{sb}(T)$, for the symmetric two-state 
system ($\varepsilon=0$) for weak ($\alpha<1/2$) and strong ($\alpha>1/2$) 
dissipation cases. The susceptibility is finite at $T=0$ with 
$\Delta_{r}(\alpha)\chi(T=0)=1/2\pi$ in all cases as seen in (a) and 
attains its free-spin value of $1/4T$ at high temperatures 
$k_{B}T \gg \Delta_{r}$, as seen in (b).
}
\label{susc-symmetric-all-alpha}
\end{figure}

\begin{figure}[t]
\centerline{\epsfysize 6.1cm {\epsffile{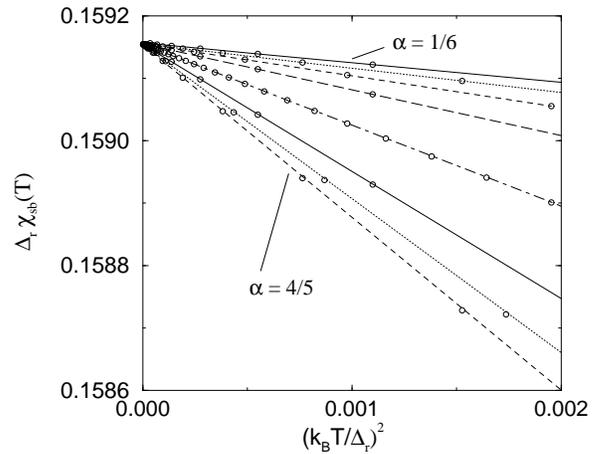}}}
\vspace{0.1cm}
\caption{The $T^{2}$ Fermi liquid corrections to the dielectric 
susceptibility at low temperatures for the symmetric case and 
$\alpha=1/6,1/5,\dots,3/4,4/5$. 
}
\label{sus-symmetric-low-t-fermi-liquid}
\end{figure}

\begin{figure}[t]
\centerline{\epsfysize 6.1cm 
{\epsffile{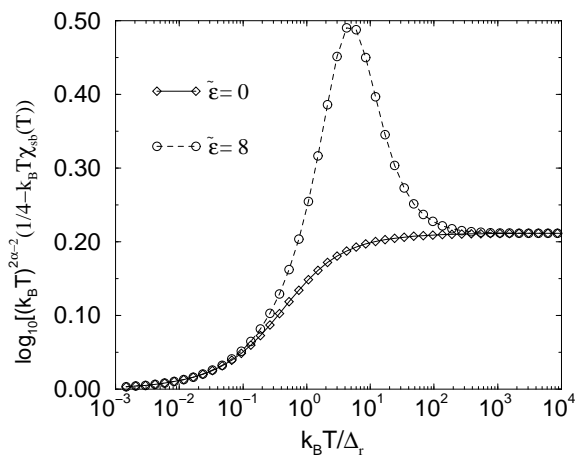}}}
\vspace{0.1cm}
\caption{
High temperature corrections to the dielectric susceptibility for 
$\alpha=2/3$ for $\tilde{\varepsilon}=0$ and $\tilde{\varepsilon}=8$. 
$\lim_{T\rightarrow \infty}\log_{10}[(k_{B}T)^{2\alpha-2}(1/4 - 
k_{B}T\chi_{sb}(T))]=\log_{10}(2B)$
}
\label{chi-sd3-high-temp-corrections}
\end{figure}

Fig.\,\ref{susc-symmetric-all-alpha}
shows that the dielectric susceptibility of the dissipative two-state 
system remains finite down to $T=0$ for all dissipation strengths 
$\alpha< 1$. This is also shown in Fig.\,\ref{sus-symmetric-low-t-fermi-liquid}
together with the Fermi liquid $T^{2}$ corrections at $k_{B}T\ll \Delta_{r}$ 
given by Eq.(\ref{eq:chi-fliq}), $\chi_{sb}(T)=
\chi_{sb}(0)(1-c(\alpha)(k_{B}T/\Delta_{r})^{2})$. 
By our definition $\chi_{sb}(T=0)=1/2\pi\Delta_{r}$, we have
that $\Delta_{r}\chi_{sb}(T=0)=1/2\pi =0.1591549$ which is reproduced by
our numerical solution to 5 decimal places in all cases (Fig.\,
\ref{sus-symmetric-low-t-fermi-liquid}).
In contrast to the specific heat, $C(T)/T$, the susceptibility is a 
monotonically decreasing function of temperature for all dissipation
strengths. There is no signature of the onset of activated behaviour in
the susceptibility as there was for $\alpha<1/3$ in $C(T)/T$. As we shall see
below, a finite temperature peak in $\chi_{sb}$ only arises when there is
a finite level asymmetry. The dielectric susceptibility looks, superficially,
like that for a non-interacting system, however the universal scaling
curves depend sensitively on $\alpha$, as can be seen in Fig.\,
\ref{susc-symmetric-all-alpha}, so that this resemblance is misleading.
The strong renormalization of the tunneling amplitude
$\Delta_{r}/\omega_{c} \sim (\Delta_{0}/\omega_{c})^{1/(1-\alpha)}$ as 
$\alpha\rightarrow 1^{-}$ gives rise to strongly renormalized dielectric
susceptibilities at low temperatures and strong dissipation 
($\chi_{sb}(T=0)=1/2\pi\Delta_{r}$). The approach of the 
susceptibility to its free spin value of
$1/4$ at $k_{B}T\gg \Delta_{r}$ at high temperatures 
(Fig.\,\ref{susc-symmetric-all-alpha}b ) is governed
by power laws with exponents which are functions of the dissipation strength
as given by Eq.(\ref{eq:chi-hightemp}), $k_{B}T\chi_{sb}(T)\approx 
1/4 -2B(\Delta_{r}/k_{B}T)^{2-2\alpha}$, and verified in 
Fig.\,\ref{chi-sd3-high-temp-corrections}.
The approach to the free spin value becomes slower as 
$\alpha\rightarrow 1^{-}$ and eventually logarithmic corrections to the 
susceptibility set in (see Fig.\,\ref{susc-symmetric-all-alpha}b ).  

\subsubsection{Asymmetric case: $\varepsilon>0$}
\label{susec-susc-asymm}
\begin{figure}[t]
\centerline{\epsfysize 6.1cm 
{\epsffile{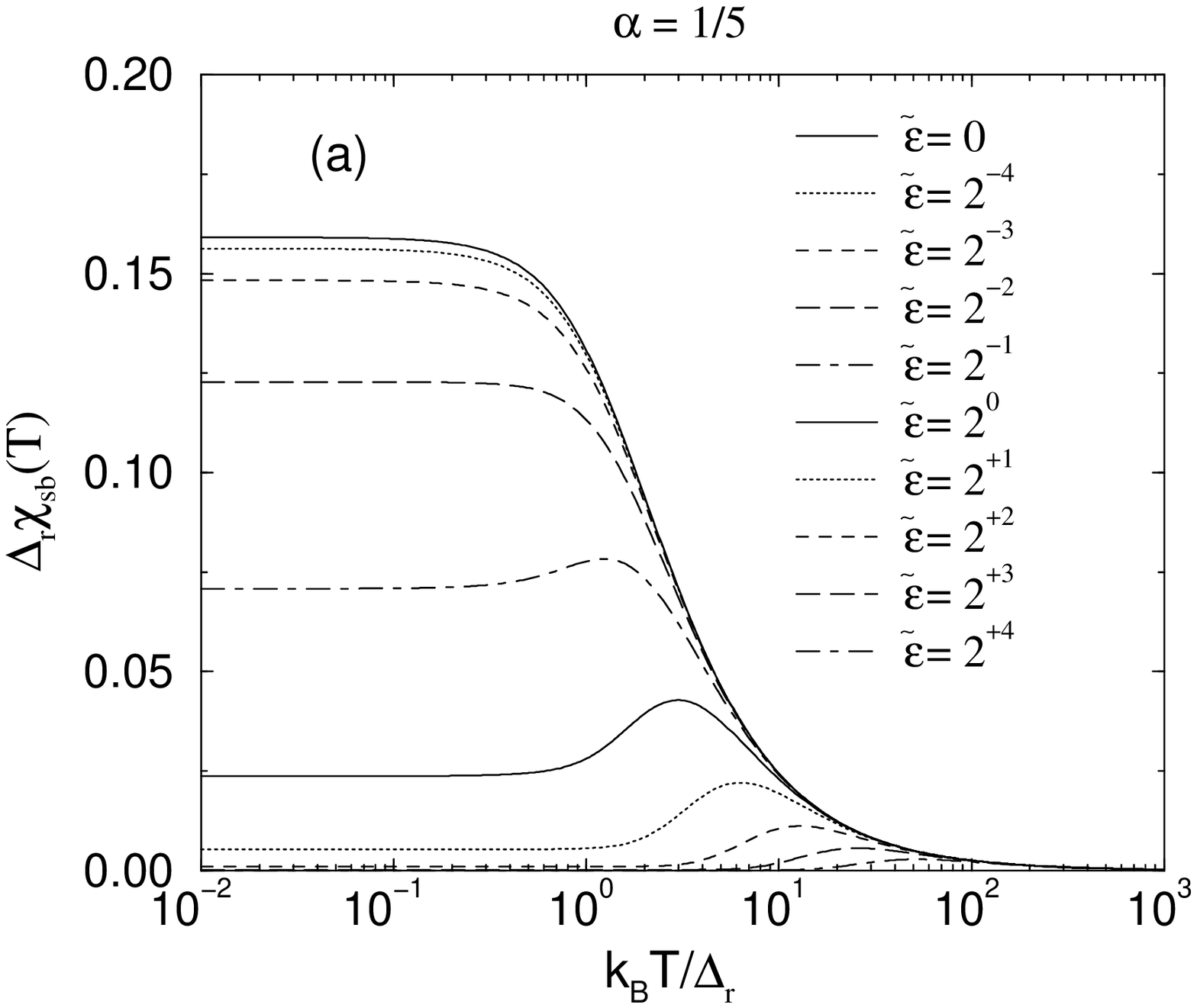}}}
\vspace{0.1cm}
\centerline{\epsfysize 6.1cm 
{\epsffile{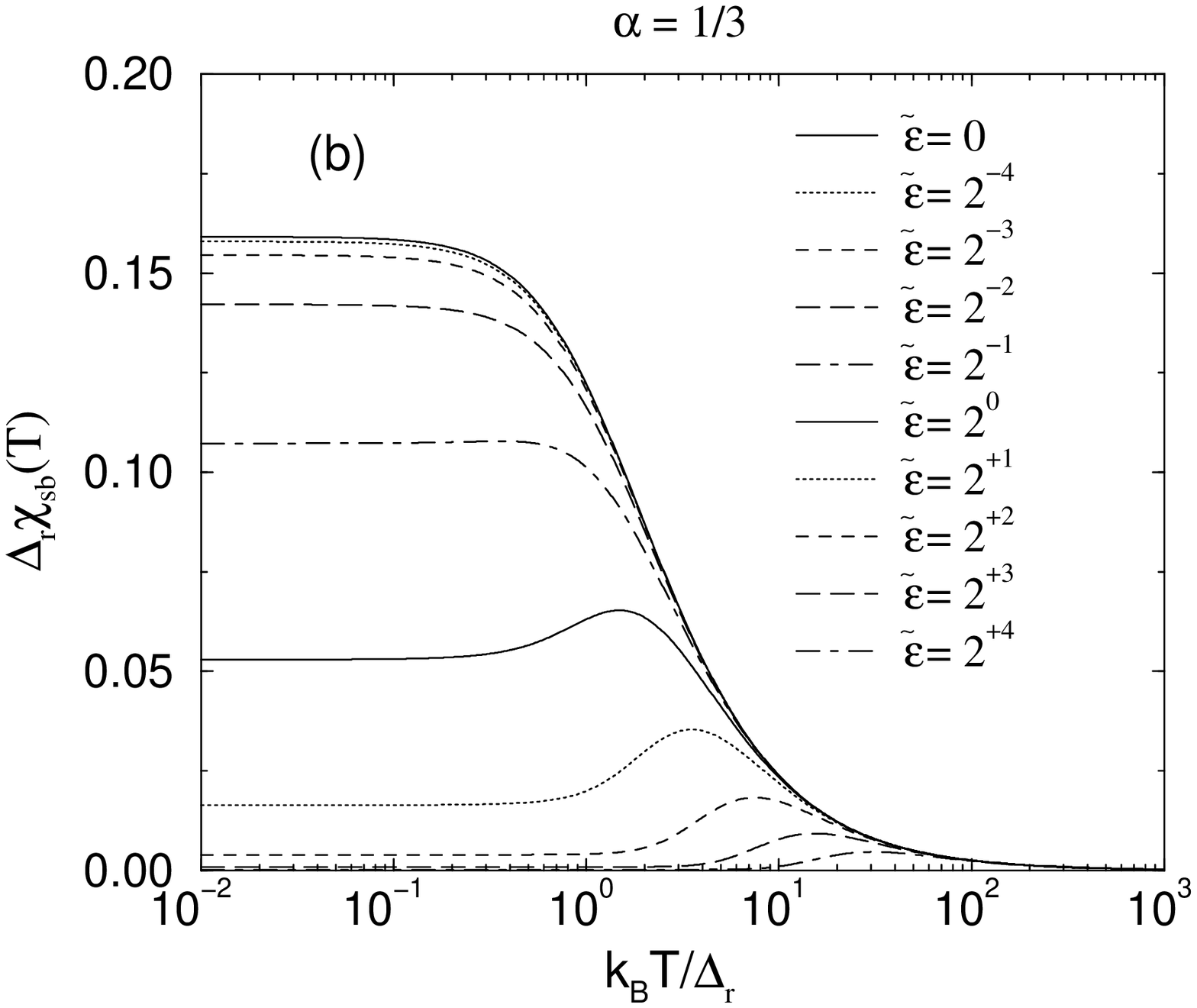}}}
\vspace{0.1cm}
\caption{
Dielectric susceptibility, $\chi_{sb}(T)$, for a range of asymmetries 
and some typical cases of weak dissipation, (a) $\alpha=1/5$, 
(b) $\alpha=1/3$.
}
\label{susc-asymmetric-all-alpha-weak}
\end{figure}

\begin{figure}[t]
\centerline{\epsfysize 6.1cm 
{\epsffile{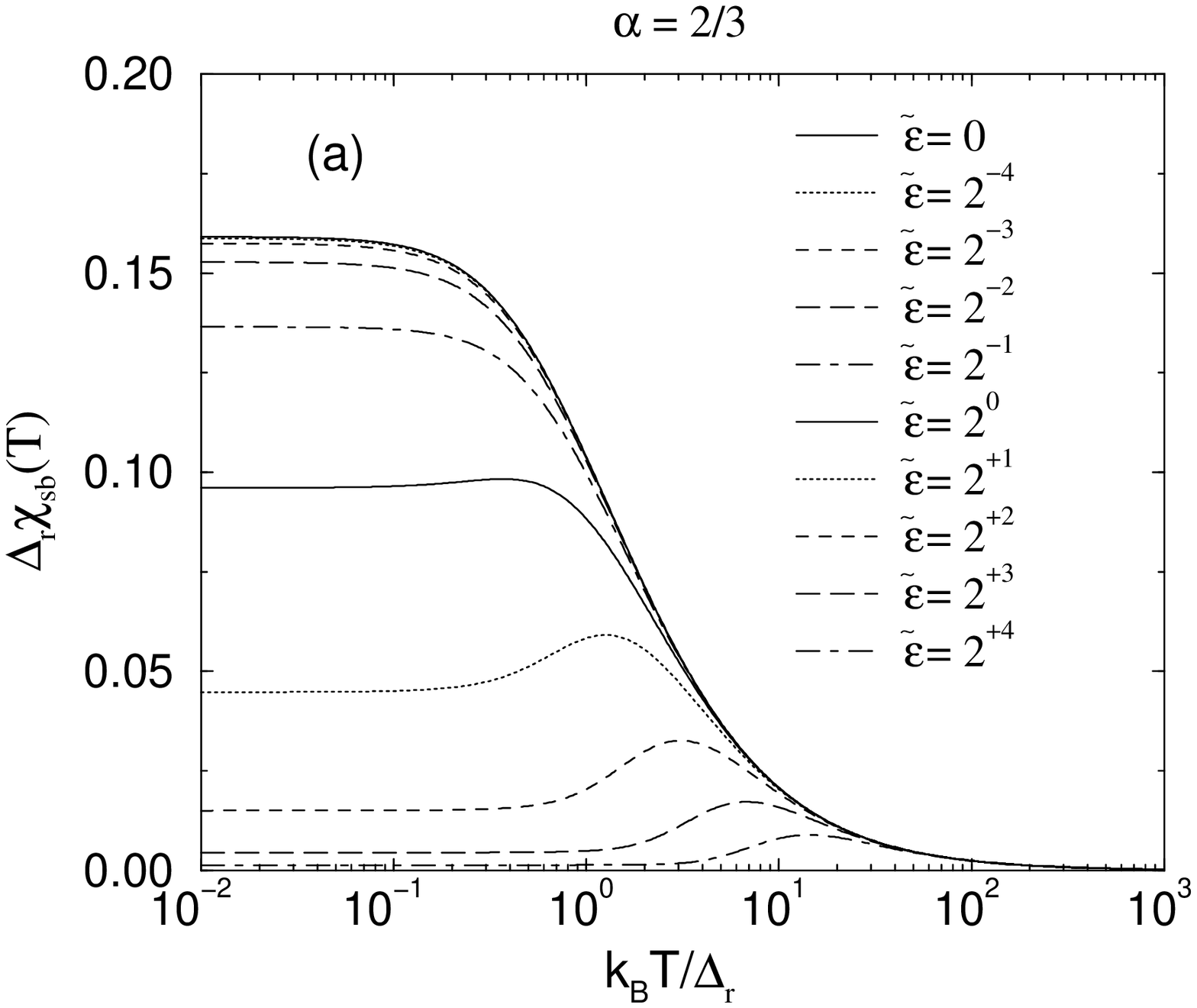}}}
\vspace{0.1cm}
\centerline{\epsfysize 6.1cm 
{\epsffile{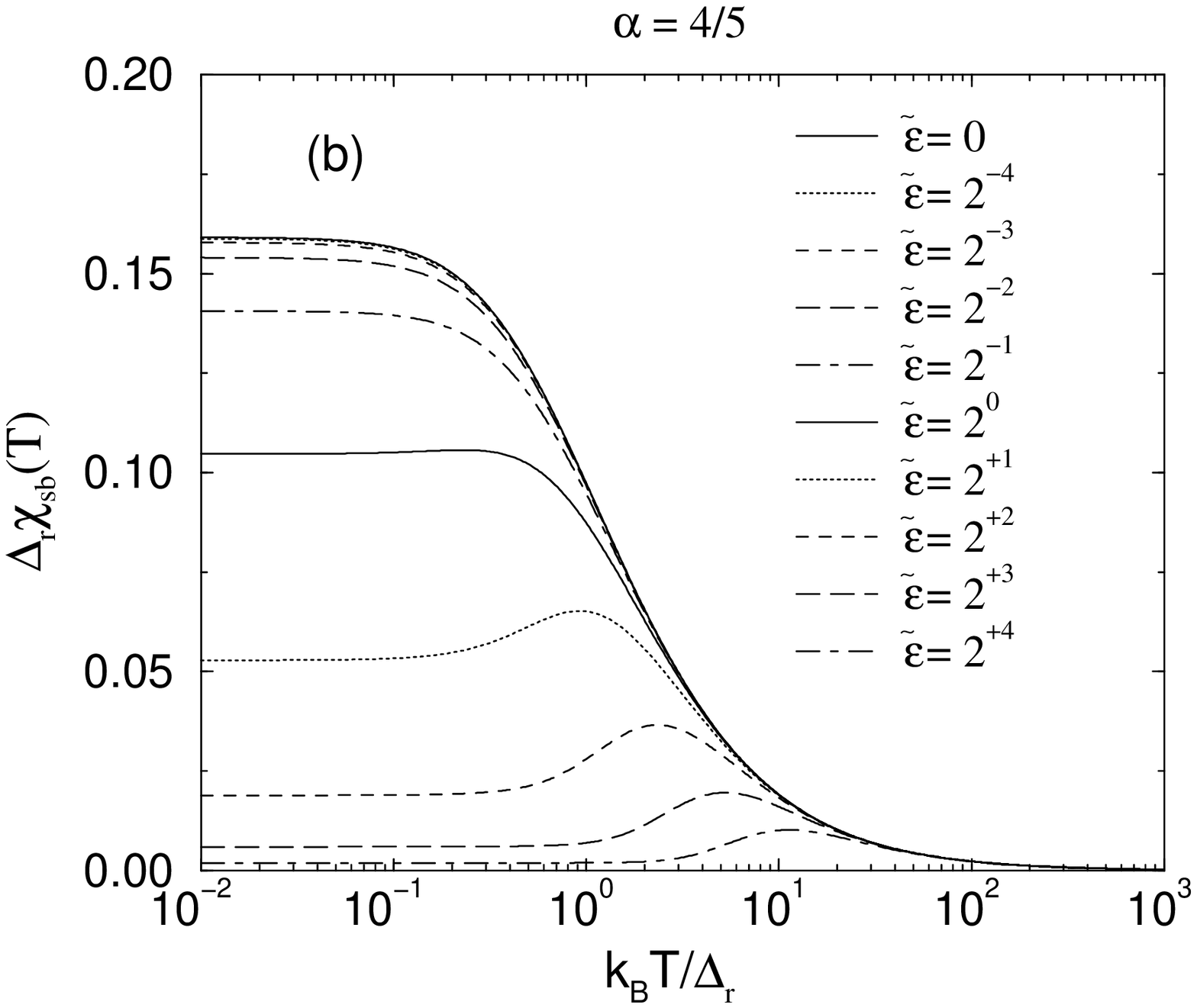}}}
\vspace{0.1cm}
\caption{
Dielectric susceptibility, $\chi_{sb}(T)$, for a range of asymmetries 
and some typical cases of strong dissipation, (a) $\alpha=2/3$,
(b) $\alpha=4/5$.
}
\label{susc-asymmetric-all-alpha-strong}
\end{figure}

The dielectric susceptibility in the presence of a level asymmetry is shown
in Fig.\,\ref{susc-asymmetric-all-alpha-weak}--\ref{susc-asymmetric-all-alpha-strong}. For all dissipation strengths we see that a sizeable asymmetry
of the order of $\Delta_{r}$ is required to give a finite temperature
peak in $\chi_{sb}$. The approach of the susceptibility to its high
temperature limit of $1/4T$ is governed by the same power laws as those 
found for the symmetric case and verified 
in Fig.\,\ref{chi-sd3-high-temp-corrections}.

\subsubsection{Wilson ratio}
The Wilson ratio for the Ohmic two-state model, $R_{sb}$, was defined earlier 
together with the usual Wilson ratio, $R_{akm}$, for the AKM. These take
the values $2/\alpha$ and $2$ respectively and are valid for both
the symmetric and asymmetric cases (i.e. in both zero and finite fields for
the Kondo model). The Wilson ratio served as a useful check on our numerical
solution, which recovered it with an accuracy of not less than 4 decimal
places for all $\alpha$ and $\varepsilon$.

\section{SUMMARY AND DISCUSSION}
\label{sec-conclusions}
In the present paper we studied the thermodynamics of a TSS  
with Ohmic dissipation by exploiting a mapping between the DTSS and
the anisotropic Kondo model, and solving the Bethe Ansatz equations
derived by Wiegman and Tsvelik for the latter. 
Treating the BA equations in a careful way we were able to 
calculate  essentially exactly 
the specific heat and the susceptibility of the DTSS for all
temperatures and level asymmetries in the delocalized phase of the DTSS,
characterized by dissipation strengths $0<\alpha<1$. The Bethe Ansatz
solution makes the universal properties of the DTSS clear: thermodynamic
quantities are universal functions of two variables, 
$k_{B}T/\Delta_{r}$ and $\varepsilon/k_{B}T$ for all $\alpha<1$. In the
limit $\alpha\rightarrow 1^{-}$ these functions reduce to those of
the usual isotropic $S=1/2$ Kondo model, which is seen as a special point 
in the parameter space of the DTSS \cite{costi.98}. The well known logarithmic
corrections to physical quantities for $\alpha\rightarrow 1^{-}$ 
at high temperatures
$k_{B}T\gg \Delta_{r}$ give way to power law corrections away from $\alpha=1$. 
We determined these power laws both analytically and numerically for all 
$\alpha<1$ at finite $\varepsilon$. In the context of the RG 
the change from logarithmic to power law corrections to physical quantities
at high temperature indicates that the tunneling term in the Hamiltonian 
changes from being marginally relevant (about the high energy fixed point) 
at $\alpha=1$ to relevant at $\alpha<1$ as discussed in 
Sec.~\ref{sec-rgflow}. At low temperature $k_{B}T\ll \Delta_{r}$ the
thermodynamics of the DTSS is that of a renormalized local Fermi liquid
with an enhanced linear specific heat $C(T)\sim \alpha k_{B}T/\Delta_{r}$, 
and an enhanced, but finite, dielectric 
susceptibility at $T=0$, $\chi_{sb}(T=0)=1/2\pi\Delta_{r}$.
The renormalizations increase dramatically as 
$\alpha\rightarrow 1^{-}$ due to the strong renormalization of the low 
energy scale $\Delta_{r}/\omega_{c}\sim (\Delta/\omega_{c})^{1/(1-\alpha)}$.

We have shown that the characteristic thermodynamic properties 
of the DTSS change smoothly as one increases the dissipation
strength from $\alpha \ll 1/2$ to $ 1 > \alpha > 1/2$, 
corresponding to weak and strong dissipations, respectively. 
In the former case, where the DTSS displays coherent oscillations between 
its two positions \cite{weiss.99}, we find the expected tendency towards
activated behaviour in the specific heat. A clear signal of this 
behaviour is the appearance of a peak in $C(T)/T$ at the renormalized 
tunneling amplitude $\Delta_{r}$. Such a peak is absent for dissipations 
$\alpha>1/3$ in the symmetric case. A finite level asymmetry accentuates 
the tendency towards activated behaviour and always gives rise to a finite
temperature peak in $C(T)/T$ at temperature
$k_{B}T=\tilde{\Delta}_{r} \sim \sqrt{\Delta_r^2 + \varepsilon^2}$ provided
$\varepsilon \ge \Delta_{r}$. For strong dissipation $\alpha>1/2$, the specific
heat is qualitatively similar to that of isotropic $S=1/2$ Kondo systems and
shows a monotonically decreasing $C(T)/T$ with increasing temperature.
The dielectric susceptibility of the DTSS was calculated at all dissipation
strengths, level asymmetries and temperatures for the first time.
For the symmetric case we found that this quantity decreases monotonically
with increasing temperature for all dissipation strengths and develops
a finite temperature peak only for sufficiently large level asymmetries
(of the order of $\Delta_{r}$).

The Ohmic two-state system is a generic model capable of describing 
a large number of different physical systems. Previous theoretical and
experimental work on such systems has, however, largely focussed on the 
dynamic properties. Given the detailed understanding which we now have 
of the thermodynamics, it may be worthwhile to consider also 
thermodynamic measurements on Ohmic two-state systems. We therefore 
briefly mention below some possibilities where our results could be 
directly tested. 

One of the possible physical realizations is provided by two-level 
systems in metals. Amorphous metals are not the best candidates, 
however, since in these materials a broad distribution
of DTSS's with very different physical parameters occur.  Therefore,
to calculate their contribution to the thermodynamic 
properties one should average  over them. This averaging is
not impossible, but would require solving the BA equations for
arbitrary values of $\alpha$, something which could be implemented by 
using further results of Ref~\cite{tsvelik.83}. In addition 
one would require the form of the TSS distribution. Better candidates 
are ${\rm H}$ tunneling in ${\rm Nb}$ \cite{wipf.84} and
metallic materials with tunneling centers 
formed by substitutional impurities, such as $Pb_{1-x} Ge_{x}
Te$ \cite{PbGeTe}.  Good quality single 
crystals can be produced from the latter alloy. 
The Germanium ions form identical eight-state
systems, which couple to the conduction electrons. 
Applying external stress on the sample one can reduce 
the degeneracy of the lowest lying states to two.
Since in this case the DTSS's have approximately identical parameters
for weak concentrations their individual thermodynamic properties can be 
observed  by measuring the sample's macroscopic properties.

Another possible candidate for thermodynamic measurements is 
provided by a SQUID. In this case the energy difference between the
two flux states of the SQUID can be easily tuned by an 
external magnetic field.  Measuring the average flux 
$ \langle\Phi\rangle$ as a function of 
the asymmetry energy one can readily determine the susceptibility
$\chi \sim \partial  \langle\Phi\rangle/\partial B$ that can directly 
be compared to our calculations. 

Finally, the results we have obtained for the thermodynamics of the
anisotropic Kondo model may have some relevance to the problem of 
an isotropic $S=1/2$ Kondo impurity in a Luttinger liquid. Schiller 
and Ingersent \cite{schiller.95} have shown that a $S=1/2$ Kondo 
impurity in a modified Luttinger liquid consisting only of left-moving 
spin-down electrons and right-moving spin-up electrons can be mapped 
exactly onto the AKM. It is clear that in a more realistic description 
of the Luttinger liquid that this exact mapping will only be approximate, 
nevertheless we expect that some of the general trends in the thermodynamic 
properties of a $S=1/2$ Kondo impurity in an interacting system should be 
captured by such a mapping onto the AKM or equivalently onto the Ohmic 
two-state system. The strength, $U$, of the Coulomb interaction can then 
be related \cite{schiller.95} to $J_{\parallel}$ in the AKM and hence to 
the dissipation strength $\alpha$ in the Ohmic two-state system.
The non-interacting system with $U=0$ corresponds to $\alpha=1$ 
and $U\rightarrow \infty$ corresponds to $\alpha=0$. Within such a 
picture we therefore expect, from the results of this paper, that the 
thermodynamic scaling functions of an isotropic $S=1/2$ Kondo impurity 
in a Luttinger liquid will change continuously with increasing Coulomb 
interaction $U$ (or decreasing dissipation strength $\alpha$). At low 
temperature the thermodynamics of such systems should be similar to that 
of a local Fermi liquid, as recently discussed in \cite{wang.98}. For 
sufficiently large $U$ (i.e. sufficiently small $\alpha$), 
$C(T)/T$ may develop a finite temperature peak at 
$k_{B}T\sim \Delta_{r}\sim T_{K}$ in analogy to our findings for the Ohmic
two-state system. Such a peak could then be taken as a signature of the
Kondo effect in a strongly interacting system. The unusual heavy fermion 
behaviour of ${\rm Nd_{2-x}Ce_{x}CuO_{4}}$, e.g. the non-monotonic 
behaviour of $C(T)/T$ \cite{brugger.93},
may be consistent with such an interpretation. Strong interactions have
also been invoked to explain this behaviour in Ref.~\cite{fulde.93}. 
Other, more conventional explanations, such as lattice coherence effects, 
cannot be ruled out however.

The approach we have developed in this paper can be extended to more 
complicated models of two-level systems coupled Ohmically to an
environment. The effect of indirect or electron-assisted tunneling 
processes \cite{Zawa,muramatsu.86} on the thermodynamics of two-level 
systems in metals will be studied in a future publication.

\acknowledgments
The authors are grateful to N. Andrei and A.M. Tsvelik for
useful discussions. We are grateful also to C. Roth for help with
Bosonization in Appendix~\ref{app:equiv}. This research has been supported by 
the Hungarian Grant Nos. OTKA~T026327, OTKA~F016604, and the 
U.S-Hungarian Joint Fund No.~587. G. Z. has been 
supported by the Magyary Zolt\'an Foundation and grant 
No.~DE-FG03-97ER45640 of the U.S DOE Office of Science, Division of 
Materials Research. T. A. C. thanks the
Deutsche Forschungsgemeinschaft for financial support and the
hospitality of the Max Planck Institut f\"{u}r Physik Komplexe Systeme,
Dresden and the Centre for Electronic Correlations and Magnetism, 
University of Augsburg, Germany, where part of this research was carried out. 

\appendix
\section{Equivalence of the AKM to the Ohmic two-state system}
\label{app:equiv}
\subsection{Bosonization of free fermions}
Consider first the free fermion Hamiltonian,
\begin{equation}
H_{0}=\sum_{k\mu}\varepsilon_{k}c_{k\mu}^{\dagger}c_{k\mu}. 
\end{equation}
We take a linear dispersion 
relation for the conduction electrons, $\varepsilon_{k}=v_{F}k$, and measure 
$k$ relative to the Fermi wavenumber $k_{F}$. 
We introduce fermion fields $\psi_{\mu}(x)$ via a Fourier series,
\begin{equation}
\psi_{\mu}(x) = L^{-1/2}\sum_{k}e^{-ikx}c_{k\mu}
\end{equation}
with wavenumbers $k=2\pi n/L, n=0,\pm 1,\dots$ for periodic boundary 
conditions appropriate to a finite system of length $L$. The density 
of states per spin direction is given by 
$\rho=1/2\pi v_{F}$. The kinetic energy can then be written as,
\begin{equation}
H_{0}=v_{F}\sum_{k,\mu} k\, c_{k\mu}^{\dagger}c_{k\mu}=
i v_{F}\int_{-L/2}^{+L/2}
\psi_{\mu}^{\dagger}(x)\partial_{x}\psi_{\mu}(x)\,dx
\end{equation}
The fields $\psi_{\mu}(x)$ are expressed in terms of hermitian bosonic
fields $\varphi(x)$ in the standard way,
\begin{equation}
\psi_{\mu}(x)=(2\pi a)^{-1/2}F_{\mu}e^{-i\varphi_{\mu}(x)}
\end{equation}
where $a$ is a lattice spacing, required for obtaining convergent 
momentum sums.
The $F_{\mu}$ are the Klein factors required to  
ladder between states with different fermion number, 
and to ensure the correct anticommutation relations for the fermion 
fields \cite{kotliar.96,vondelft.98}, and
\begin{equation}
\varphi_{\mu}(x)=\phi_{\mu}(x)+\phi_{\mu}^{\dagger}(x),
\end{equation}
with the bosonic fields $\phi,\, \phi^{\dagger}$ given by
\begin{equation}
\phi_{\mu}^{\dagger}(x)=(\phi_{\mu}(x))^{\dagger}\equiv 
=-\sum_{q>0}n_{q}^{-1/2}e^{iqx}a_{q\mu}^{\dagger}e^{-aq/2}.
\end{equation}
The $a_{q},\, a_{q}^{\dagger}$, defined for $q>0$, satisfy boson commutation 
relations with $n_{q}=(qL/2\pi)^{1/2}$, and are given by,
\begin{equation}
a_{q\mu}^{\dagger}=(a_{q\mu})^{\dagger}=
i\,n_{q}^{-1/2}\,\sum_{k}c_{k+q\mu}^{\dagger}c_{k\mu}
\end{equation}
It is convenient to introduce spin and charge density operators in place
of $a_{q\uparrow},a_{q\downarrow}$ as follows:
\begin{eqnarray*}
a_{qC}=\frac{1}{\sqrt{2}}\left(a_{q\uparrow}+a_{q\downarrow}\right)\\
a_{qS}=\frac{1}{\sqrt{2}}\left(a_{q\uparrow}-a_{q\downarrow}\right).
\end{eqnarray*}
The corresponding hermitian fields are,
\begin{eqnarray*}
\varphi_{C}=\frac{1}{\sqrt{2}}\left(\varphi_{\uparrow}+\varphi_{\downarrow}
\right)\\
\varphi_{S}=\frac{1}{\sqrt{2}}\left(\varphi_{\uparrow}-\varphi_{\downarrow}
\right),
\end{eqnarray*}
with commutation relations,
\begin{eqnarray}
\left[\varphi_{C}(x),\,\varphi_{S}(x')\right] &=& 0,\\
\left[\varphi_{C,S}(x),\,\partial_{x'}\varphi_{C,S}(x')\right] 
& = & 2\pi i\delta(x-x')
\end{eqnarray}
In terms of these we have for the fields and densities,
\begin{eqnarray}
\psi_{C,S}(x) &\equiv& \frac{F_{C,S}}{\sqrt{2\pi a}}e^{-i\varphi_{C,S}(x)}\\
\rho_{C,S}(x) &\equiv& \frac 1{\sqrt{2}} \psi_{C,S}^{\dagger}(x)\psi_{C,S}(x) =
\frac{1}{2\pi \sqrt{2}}
\partial_{x}\varphi_{C,S}(x) 
\end{eqnarray}
\subsection{Transformation to the Ohmic spin-boson model}
We start with the anisotropic Kondo model
\begin{equation}
H_{AKM}=H_{0}+H_{\perp}+H_{\parallel}
\end{equation}
with $H_{0}$ as above and,
\begin{eqnarray*}
H_{\perp} &=&
\frac{J_{\perp}}{2}\sum_{k,k'}\,(c_{k\uparrow}^{\dagger}c_{k'\downarrow}\,S^{-}+c_{k\downarrow}^{\dagger}c_{k'\uparrow}\,S^{+})\\
H_{\parallel}&=&\frac{J_{\parallel}}{2}\sum_{k,k'}\,(c_{k\uparrow}^{\dagger}c_{k'\uparrow}-c_{k\downarrow}^{\dagger}c_{k'\downarrow})S_{z}
\end{eqnarray*}
In terms of $\varphi_{C,S}$, we can write,
\begin{eqnarray*}
 H_{0} &=& v_{F}\sum_{q>0}q\,(a_{q\uparrow}^{\dagger}a_{q\uparrow}
+a_{q\downarrow}^{\dagger}a_{q\downarrow})\\
&=& \frac{v_{F}}{2}\int_{-L/2}^{+L/2}\frac{dx}{2\pi}\, :
\left(\partial_{x}\varphi_{C}(x)\right)^{2}+
\left(\partial_{x}\varphi_{S}(x)\right)^{2}:\\
H_{\parallel} & = & \frac{J_{\parallel}}{2} \,S_{z}\,
( \psi_{\uparrow}^{\dagger}(0)\psi_{\uparrow}(0) - 
\psi_{\downarrow}^{\dagger}(0)\psi_{\downarrow}(0) )\\
& = & \frac{J_{\parallel}}{2} S_{z}\, \frac{1}{2\pi}\sqrt{2}\,\partial_{x}\varphi_{S}(0)\\
H_{\perp} &=& \frac{J_{\perp}}{2}\,(\psi_{\uparrow}^{\dagger}(0)
\psi_{\downarrow}(0)\,S^{-}+\psi_{\downarrow}^{\dagger}(0)\psi_{\uparrow}(0)\,S^{\dagger})\\ 
& = & \frac{J_{\perp}}{4\pi a}
\left( e^{i\sqrt{2}\varphi_{s}(0)}F_{\uparrow}F_{\downarrow}^{\dagger}\,S^{-}+
e^{-i\sqrt{2}\varphi_{s}(0)}F_{\downarrow}F_{\uparrow}^{\dagger}\,S^{+}\right)
\end{eqnarray*}

We note that $\varphi_{C}$ (which commutes with $\varphi_{S}$) 
does not couple to the impurity and only gives
a contribution to the kinetic energy. 

\subsubsection{Canonical transformation on the bosonized Model}

We show that the unitary transformation 
$U=\exp(i\sqrt{2}S_{z}\varphi_{S}(0))$ applied to $H_{AKM}$ gives 
the spin-boson Hamiltonian, $H_{SB}$, for Ohmic dissipation, i.e. that
$U H_{AKM} U^{\dagger}=H_{SB}$. We use the Baker-Hausdorff formula 
$e^{-B}Ae^{B}=A+[A,B]$ with $[A,B]$ a c-number and the commutation 
relations for $\varphi_{C},\varphi_{S}$, to obtain,
\begin{eqnarray*}
U H_{0} U^{\dagger} &=&
H_{0}-\sqrt{2}v_{F}S_{z} \left. \frac{\partial\varphi_{S}}{\partial x}
\right| _{x=0}\\
&=& H_{0}-\sqrt{2}v_{F}S_{z}\\
&\times&\sum_{q}\sqrt{\frac{2\pi q}{L}}\,
\left(ia_{qS}+(ia_{qS})^{\dagger}\right)\,e^{-aq/2},\\
U H_{\parallel} U^{\dagger} &=& H_{\parallel} + \mbox{constant},\\
U H_{\perp} U^{\dagger} &=& \frac{J_{\perp}}{4\pi a}\,
(e^{i\sqrt{2}\varphi_{s}(0)}
F_{\uparrow}F_{\downarrow}^{\dagger}\,U S^{-} U^{\dagger}\\
&+& e^{-i\sqrt{2}\varphi_{s}(0)}
F_{\downarrow}F_{\uparrow}^{\dagger}\,U S^{+} U^{\dagger}.
\end{eqnarray*}
On using the identities,
\begin{eqnarray*} 
U S^{-} U^{\dagger} &=& e^{-i\sqrt{2}\varphi_{S}(0)}S^{-},\\
U S^{+} U^{\dagger} &=& e^{i\sqrt{2}\varphi_{S}(0)}S^{+},
\end{eqnarray*}
and the representation, 
$\frac{1}{2}\sigma^{+}=F_{\downarrow}F_{\uparrow}^{\dagger}\,S^{+}$,\, 
$\sigma^{-}=(\sigma^{+})^{\dagger}$ and $\frac{1}{2}\sigma_{z}=S_{z}$,
of the Pauli spin operators, the term $UH_{\perp}U^{\dagger}$ becomes,
\begin{equation}
U H_{\perp} U^{\dagger} = J_{\perp}\,\frac{1}{4\pi
a}\,\sigma_{x}.
\end{equation}

\subsubsection{Identification of the parameters of the Kondo Hamiltonian with
those of the Spin-boson Hamiltonian}

We notice that $a_{qC}$ does not couple to the
impurity so we write $a_{q}=i a_{qS}$ and omit the charge density operators
to obtain the Hamiltonian for the spin density excitations, 
\begin{eqnarray*}
H_{SB} &=& \frac{J_{\perp}}{4\pi a}\sigma_{x} + 
v_{F}\sum_{q}q\,a_{q}^{\dagger}a_{q}\\
&+&\left(\frac{J_{\parallel}}{4\pi}-v_{F}\right)\sqrt{2}\frac{\sigma_{z}}{2}
\sum_{q>0}\sqrt{\frac{2\pi q}{L}}\,\left(a_{q}+a_{q}^{\dagger}\right)\,e^{-aq/2}.
\end{eqnarray*}
This is precisely the spin-boson model,
\[H_{SB}=\sum_{q>0}\omega_{q}\,a_{q}^{\dagger}a_{q}-\frac{\Delta}{2}\sigma_{x}+\frac{q_{0}}{2}\sigma_{z}\sum_{q}\frac{C_{q}}{\sqrt{2m_{q}\omega_{q}}}\,\left(a_{q}+a_{q}^{\dagger}\right)
\]
with $\omega_{q}=v_{F}q$ and a spectral function for the harmonic oscillators,
\[J(\omega)=\frac{\pi}{2}\sum_{q}\frac{C_{q}^{2}}{m_{q}\omega_{q}}\delta(\omega-\omega_{q})
=\frac{2\pi\alpha}{q_{0}^{2}}\,\omega\,
e^{-\frac{\omega}{\omega_{c}}},\]
provided one chooses, 
\begin{eqnarray*}
\frac{C_{q}}{\sqrt{m_{q}}}&=&-\sqrt{\alpha}\frac{2}{q_{0}}\,\left(
\frac{2\pi v_{F}}{L} \right) ^{1/2}\,\omega_{q}\,e^{-\frac{\omega_{q}}{2\omega_{c}}},
\end{eqnarray*}
with a cutoff,
\[\omega_{c}=\frac{v_{F}}{a}.\]
One can identify the parameters,
\[ -\frac{\Delta}{2}=\frac{J_{\perp}}{4\pi a}\]
and
\[-\sqrt{\alpha}=\frac{J_{\parallel}}{4\pi v_{F}}-1,\]
which, together with the density of states (per spin) of the conduction 
electrons, $\rho=1/2\pi v_{F}$, and the above definition of $\omega_{c}$, 
result in the following identification of the dimensionless couplings of 
the two models,
\begin{eqnarray}
\frac{\Delta}{\omega_{c}}=-\rho J_{\perp}\\
\alpha=(1-\frac{1}{2}\rho J_{\parallel})^{2}
\label{eq:alpha0}
\end{eqnarray}
The sign of $J_{\perp}$ (or $\Delta$) plays no role and we may
choose $\Delta/\omega_{c}=+\rho J_{\perp}>0$. 

We remark here that the precise form of Eq.~(\ref{eq:alpha0}) depends on
the specific regularization scheme used. Within the framework of 
Abelian bosonization the coupling $J_\parallel$ is directly proportional 
to the phase shifts,  $\rho J_\parallel = 4\pi/\delta$.\cite{EKrev} 
For a finite-band  model with cut-off $\omega_{c}$, the expression for 
$\alpha$ in terms of $\rho J_{\parallel}$ the phase shift is given by 
$\delta=-\arctan(\pi\rho J_{\parallel}/4)$, and Eq.~(\ref{eq:alpha0})
takes the more general form:
\begin{equation}
\alpha = (1+\frac{2}{\pi}\delta)^{2}.
\label{eq:alpha} 
\end{equation}

\section{Derivation of the thermodynamical Bethe Ansatz equations}
\label{tba-derivation}
\subsection{Bethe Ansatz equations for finite system sizes}
To construct the Bethe Ansatz (BA) solution of the model one first
rewrites  the AKM Hamiltonian Eq.~(\ref{eq:AKM}) in coordinate space as:
\begin{equation}
H = \sum_{j = 1}^{N_e} \Bigl\{ 
-i {\partial_j} +\frac12 \delta(x_j)\bigl[ 
I_z \sigma_j^z S^z  
  +  I_\perp (\sigma_j^+ S^- + \mbox{h.c})\bigr]\Bigr\} \;,
\label{eq.H}
\end{equation}
where $x_j$ and $\vec {\sigma_j}$ denote the coordinate and spin of conduction 
electron $j$, ($j = 1,..,N_e$), and $\vec S =  \vec {\sigma_0}/2$ is the 
impurity spin. The explicit relationship between the dimensionless couplings 
$I_{z,\perp}$ and the $J$'s in Eq.~(\ref{eq:AKM}) is not straightforward and depends 
on the specific regularization and cutoff scheme  used in the original model 
and the BA approach. For small couplings $I_z \approx J_z$ and $I_\perp 
\approx  J_\perp$. 

As a next step one constructs the impurity-conduction electron scattering 
matrix from Eq.~(\ref{eq.H}). Introducing
the impurity and the conduction electron 'rapidities', 
$\lambda_0 = -1$ and $\lambda_j = 0$, 
the s-matrix between the $j$'th electron and the local impurity spin
can be written as:
\begin{equation}
 S_{0j}(\lambda_0 - \lambda_j)_{\sigma_0 \sigma_0';
\sigma_j \sigma_j'}\;,
\end{equation}
where --- apart from some unimportant phase factors --- $ S_{0j} $ can 
be expressed as
\begin{eqnarray}	
\label{eq.S}
S_{0j}(\lambda)& = &
a(\lambda) P_{\uparrow\uparrow} + 
b(\lambda) P_{\uparrow\downarrow} +
\frac12 { c(\lambda)}
\bigl(\sigma_0^+\otimes\sigma_j^- + \mbox{h.c.}\bigr)\;,
\nonumber\\
&& {a(\lambda)\over c(\lambda)} = 
{\sinh(i\mu + \lambda f ) \over \sinh(i\mu)}\;, 
 \label{eq.smatrix}\\
&& {a(\lambda)\over b(\lambda)} = 
{\sinh(i\mu + \lambda f) \over \sinh(\lambda f)}\;,
\nonumber 
\end{eqnarray}
with $P_{\uparrow\uparrow}$ and $P_{\uparrow\downarrow}$ being the 
projection operators for parallel and opposite spins.

The parameters $\mu$ and $ f $ are connected to the bare couplings
via:
\begin{eqnarray}
\cos(\mu) &=& {\cos(I_{||} /2) \over \cos(I_\perp / 2 )}, 
\;\;\;\; \\
{\rm cth}^2(f) &=&  {\sin^2(I_{||} /2) \over \sin((I_{||} + I_\perp)/ 2 )
\sin((I_{||} - I_\perp)/ 2 )}\;. 
\end{eqnarray}
The electron-electron s-matrix is not determined by Eq.~(\ref{eq.H}) but is 
fixed by the requirement of integrability, and is simply given by 
$S_{ij}(\lambda_i-\lambda_j)$. With this choice  the impurity-electron 
and electron-electron s-matrices belong to the same family  of
s-matrices, Eq.~(\ref{eq.smatrix}), and satisfy the following
Yang-Baxter relations: 
\begin{equation}
S_{ij}(\alpha) S_{ik}(\alpha+\beta) S_{jk}(\beta) = S_{jk}(\beta) 
S_{ik}(\alpha + \beta)  S_{ij}(\alpha)\;
\label{eq.triangle}
\end{equation}
insuring the integrability of the model \cite{tsvelik.83,lec}.
The Bethe Ansatz (BA) equations are derived using standard algebraic methods,
based on Eq.~(\ref{eq.triangle}) by creating 'spin waves'  from the 
completely spin-polarized state 
\cite{tsvelik.83,lec}.
The spin  waves are characterized by rapidities 
$\lambda_\beta$ ($\beta = 1,..,M$) and the obtained state is an 
eigenstate of $H$ if these rapidities satisfy the following BA equations:
\begin{eqnarray}
&&\prod_{\beta = 1}^{M} 
{\sinh(\mu (\lambda_\alpha - \lambda_\beta +i)) \over 
\sinh(\mu (\lambda_\alpha - \lambda_\beta - i))} = 
-{\sinh(\mu (\lambda_\alpha + f/\mu +i/2)) \over 
\sinh(\mu (\lambda_\alpha  + f/\mu - i/2))} \nonumber \\
&&\phantom{nnnnn}\times \left[
{\sinh(\mu (\lambda_\alpha +i/2)) \over 
\sinh(\mu (\lambda_\alpha  - i/2))}\right]^{N_e}   \;.
\label{eq:spinBA}
\end{eqnarray}
Thus an eigenstate of the Hamiltonian is characterized by the momenta $k_j$ 
($j=1..N$) of the conduction electrons and the rapidities $\{\lambda_\beta\}$,
corresponding to the charge and spin degrees of freedom, respectively.
The energy of a state  is determined through the periodic boundary conditions:
\begin{eqnarray}
e^{i k_j L} & = &
\prod_{\beta = 1}^{M} 
{\sinh(\mu (\lambda_\beta  +i/2)) \over 
\sinh(\mu (\lambda_\beta - i/2))}\;, \\
E & = & \sum_{j=1}^N k_j- h g S_z^{\rm (tot)}\;.
\label{eq:energy}
\end{eqnarray}
Since  the $\lambda_\beta$'s  ($\beta=1,..,M$) are spin $-1$ excitations
\cite{tsvelik.83} the  total spin  is simply  $S_z^{\rm (tot)} = S^{\rm (tot)} = 
(N_e + 1 - 2M)/2$.

\subsection{Thermodynamic BA equations} 
To derive the thermodynamic BA equations (TBA) one takes the limit
$L,N\to\infty$, $N/ L =D$. In this limit the 
rapidities $\lambda_\alpha$ in Eq.~(\ref{eq:spinBA}) can be shown to 
organize into $r$-strings \cite{tsvelik.83}, $\lambda_\alpha \to \lambda^{(r,v)}_\beta$ ($\beta = 1,.., M_{(r,v)}$),  with  integer ranks $r$ and 
parity $v=\pm$: 
\begin{equation}
\lambda^{(r,v)} \Leftrightarrow  \lambda^{(r,v)}_{q=1,..,r} = 
\lambda^{(r,v)} + \bigl[ {\textstyle \frac {r+1}2} -q\bigr] + i {\textstyle{\pi\over 4\mu}}
(1-v)\;.
\end{equation}
The allowed  $(r,v)$ values  must satisfy the stability condition
\begin{equation}
v \sin(\mu q) \sin(\mu(r-q)) > 0\;,\;\;\;q=1,..,[r/2]\;,
\label{eq.stability}
\end{equation}
and can be determined by expressing 
$\mu/\pi$ as an infinite (finite) fraction. 
For rational  $\mu/\pi$'s only a finite number of possible ranks 
and parities survive  \cite{takahashi.73}. Specifically, for 
$\mu = \pi/\nu$ only the pairs
\begin{equation}
n=(r,v)= (1,+), (2,+),..,(\nu-1,+) \mbox{ and } (1,-)
\end{equation}
are allowed, which we label by $n=1,..,\nu$ in what follows.

After classifying the possible strings one  introduces the 
density of rapidities  of the $n=(r,v)$-strings, $\varrho_n(\lambda)$,
and the density of 'holes', $\tilde \varrho_n(\lambda)$. These are 
connected by a linear integral equation derived from Eq.~(\ref{eq:spinBA})
as 
\begin{equation}
\varrho_n(\lambda) = 
a_n(\lambda) - \sum_{n'} \int d\lambda' K_{nn'}(\lambda-\lambda')\varrho_{n'}
(\lambda')\;,
\label{contconstr}
\end{equation}
where the functions $a_n$ and the Kernel $K_{nn'}$ depend only on
$\mu$. The spin part of the energy, Eq.~(\ref{eq:energy}), and the
entropy can be expressed as: 
\begin{eqnarray}
E^{(spin)} &= &\sum_n\int d\lambda f_n(\lambda)\varrho_n(\lambda)\;, \\
S^{(spin)} &= & k_B \sum_n \int d\lambda \bigl[
\varrho_n \ln (1 + \frac {\tilde \varrho_n}{ \varrho_n})
+ \tilde \varrho_n \ln (1 + \frac {\varrho_n}{\tilde \varrho_n}) \bigr]\;.
\nonumber
\end{eqnarray} 
Then Eqs.~(\ref{eq:BAepsilonint}) and (\ref{eq.impfreeen}) can be  
derived simply by minimizing the free energy
$F^{(spin)} = E^{(spin)} - T S^{(spin)}$ with respect to 
$\delta\varrho_n(\lambda)$ using the constraint Eq.~(\ref{contconstr}) and  
introducing the 'excitation energies'  $\varepsilon_n(\lambda)/T
\equiv \ln(\tilde \varrho_n/\varrho_n)$.

\section{Numerical solution the the thermodynamic Bethe Ansatz equations}
\label{num-procedure}
In this appendix we describe in detail a new numerical procedure
for solving thermodynamic Bethe Ansatz equations of the form 
(\ref{eq:tba-ksi}) for the
quantities $\xi_{j},j=1,\dots,\nu$. Such equations arise also in
other contexts, for example, in the context of the anisotropic Heisenberg 
model\cite{takahashi.73}. 
The method we describe has the advantage that the thermodynamics is
calculated without the need to take numerical derivatives of the
free energy (see Sec.IVa). An implementation of this for the special
case of zero field static susceptibilities has been given in 
\cite{desgranges.85} for the multichannel Kondo model and in \cite{jerez.98} 
for the $SU(N)\times SU(f)$ 
Coqblin-Schrieffer model. We have gone further and have shown that other
quantities, such as the entropy and in principle also the specific heat,
can be calculated in the same way and for finite fields, as explicitly 
described in Sec.IVa. However, more importantly we show how to obtain with
the same numerical procedure uniformly accurate results for thermodynamics 
in the presence of an {\em arbitrarily} large level asymmetry, 
$\varepsilon/T$, (or equivalently an arbitrarily large magnetic field, 
$h/T$, in the Kondo model). This was particularly important for obtaining 
the very low temperature thermodynamics of the asymmetric two-level 
system in this paper. Previous techniques, for related TBA equations where
the same technical difficulty arises, have either not been able to 
access large $\varepsilon/T$ ($h/T$) \cite{rajan.82} or, 
\cite{sacramento.91}, have required solving a 
separate set of approximate TBA equations suitable for large 
$\varepsilon/T$ ($h/T$). The latter requires matching the resulting 
thermodynamics at the boundary of the two temperature ranges. 

The iterative procedure to solve the TBA equations (\ref{eq:tba-ksi}) 
is as follows. An 
evenly spaced grid of 1000 points $\{\lambda_{i},i=1,\dots,1000\}$ is 
used in the interval $-40\leq\lambda \leq +40$, the functions 
$\xi_{j}(\lambda_{i}),j=1,\nu$ are given initial values 
$\xi_{j}(\lambda_{i})=\xi_{j}^{m=1}(\lambda_{i})$, and are then
represented by least squares cubic splines\cite{NAG}. 
The next iteration, $\xi_{j}^{m=2}(\lambda_{i})$, was obtained by 
evaluating the integrals to a relative error of less than $10^{-5}$ at 
each point for each function in turn. This procedure was repeated until the
$\epsilon_{j}\equiv ||\xi_{j}^{N}-\xi_{j}^{N-1}||/||\xi_{j}^{N}||$,
$j=1,\dots,\nu$ with $||\xi_{j}||\equiv
\sqrt{\sum_{i}|\xi_{j}(\lambda_{i})|^{2}}$ 
(when non-zero) reached machine precision (approximately $10^{-16}$). 
Convergence depended on $\nu$, with the procedure converging for a typical
case after approximately $30,44$ and $61$ iterations for 
$\alpha=(1-1/\nu)>1/2$ and $\nu=3,4$ and $5$ respectively.
For fixed level asymmetry, $\varepsilon$, we solved the above equations
at each temperature. Similarly for $\alpha=1/\nu<1/2$ the number of
iterations required to reach convergence was typically $52,79$ and $133$ for 
$\nu=3,4$ and $5$ respectively. 
\begin{figure}[t]
\centerline{
{\epsfysize 3.7cm \epsffile{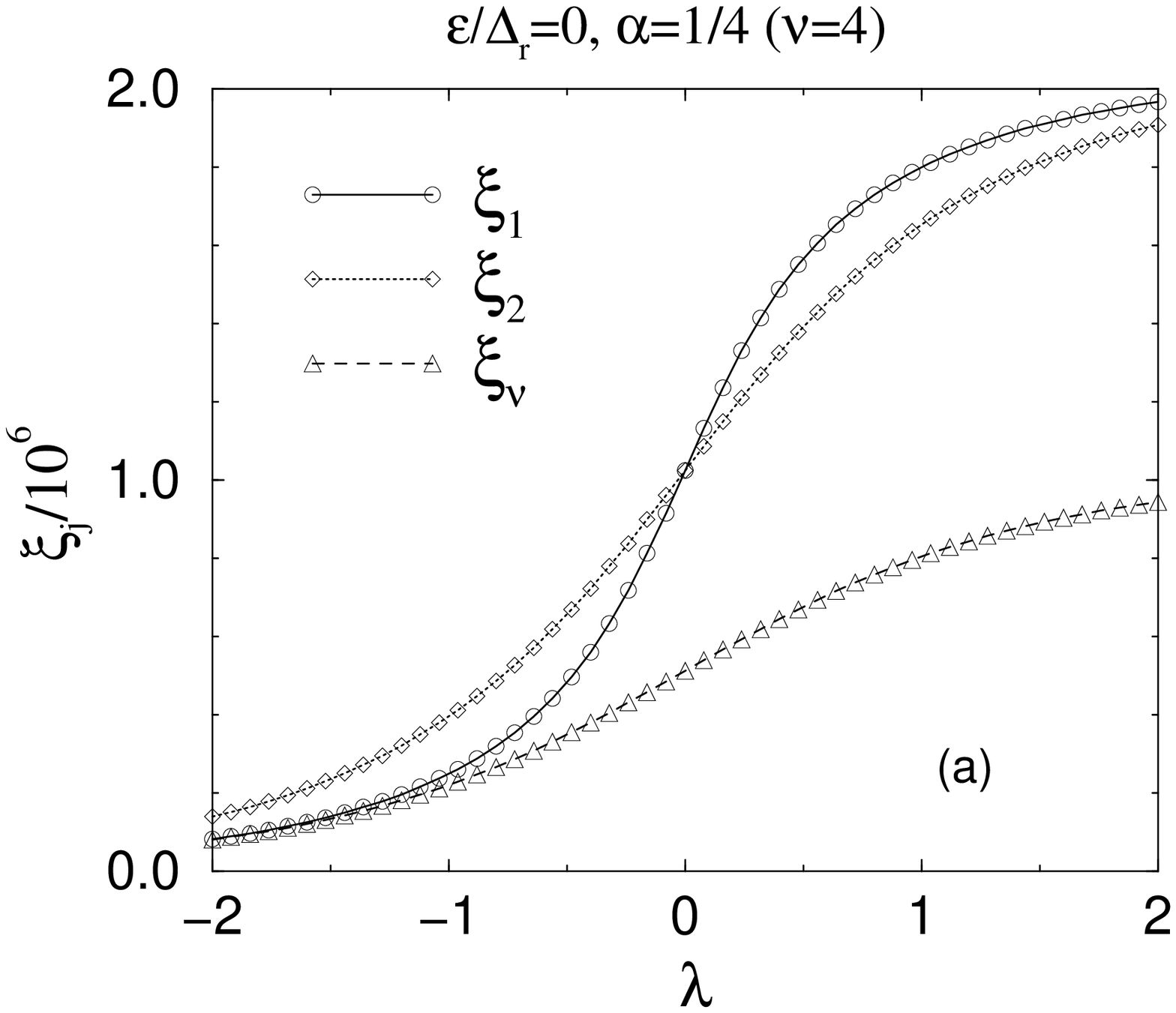}\epsfysize 3.7cm 
\epsffile{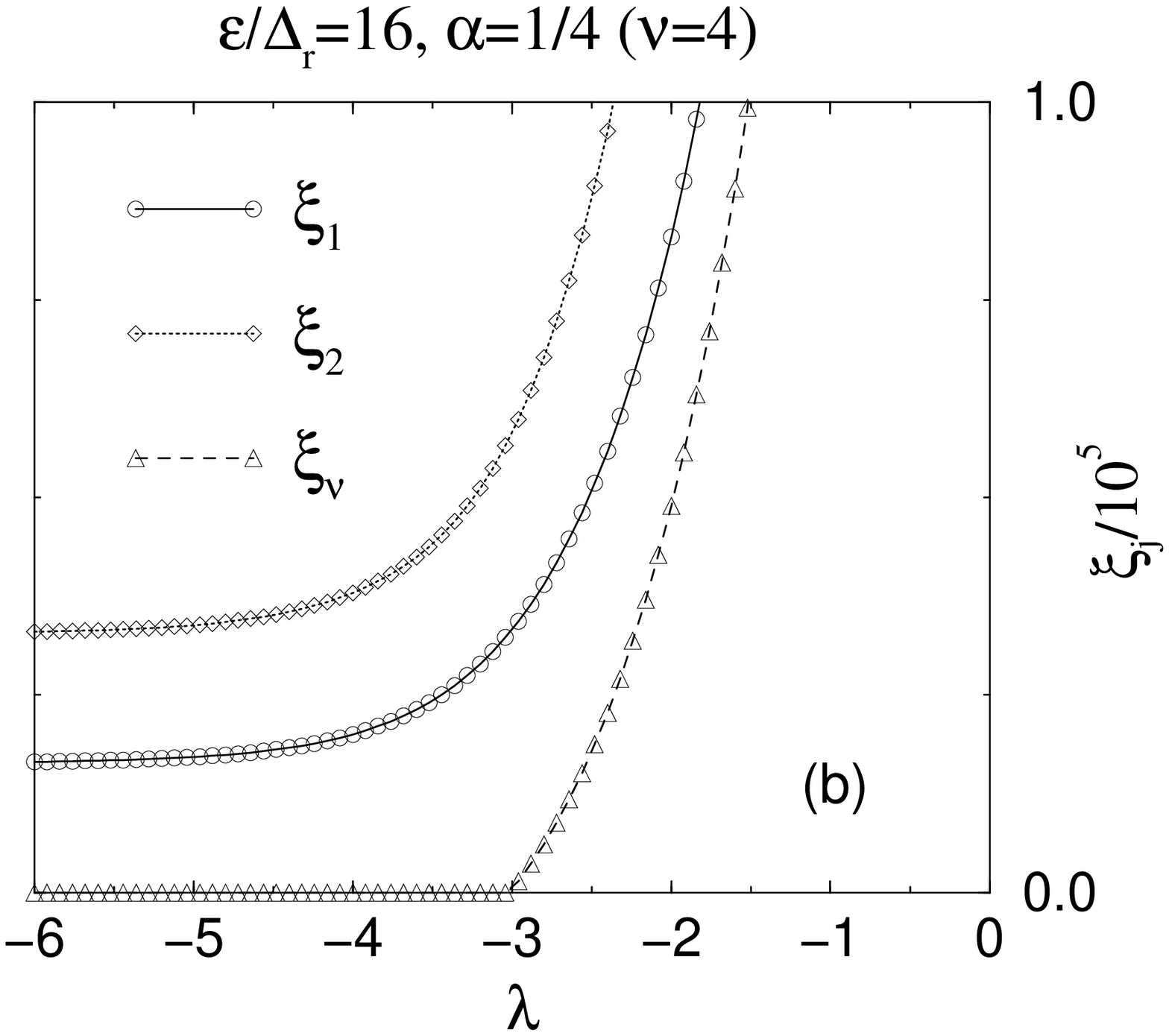}}
}
\vspace{0.1cm}
\centerline{
{\epsfysize 3.7cm \epsffile{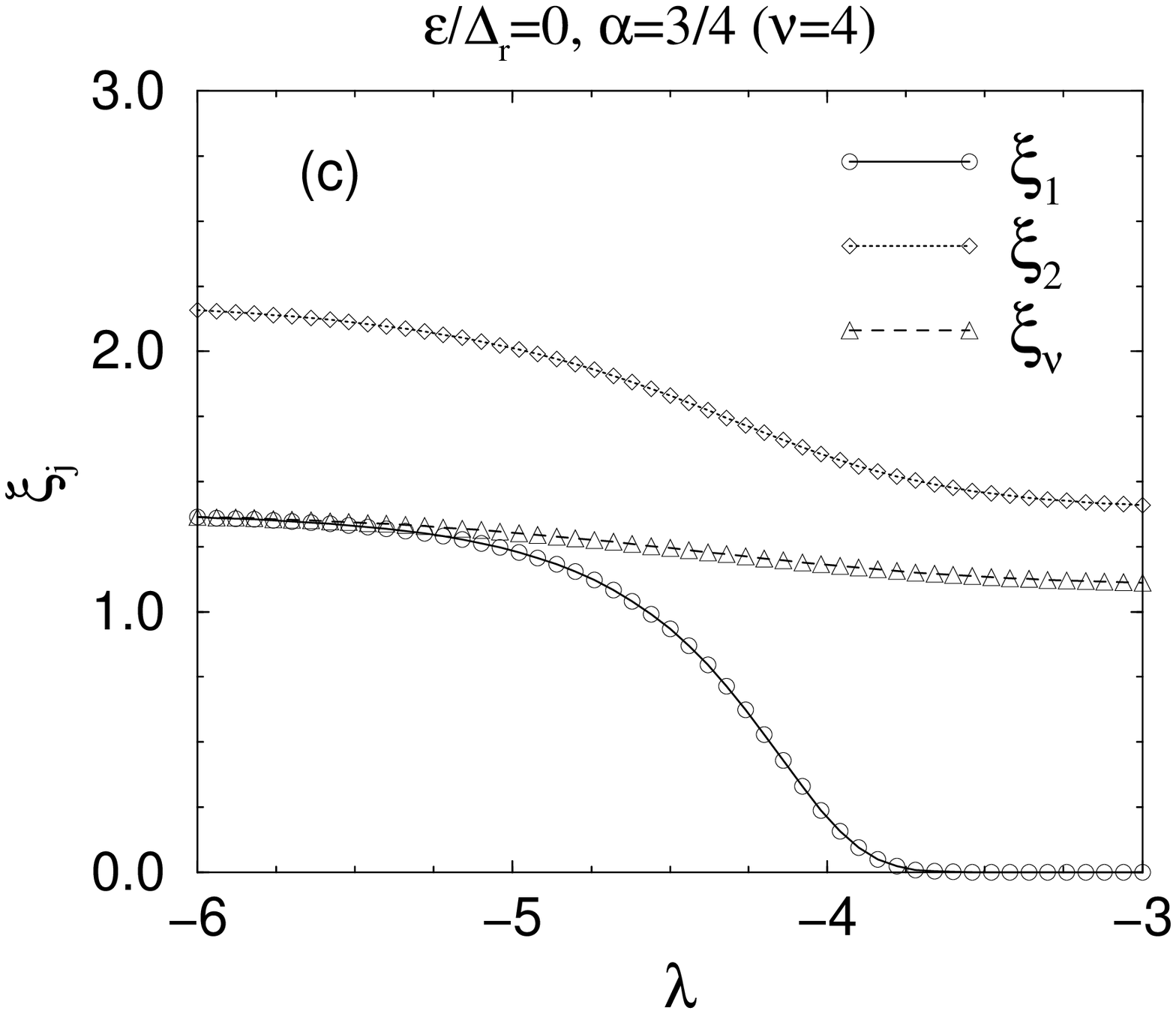}\epsfysize 3.7cm 
\epsffile{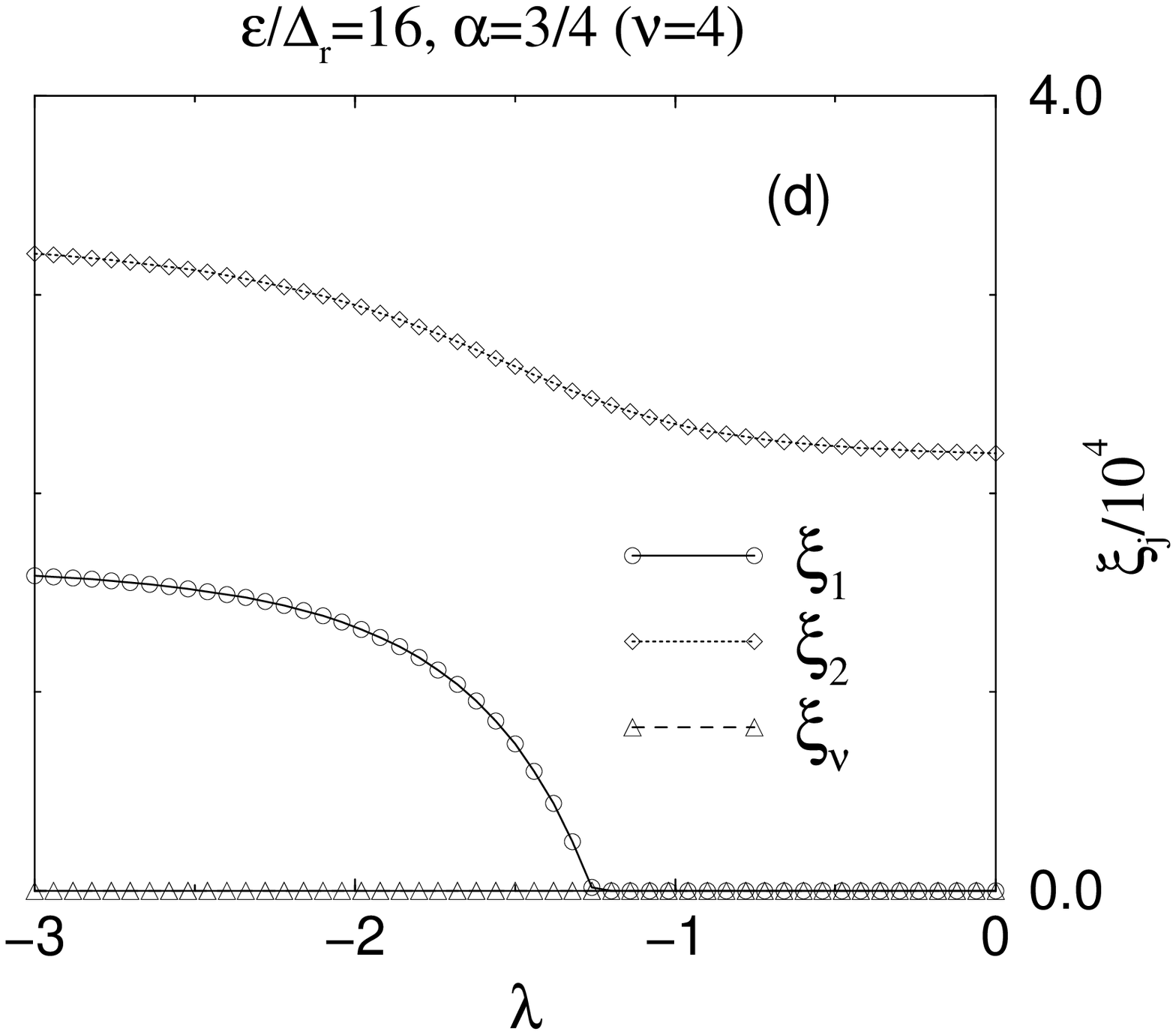}}
}
\vspace{0.1cm}
\caption{
Solutions to the TBA equations (\ref{eq:tba-ksi}) at $\alpha 
T/\Delta_{r}=2^{-10}$ for $\alpha=1/4<1/2$ (a) and (b) (top) 
and $\alpha=3/4>1/2$ (c) and (d) (bottom)
and for zero (left) and large (right) level asymmetry.
The exponential decay in $\xi_{1}$ ($\xi_{\nu}$) for strong (weak)
dissipation at large level asymmetries is clearly seen. 
}
\label{ksi-functions}
\end{figure}
A useful check on the correctness of
the solution of the TBA equations was the value of the functions 
$\xi_{j},E_{j},F_{j}$ and $G_{j}$ at $\lambda\rightarrow\pm\infty$. These
satisfy transcendental equations which can be solved analytically or
numerically\cite{tsvelik.83}. We checked that our numerical solution 
for the above functions satisfied these boundary conditions at all 
temperatures and level asymmetries.

In order to indicate the difficulty one encounters in implementing 
the above procedure for large asymmetries and low temperatures we show
some results in Fig.\,\ref{ksi-functions}. 
For zero asymmetry, the functions are well behaved
and have variations on scales $\lambda\sim 1$ so the interpolation scheme
used above, which is crucial for carrying out the integrations accurately
and ensuring the accuracy of the solution, is highly accurate. For large
$\varepsilon/T$ we see however that for $\alpha>1/2$ ($\alpha<1/2$) 
the function $\xi_{1}$ ($\xi_{\nu}$) decays exponentially  to a 
vanishingly small value above (below) a characteristic rapidity 
$\lambda = B$ within
an interval which vanishes as $x_{0}=g\varepsilon\nu/2T$ increases
\cite{note-exp-decay}. In
this case a uniform grid of points is inadequate and any interpolation
will give wrong results. One could increase the density of points
close to $B$, but a much simpler solution is not to solve explicitly for the
functions $\xi_{1}$ ($\xi_{\nu}$) but to solve instead for
the quantities upon which they depend on (see Eq.(\ref{eq:tba-ksi})), 
namely $s_f(\lambda)=s*\xi_{2}(\lambda)$ 
($s_f(\lambda)=s*\xi_{\nu-2}(\lambda)$) 
where $s*$ is the integral operator defined in 
Eq.(\ref{eq:integral-operator}). The latter functions
$s_f(\lambda)$ vary on a scale of $O(1)$ and can safely be interpolated. From
these one can then obtain $\xi_{1}$ ($\xi_{\nu}$) via 
its definition (\ref{eq:tba-ksi}).
The characteristic energy $B$ is determined at each iteration and
used to split up the integration ranges for those integrals containing
$\xi_{1}$ ($\xi_{\nu}$) so that the relative error of $10^{-5}$ for
each integration can be maintained. 

The integral equations for the $E_{j},F_{j}$ and $G_{j}$ 
(\ref{eq:ejba}--\ref{eq:gjba}), apart from being
linear inhomogeneous integral equations as opposed to the non-linear TBA
equations for the $\xi_{j}$, have an identical structure to those of 
the latter. As above, for large $x_{0}$ and 
$\alpha>1/2$ ($\alpha<1/2$), one has to solve explicitly not for the
$E_{1},F_{1},G_{1}$ ($E_{\nu},F_{\nu},G_{\nu}$), which vary rapidly
near $B$ as a result of the factors $(1-e^{-\xi_{1}})$ ($(1-e^{-\xi_{\nu}})$)
appearing in the corresponding integral equations 
(\ref{eq:ejba}--\ref{eq:gjba}), but for the quantities
$s*E_{2},s*F_{2},s*G_{2}$ ($s*E_{\nu-2},s*F_{\nu-2},s*G_{\nu-2}$) with
splitting of the integration range for integrals involving the former.
The same algorithm is therefore used as for the $\xi_{j}$. Since this turned
out to be a highly stable and accurate algorithm equally applicable without
changes to both sets of equation we did 
not consider the possibly simpler procedure of solving the linear
equations for the $E_{j},F_{j}$ and $G_{j}$ by Fourier transformation. 
This would involve numerically Fourier transforming the inhomogeneous terms, 
solving the resulting equations and then Fourier transforming the functions 
back to $\lambda$-space. Computationally, this would be more efficient, 
but may not be as accurate as the procedure we have used. 
In considering larger systems of coupled integral equations, 
$\nu\gg 1$, for which computational time is a factor, the more efficient
procedure may have to be implemented.

\begin{figure}[t]
\centerline{\epsfysize 6.1cm {\epsffile{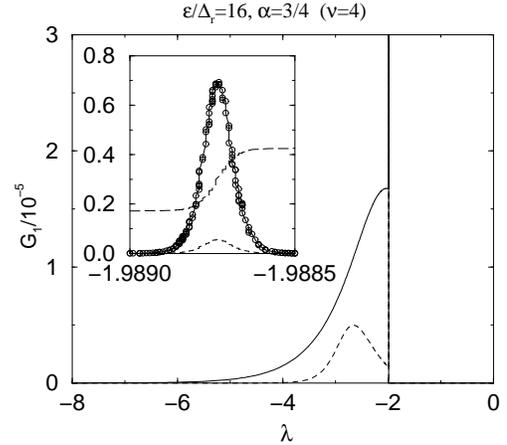}}}
\vspace{0.1cm}
\caption{
$G_{1}(\lambda,\varepsilon,T)$ (solid line), and the integrand 
$s(\lambda+f/\mu)G_{1}(\lambda,\varepsilon,T)$ (dashed line) of the 
expression for the entropy in Sec.IVa. 
The inset shows that the spike close to $\lambda=B\approx -2$, due
to the exponential decrease of $\xi_{1}$ for large 
level asymmetry and low temperature,
is a smooth structure on a small scale which is resolved by our numerical
procedure. The integrated weight $\tilde{S}(\lambda)=\int_{-40}^{\lambda}d\lambda'
s(\lambda'+f/\mu)G_{1}(\lambda',\varepsilon,T)$ (multiplied by 40000),
giving the entropy, is also shown in the inset 
(long dashed line). The contribution coming from the spike is seen to
be approximately the same as that coming from the broad feature.
}
\label{g1function}
\end{figure}

The above technique, which resolves the problem of the exponential decay
of certain functions to zero and thus allows the integrations at all
points to be carried out with uniform accuracy, gives a controlled way 
of solving the TBA equations. The thermodynamics is then calculated 
without taking any numerical derivatives. Integrating the functions  
$\xi_{1},E_{1},F_{1}$ and $G_{1}$, weighted with $s(\lambda+f/\mu)$ 
as described in Sec.IVa, gives the free energy, polarization, dielectric 
susceptibility and entropy respectively. Fig.\,\ref{g1function} 
illustrates how the above numerical difficulties were overcome for
$G_{1}(\lambda,\varepsilon,T)$ at low temperature and large level asymmetry. 

\section{Universal thermodynamic Bethe Ansatz equations for $\alpha<1/2$}
\label{wd-univ-eq}
To derive the universal Bethe Ansatz equations one has to write  
Eq.~(\ref{eq:BAepsilonint}) in a form where the cutoff $\omega_c$,
explicitly only present in the driving term Eq.~(\ref{eq:driving}),  
can safely be removed. A simple analysis of
Eq.~(\ref{eq:BAepsilonint}) shows that
for $\alpha>1/2$ in the $\lambda\to \infty$ limit 
 all the  $\epsilon_j$'s tend to a finite value determined 
by  $T$ and $\varepsilon$ (or $h$) except for  $\epsilon_1$ 
which approaches $\epsilon_1 \sim -\omega_c$ and thus 
diverges in the $\omega_c\to\infty$
limit. This divergence is, however, harmless, since  $\epsilon_1$
occurs  only in the form $\sim \exp(\epsilon_1(\lambda))$ at the 
 r.h.s. of Eq.~(\ref{eq:BAepsilonint}), and therefore all the 
'interaction terms' behave nicely  even in the $\omega_c\to
\infty$  limit.  

For $\alpha < 1/2$ the situation drastically changes
due to the sign change of the driving term. In this
case one can easily convince himself that  $\epsilon_j \sim \omega_c$ 
for $j=1,..,\nu-1$ while
$\epsilon_\nu \sim -\omega_c$ as $\lambda\to\infty$, i.e. the
interaction terms blow up as one tries to remove the cutoff. The basic
idea to obtain the universal equations is therefore to transform 
Eqs.~(\ref{eq:BAepsilonint}) into a form where only
$\exp(-\epsilon_j/T)$ ($j=1,..,\nu-1$) and $\exp(\epsilon_\nu/T)$
occur in the interaction terms. After a tedious calculation one
finally arrives at the following equations:
\begin{eqnarray}
\frac1T {\tilde g}_j & = &  \ln(1+e^{\epsilon_j/T}) -  
	\sum_{l=1}^{\nu-1} B_{jl} * \ln(1+e^{-\epsilon_l/T})
	\nonumber \\
&- & Q_j* \ln(1+e^{\epsilon_\nu/T})\;, \phantom{nn} (j=1,..,\nu-1)
\\
\frac1T {\tilde g}_\nu &=&    \epsilon_\nu/T +
	\sum_{l=j}^{\nu-1} Q_{j} * \ln(1+e^{-\epsilon_j/T})
	\nonumber \\
& + & K * \ln(1+e^{\epsilon_\nu/T})\;,
\end{eqnarray}
where the integral operators $K$, $Q_j$ and $B_{jl}$ are defined by 
Eqs.~6.2.27 and 6.2.6 of Ref.~\onlinecite{tsvelik.83}. (Note that the last one
of Eqs. 6.2.27 is written in Fourier space, and that  the
$a_j(\lambda)$'s occurring in the definition of $A_{jk}$ are the
functions given by the {\it first line} in Eq. 6.2.6 for
all indices, $j$. Furthermore, in  Eq.~6.2.6  the last  $\delta_{jk}$ 
term must be dropped in the  definition of $A_{jk}$.)

The driving terms, $\tilde g_j$ in the equations above are defined
as
\begin{eqnarray}
\tilde g_\nu & =&  -\tilde g_{\nu-1} = -\omega_c
{\rm artan}\;e^{\pi\lambda/(\nu-1)}\;, \\
  \tilde g_{j=1,..,\nu-2} & = & \omega_c \left[ {\pi\over 2}
+ {\rm artan}\Bigl\{ {\sinh\bigl({\pi \lambda\over \nu-1}\bigr)
\over \sin\bigl({\pi j \over 2(\nu-1)}\bigr)} \Bigr\}\right]\;,
\end{eqnarray}
where we corrected a factor of two with respect to Eqs. 6.2.28 of 
Ref.~\onlinecite{tsvelik.83}. Finally, approximating $\tilde g_j$ as 
$\tilde g_j = c_j \exp\{ \pi\lambda/(\nu - 1)\}$ and defining the 
dimensionless functions $\varphi_j(\lambda)\equiv 
\epsilon_j\bigl(\lambda + {\nu - 1\over \pi} \ln {T\over \omega_c}
\bigr)/T$ one arrives at the 
following cutoff-independent universal equations:
\begin{eqnarray}
d_{j}  & = & 
	 \ln (1+e^{\varphi_j})
	- \sum_{l=1}^{\nu-1} B_{jl} * \ln(1+e^{-\varphi_l})
	\nonumber \\
& - &  Q_j* \ln(1+e^{\varphi_\nu})\;, \phantom{nnn} (j=1,..,\nu-1)
\\
d_\nu  &=& \varphi_\nu  + 
	\sum_{l=j}^{\nu-1} Q_{j} * \ln(1+e^{-\varphi_j})
	\nonumber \\
& + & K * \ln(1+e^{\varphi_\nu})\;,
\end{eqnarray}
where the 'universal' driving terms are defined as
\begin{eqnarray}
&&d_{j=1,..,\nu-2}(\lambda) = 2
\left(\sin{\pi j\over 2(\nu-1)}\right)
\exp\left({\lambda \pi\over\nu-1}\right)\;,
\nonumber\\
&& d_\nu = -d_{\nu-1} =   - \exp\left({\lambda \pi\over\nu-1}\right)\;.
\nonumber
\end{eqnarray}
These universal equations, apart from some sign changes, coincide 
with Eqs.~6.2.29 of  Ref~\onlinecite{tsvelik.83}. Finally, after some 
manipulations the impurity contribution to the free energy can be 
written as 
\begin{eqnarray}
F^{(i)} = -{T\over\pi \omega_c}  \sum_{j=1}^\nu \int_{-\infty}^\infty v_j 
{d\tilde g_j (\lambda + {\nu -1 \over \pi}\ln{T\over T_K}) \over
d\lambda} \nonumber \\
\ln\bigl(1+ e^{-v_j \varphi_j(\lambda)}\bigr) d\lambda \; ,
\end{eqnarray}
where $v_j = +$ for $j=1,..,\nu-1$ and $v_{\nu} = -$.  
The Kondo temperature $T_K$ 
(renormalized tunneling  amplitude) emerges naturally in course of 
the calculations, and is given by
\begin{eqnarray}
T_K(\alpha<1/2) & = & 2D e^{- \nu f /  (\nu-1) }
\;\nonumber \\
&\equiv & \Delta_{r}/\alpha \;,
\end{eqnarray}
which is identical with  Eq.~(\ref{eq:T_K}).

\begin{figure}[b]
\centerline{\epsfysize 6.1cm {\epsffile{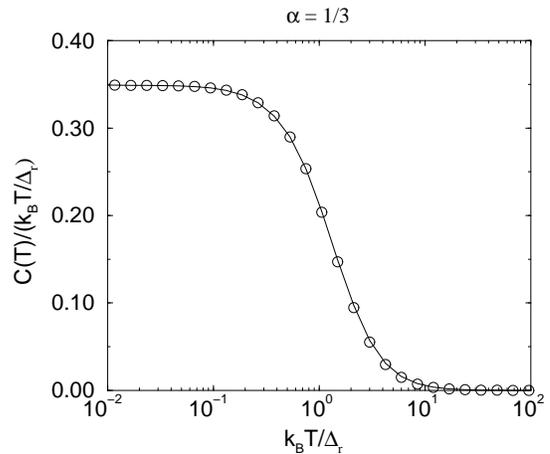}}}
\vspace{0.1cm}
\caption{
Specific heats, $C(T)/T$, for the symmetric two-state 
system ($\varepsilon=0$) at $\alpha=1/3$ from (a) the universal form of the
TBA equations (circles) and (b) the original untransformed TBA equations 
(solid line).
}
\label{comparison-uni+nuniv}
\end{figure}

We have checked that the solution of the above thermodynamic Bethe Ansatz
equations gives identical results for the specific heat and static 
susceptibility as the much simpler equations from which they are derived.
Fig.\,\ref{comparison-uni+nuniv} shows a comparison for the specific heat 
calculated using the two sets of equations.

\begin{references}
\bibitem{note-coherence} We define ``coherence'' to mean the presence
of oscillatory contributions to real time dynamical quantities, such
as $\Im \langle\sigma_{z}(t)\sigma_{z}(0)\rangle$ and
$\Re \langle\sigma_{z}(t)\sigma_{z}(0)\rangle$. This implies that the
system exhibits phase coherence in time.
\bibitem{note-exponential-decay} Certain correlations, such as 
$\Re \langle\sigma_{z}(t)\sigma_{z}(0)\rangle$, have long time $1/t^{2}$ 
tails at $T=0$ as their leading contribution. The exponential decay then
refers to the oscillatory contribution which is a subleading contribution
\bibitem{problematical-nrg-chi} This difficulty is due to the fact that
the local $S_{z}$ in the AKM is not conserved as opposed to total 
z-component of the spin. A discussion where this problem arises for 
similar quantities can be found in T. A. Costi, 
Phys. Rev. B {\bf 55}, 3003 (1997).
\bibitem{note1} Note, however, that in region $B$ $\Delta/\omega_{c}$ can be
irrelevant to start with for $\alpha_{0}>1$, but will eventually become
relevant once $\alpha$ decreases to $\alpha < 1$. This does not
happen in regions $A$ and $C$.
\bibitem{note-factor-alpha} The appearance of the factor of $\alpha$ 
relating the two scales $\Delta_{r}$ and $T_{K}$ stems from the 
introduction of electron-electron interactions into the AKM 
to ensure integrability within the Bethe Ansatz (without such
interactions we could simply write $\Delta_{r}=T_{K}$). 
This results in susceptibilities being renormalized
by a factor $1/\alpha$ within the Bethe Ansatz calculations (see
Sec.~\ref{Wilsonr} for details), so in order to relate the scales of the 
AKM and the Ohmic two-state system from the $T=0$ susceptibility of the 
Bethe Ansatz calculation one needs to include this factor.
\bibitem{note-SCREAM} We emphasize the distinction we are making between the
susceptibility of the AKM and the susceptibility of the AKM calculated from
the BA which involves modifications to the model in order to make it 
integrable. The distinction, as shown below, is important.
\bibitem{note-exp-decay} The reason for the exponential decay of $\xi_{1}$
($\xi_{\nu}$) for $\alpha>1/2$ ($\alpha<1/2$) is clear from the form
of $\xi_{1}$ ($\xi_{\nu}$) and the values which the functions $\xi_{j}$ take
at the boundaries $\lambda\rightarrow\pm\infty$. For definiteness consider
$\alpha>1/2$, then $\xi_{1}=\ln[1+\exp(D(\lambda)+s*\xi_{2}+
\delta_{1,\nu-2}s*\xi_{3})]$. It can be shown that the $\xi_{j}$ 
are monotonically decreasing
and for $\varepsilon/T\gg 1$ take on values at $\pm\infty$ of order 
$\varepsilon/T$. Due to the negative driving term in $\xi_{1}$ 
the sign of the term in the exponential will change at some 
$\lambda=\lambda_{0}$ from a very large positive value to a 
very large negative value thereby giving rise to the exponential 
decay in $\xi_{1}$. A similar reason holds for the function $\xi_{\nu}$
in the case $\alpha<1/2$ for $\varepsilon/T \gg 1$ (the large terms in the
exponential now being $x_{0}$ and $s*\xi_{\nu-2}$).
\bibitem{note-ct-peak} The appearance of a well defined oscillatory mode,
in the sense we have described, should be distinguished with the appearance
of tunneling oscillations or ``coherence'' which happens at $\alpha=1/2$
\cite{note-coherence}.
\end{references}
\end{document}